\documentclass[11pt,aps,prd,preprintnumbers,nofootinbib,showkeys,superscriptaddress]{revtex4-2}
\usepackage{float}
\usepackage{cancel}
\usepackage{amsfonts}
\usepackage{amsmath}
\usepackage{amssymb}
\usepackage{color}
\usepackage{tikz}
\usepackage{graphicx}
\usepackage{mathrsfs}
\usepackage{epsfig}
\usepackage{hyperref}
\usepackage{bm}
\usepackage{soul}
\usepackage{multirow}

\def\d{{\rm d}}
{}
{}

\def \nchi0{\widetilde\chi^0}

\newcommand{\newc}{\newcommand}
\newc{\ba}{\begin{array}}
\newc{\ea}{\end{array}}
\newc{\bea}{\begin{eqnarray}}
\newc{\eea}{\end{eqnarray}}
\newc{\beastar}{\begin{eqnarray*}}
\newc{\eeastar}{\end{eqnarray*}}
\newc{\beq}{\begin{equation}}
\newc{\eeq}{\end{equation}}
\newc{\bestar}{\begin{equation*}}
\newc{\eestar}{\end{equation*}}
\newc{\ben}{\begin{enumerate}}
\newc{\een}{\end{enumerate}}

\begin{document}

\title{Radiative Corrections to Aid the Direct Detection of the Higgsino-like Neutralino 
Dark Matter: Spin-Independent Interactions}

\author{Subhadip Bisal}
\email{subhadip.b@iopb.res.in}
\affiliation{ Institute of Physics, Sachivalaya Marg, Bhubaneswar, 751 005, India}

\affiliation{Homi Bhabha National Institute, Training School Complex, Anushakti Nagar, Mumbai 400 094, India}

\author{Arindam Chatterjee}
\email{arindam.chatterjee@snu.edu.in}
\affiliation{Shiv Nadar IoE Deemed to be University, Gautam Buddha Nagar, Uttar Pradesh, 201314, India}

\author{Debottam Das}
\email{debottam@iopb.res.in}
\affiliation{ Institute of Physics, Sachivalaya Marg, Bhubaneswar, 751 005, India}
\affiliation{Homi Bhabha National Institute, Training School Complex, Anushakti Nagar, Mumbai 400 094, India}

\author{Syed Adil Pasha}
\email{sp855@snu.edu.in}
\affiliation{Shiv Nadar IoE Deemed to be University, Gautam Buddha Nagar, Uttar Pradesh, 201314, India}


\begin{abstract}
The lightest neutralino ($\tilde{\chi}_1^0$) is a good Dark Matter (DM) candidate 
in the R-parity conserving Minimal Supersymmetric Standard Model (MSSM). In this 
work, we consider the light higgsino-like neutralino as the Lightest 
Stable Particle (LSP), thanks to a rather small higgsino mass parameter  $\mu$. We then 
estimate the prominent radiative corrections to the neutralino-neutralino-Higgs boson 
vertices. We show that for higgsino-like $\tilde{\chi}_1^0$, these corrections can
significantly influence the spin-independent direct detection cross-section, even  
contributing close to 100\% in certain regions of the parameter space. These 
corrections, therefore, play an important role in deducing constraints on the mass of the 
higgsino-like lightest neutralino DM, and thus the $\mu$ parameter.
\end{abstract}

\maketitle
\section{Introduction}
\label{sec:intro}
 
A prime motivation for supersymmetric extensions of the standard model of particle 
physics (SM) have been to address the ``naturalness" concerns. While there are several 
studies in literature to quantify ``naturalness" in a supersymmetric framework, the 
measure of naturalness is often debated \cite{Barbieri:1987fn, Ellis:1986yg, Chan:1997bi, 
Feng:2013pwa, Giudice:2013nak, Baer:2012cf, Mustafayev:2014lqa}. In the minimal supersymmetric 
extension of the standard model (MSSM), a small value of the higgsino mass parameter $\mu$ 
\cite{Barbieri:1987fn,Ellis:1986yg,Feng:2013pwa,Giudice:2013nak} and possibly rather light 
stop squarks and gluinos ($\lesssim 2-3$ TeV) \cite{Baer:2012up, Baer:2012cf, Baer:2013ava, 
Mustafayev:2014lqa, Bae:2019dgg} remain desirable in ``natural" scenarios at the 
electro-weak (EW) scale. 

At the Large Hadron Collider (LHC), the discovery of the Higgs boson has established 
the Standard Model (SM) physics. However, in spite of strong motivation 
for physics beyond the standard model, no hints for the same have been observed so far. The 
constraints on the supersymmetric spectrum \cite{ATLAS:2021yqv, ATLAS:2019wgx, ATLAS:2022zwa,ATLAS:2022hbt,ATLAS:2021moa,CMS:2022sfi,CMS:2023xlp}, while generally dependent on the nature of 
the low-lying states, have been raising concerns about the ``naturalness" 
requirements \cite{Baer:2012up, Baer:2012cf, Baer:2013ava, 
Mustafayev:2014lqa, Bae:2019dgg}. In this regard, within the minimal supersymmetric 
standard model (MSSM) paradigm, the constraint on the $\mu$ parameter is of particular interest. 
This has been widely studied in literature in the light of LHC.\footnote{Within 
the high scale supersymmetric models, a moderate value of the $\mu$ parameter may be 
realized  in the focus point region \cite{Feng:1999mn,Chan:1997bi,Baer:2003wx,Akula:2011jx,
Feldman:2011ud,Ross:2017kjc}.} In the minimal construct, a small higgsino mass parameter 
$\mu$  (of \(\mathscr{O}(100) \) GeV) is of relevance.\footnote{Note that, the fine-tuning 
measure, estimated following the ``electroweak" naturalness criteria, is stated to be about
$\mathcal{O}(10-100)$ assuming the masses of the third generation squarks and gluons  
in the ballpark of several TeV \cite{Baer:2012cf}.} Assuming that the gaugino mass 
parameters $M_1$ and $M_2$ are in the ballpark of multi-TeV, such a scenario leads to 
a compressed higgsino spectrum. In the R-parity conserving scenario, where the lightest 
supersymmetric particle (LSP) is stable. This leads to the higgsino-like lightest neutralino 
being a Dark Matter (DM) candidate. Such a scenario attracts rather weak constraints from the 
electroweakino searches at the LHC, as the decay of the next two heavier (higgsino-like) 
states lead to soft SM particles in the final states at the collider.  

Several studies have considered the prospects and constraints on the lightest neutralino 
in the light of collider and DM searches. Also, the implications of a compressed 
higgsino-like spectrum at the LHC have been studied widely \cite{Chakraborti:2015mra,
Chattopadhyay:2005mv, Baer:2011ec,Han:2013usa,Han:2014kaa, Drees:2015aeo, 
Barducci:2015ffa,Baer:2016usl,Fukuda:2017jmk,Mahbubani:2017gjh, Chakraborti:2017dpu,  
Delgado:2020url}. While the neutralino and chargino pair production cross-sections are 
sizable, especially for small $\mu$, the soft decay products in the compressed spectrum 
ensure that the constraints are rather weak. With higgsino-like neutralinos as DM candidates, 
a complimentary set of constraints on $\mu$ parameter \cite{Baer:2013vpa,
Chakraborti:2017dpu,Dessert} also follow from the direct \cite{LZ:2022lsv, XENON:2023cxc, 
PandaX-II:2020oim, PICO-60} and indirect \cite{MAGIC:2016xys,Fermi-LAT:2016uux,FermiLAT} 
searches for DM. 

In the present work, we revisit the implications of the spin-independent 
direct detection constraints on the higgsino-like ($\tilde{\chi}_1^0$). As the coupling 
of CP-even neutral Higgs bosons 
with a pair of $\tilde{\chi}_1^0$ is vanishingly small at the tree-level, the contribution to 
the spin-independent neutralino-nucleus interaction process is suppressed.\footnote{
We note in passing that for a mixed LSP, significant parts of the MSSM parameter 
space have been ruled out by direct directions constraints unless one hits the ``blind 
spot" \cite{Baer:2014eja,Huang:2014xua}; a similar situation arises if one tunes the 
Yukawa parameters leading to cancellation among different contributing processes 
\cite{Das:2020ozo}.}
Consequently, the radiative contributions to the scattering process need to be considered 
in order to accurately estimate the relevant cross-sections.  While such a scenario has been 
previously considered in the literature, in the context of pure-higgsinos, the importance 
of radiative corrections to the direct detection process has received some attention 
\cite{Drees:1996pk,Hisano:2004pv,Hisano:2011cs,Hisano:2012wm}.\footnote{
Certain classes of radiative corrections to the direct detection process and the relic 
abundance in the context of neutralino DM have been studied in 
refs.\cite{Baro:2007em, Baro:2009na, Chatterjee:2012hkk, Harz:2014gaa,
Harz:2014tma,Klasen:2016qyz, Harz:2023llw, Bisal:2023iip}.}

We improve the estimation of the radiative corrections to the spin-independent direct 
detection cross-section, by incorporating the contributions from the gauge bosons, the 
Higgs bosons, the respective superpartners and the third generation (s)quarks to the 
relevant vertices involving neutralino and Higgs bosons. Further, we renormalize the 
chargino-neutralino sector using the use the on-shell renormalization scheme and 
estimate the relevant vertex counterterms, thus paving the way towards a full 
one-loop treatment to the neutralino-Higgs boson vertices. 

In order to satisfy the thermal relic abundance of $\Omega_{\rm DM} h^2 \simeq 
0.12$, as required by cosmological considerations \cite{Aghanim:2015xee, Planck:2018vyg}, 
the higgsino-like neutralino LSP must be around 1 TeV and can be lowered further 
in the presence of co-annihilation \cite{Chakraborti:2017dpu}.
Below this mass scale, it is generally under abundant. However, there are viable non-thermal 
production scenarios, where adequate production of such DM may be possible in 
the early Universe \cite{Allahverdi:2012wb, Aparicio:2016qqb}. Further, the presence of 
additional DM components, e.g., axions, may contribute to the DM abundance 
\cite{Tegmark:2005dy, Baer:2011hx, Baer:2014eja}.  
In this work, we will not concern ourselves with satisfying the thermal relic abundance 
in the early Universe. We will only focus on the impact of certain radiative corrections 
on such a DM candidate in the light of direct DM searches. Note that if the LSP 
constitutes only a fraction of the required DM relic abundance (and, therefore, the 
local DM density),  the constraint on the DM-nucleon scattering cross-section from 
direct searches will be relaxed in the same proportion. 

This article is organized as follows: in \ref{sec:framework}, the chargino-neutralino spectrum 
of interest has been described, and the tree-level interactions between $\tilde{\chi}^0_1$ 
and the CP-even Higgs bosons are described. Following this, in \ref{sec:DD}, the generalities 
of (spin-independent direct detection of DM and the implications in the context of a 
higgsino-like DM candidate have been discussed. Subsequently, in \ref{sec:results}, we present 
the important electroweak radiative corrections to the vertices involving neutralino 
and the Higgs bosons and study its impact on the spin-independent DM-nucleon 
cross-sections. We also reflect on the parameter region where such 
corrections are significant and comment on the implications on 
the viable region for the higgsino mass parameter $\mu$. 
Finally, in \ref{sec:Conclusion}, we summarize the results and conclude.

\section{The Framework}
\label{sec:framework}
In this section, we briefly discuss the chargino-neutralino sector in the MSSM; 
in particular, we focus on the parameter region with rather small higgsino mass 
parameter $ \mu$ and light higgsino-like states.
\subsection{The Spectrum: Compressed Higgsinos} 
In the gauge eigenbasis expressed in terms of the Weyl spinors (the charged wino $(\widetilde{W}^{\pm})$, and charged higgsinos $(\tilde{h}_{i}^{\pm})$, for $i \in \{1,2\}$) with 
$\psi^+ = (\widetilde{W}^{+},~~\tilde{h}^{+}_{2})^{T}$ and $ \psi^- = (\widetilde{W}^{-},~~\tilde{h}^{-}_{1})^{T}$, the tree-level mass term for the charginos is given by 
\cite{Drees:2004jm}
\beq
-\mathscr{L}^{\rm c}_{\rm mass}  =  \psi^{-T} M^c \psi^{+} + h.c.
\eeq
The mass matrix $M^{\rm c}$ can be expressed as, 
\beq \label{mc}
M^{\rm c}= \left( \begin{array}{cc}
M_{2} & \sqrt{2} M_W\sin \beta \\
\sqrt{2} M_W \cos \beta & \mu \\
\end{array} \right).
\eeq
Here $M_2$ and $\mu$ stand for the supersymmetry breaking $SU(2)$ wino mass 
parameter and the supersymmetric higgsino mass parameter, respectively. $M_W$ is 
the mass of the $W$ boson, and $\tan\beta$ is the ratio of the $vevs$ of 
the up-type and the down-type CP-even neutral Higgs bosons. The matrix 
$M^{\rm c}$ can be diagonalized with a bi-unitary transformation using the 
unitary matrices $U $ and $V$ to obtain,
\beq \label{mcd}
M^{\rm c}_{D} = U^*M^{\rm c}V^{-1} = {\rm Diagonal} (m_{\tilde{\chi}_{1}^{+}}
~m_{\tilde{\chi}_{2}^{+}}) , 
\eeq
The eigenstates are ordered such that $m_{\tilde \chi_{1}^{+}} \leq m_{\tilde 
\chi_{2}^{+}}$. The left-- and right--handed components of these mass eigenstates, 
the charginos ($\tilde \chi^+_i$ with $i \in \{1,2\} $), are 
\beq \label{ec}
P_L \tilde \chi^{+}_i = V_{ij} \psi^+_j,~~ P_R \tilde \chi^+_i = U^*_{ij}
  \overline{\psi^-_j}\,,
\eeq
where $P_L$ and $P_R$ are the usual projectors, $\overline{\psi^-_j} = \psi^{- 
\dagger}_j$, and summation over $j$ is implied. 

For the neutralino states, in the gauge eigenbasis [consisting of the bino 
$(\widetilde{B}^{0})$, neutral wino $(\widetilde{W}^{3})$, and down-type and up-type 
neutral higgsinos ($\tilde h_1^0$ and $\tilde h_2^0$ respectively)], 
$\psi^{0} = \left( \begin{array}{cccc} \widetilde{B}^{0}, & 
\widetilde W^3, & \tilde h_1^0, & \tilde h_2^0 \end{array}\right)^{T}$, the 
mass term takes the following form \cite{Drees:2004jm}:
\beq
-\mathscr{L}^{\rm n}_{\rm mass}  =  \frac{1}{2} \psi^{0T} M^n
\psi^{0} + h.c.  
\eeq
The neutralino mass matrix $M^{\rm n}$  is given by,
\beq \label{mn}
M^{\rm n} =  \left( \begin{array}{cccc}
M_{1} & 0 & -M_{Z}s_W c_\beta & M_{Z}s_W s_\beta\\
0 &  M_{2} & M_{Z}c_W c_\beta & -M_{Z}c_W s_\beta \\
-M_{Z}s_W c_\beta & M_{Z}c_W c_\beta & 0 & -\mu \\ 
M_{Z}s_W s_\beta & -M_{Z}c_W s_\beta & -\mu & 0 \end{array} \right)\,.
\eeq
In the above equation $s_W, s_\beta, c_W$ and $c_\beta$ stand for $\sin \theta_W, 
\sin \beta, \cos \theta_W$ and $\cos \beta$ respectively while $\theta_W$ is the 
weak mixing angle. $M_Z$ is the mass of the $Z$ boson, and $M_1$ is the 
supersymmetry breaking $U(1)_Y$ gaugino (bino) mass parameter. $M^{\rm n}$ can be 
diagonalized by a unitary matrix $N$ to obtain the masses of the neutralinos 
as follows, 
\beq \label{mnd}
M^{\rm n}_{D}=N^* M^{\rm n} N^{-1}= {\rm  Diagonal} (m_{\tilde{\chi}^{0}_{1}} ~
m_{\tilde{\chi}^{0}_{2}} ~m_{\tilde{\chi}^{0}_{3}} ~m_{\tilde{\chi}^{0}_{4}})
\eeq
The eigenstates ($\tilde{\chi}_i^0$) are ordered according to the respective 
mass eigenvalues as follows,\footnote{Note that some of the eigenvalues may 
be negative depending on the input parameters. In such cases, the chiral rotation of the 
corresponding mass eigenstate may be performed to change the sign of the eigenvalue.}   
$m_{\tilde{\chi}^{0}_{1}} \leq m_{\tilde{\chi}^{0}_{2}} \leq m_{\tilde{\chi}^{0}_{3}} 
\leq m_{\tilde{\chi}^{0}_{4}}.$
These eigenstates satisfy $\tilde \chi_i^{0 c} = \tilde \chi_i^0$, where the superscript 
$c$ stands for charge conjugation. The left--handed components of these mass 
eigenstates, the Majorana neutralinos, $\tilde \chi_i^0$ ($i \in \{1,2,3,4\}$), 
may be obtained as,
\beq
P_L \tilde \chi^{0}_i = N_{ij}  \psi_j^{0},
\eeq
where summation over $j$ is again implied. 

The analytical expressions corresponding to the chargino and the neutralino mass 
eigenvalues have been obtained in the literature \cite{Choi:2001ww, Bertone:2004pz}. 
However, a numerical estimation of the eigenvalues is straightforward and convenient, 
especially in the case of the neutralinos. 

In the region of interest in this article, the higgsino mass parameter is rather 
small in comparison with the gaugino and the wino mass parameters, i.e., $|\mu| 
\ll |M_1|, M_2$; the masses of the light higgsino-like particles may be approximately 
given by \cite{Drees:1996pk,Giudice:1995np}\footnote{An exact analytical result may be 
found in refs. \cite{Barger1994,Kheishen}.} 
\bea
m_{\tilde\chi^{\pm}_1} & = & |\mu |\left(1- \frac{M_W^2 \sin2\beta}{\mu  M_2}\right) + 
\mathscr{O}(M_2^{-2})+ {\rm rad. corr.} \nonumber\\
m_{\tilde\chi^{0}_{a,s}} & = & \pm \mu - \frac{M_Z^2}{2}(1\pm \sin2\beta)
\left(\frac{\sin\theta_W^2}{M_1}+\frac{\cos \theta_W^2}{M_2}\right) + {\rm rad. corr.} 
\label{eq:mhiggsino}
\eea
In the above expression, subscripts $a$($s$) refer to anti-symmetric (symmetric) combinations 
of up-type $(\tilde h_2^0)$ and down-type $(\tilde h_1^0)$ higgsinos constituting the respective 
mass eigenstates. Here, the symmetric and anti-symmetric states refer to the higgsino-like 
states with compositions without and with a relative sign between $N_{i3}$ and $N_{i4}$ 
respectively. It has been pointed out in the literature that, due to the mixing effects, the mass 
differences $\Delta m_1 = m_{\tilde\chi^{\pm}_1} - m_{\tilde\chi^{0}_1} $ may become very small 
in certain regions of the parameter space \cite{Kribs:2008hq, Han:2014kaa, Barducci:2015ffa, Chatterjee:2017nyx}. In the present context, we will consider the mass differences 
$\Delta m_1, ~\Delta m_2  \gg \mathscr{O}(1~ {\rm MeV})$. Thus, as we will discuss in the 
next section, in the direct detection experiments, only the elastic scattering of 
$\tilde\chi^{0}_1$ with the nucleon will be relevant.

\subsection{Neutralino-Higgs boson(s) Interaction : Tree-level and at One-loop}
As we will elaborate in \ref{sec:DD}, for the spin-independent direct detection of $\tilde
\chi^{0}_1$, the relevant vertices involve the lightest neutralino and the CP-even Higgs 
bosons. The gauge symmetry of the MSSM, particularly the electroweak gauge group, 
prohibits any superpotential term with two higgsino states and a Higgs boson in the gauge 
eigenbasis. Therefore, the tree-level interaction term, in the gauge eigenbasis, involves 
one higgsino, one gaugino, and a Higgs boson. Consequently, in the mass eigenbasis, 
the tree-level vertex takes the following form \cite{Drees:2004jm}, 
\beq
\mathscr{L}  \supset   - \dfrac{1}{2} h_1 \bar{\tilde{\chi}}_1^0 (\mathscr{C}^{R}_1 P_R +  
\mathscr{C}^{L}_1 P_L) \tilde{\chi}_1^0   - \dfrac{1}{2}    h_2 \bar{\tilde{\chi}}_1^0
(\mathscr{C}^{R}_{2} P_R +  \mathscr{C}^{L}_{2} P_L) \tilde{\chi}_1^0
\label{eq:hchi_lgrng}
\eeq
where $\mathscr{C}^{L}_{i} = \mathscr{C}^{R*}_{i}$ for $i \in \{1,2\}$, and, 
\bea
\mathscr{C}^{R}_{1} & = &  (S_1 \sin\alpha + S_2 \cos\alpha),  \\
\mathscr{C}^{R}_{2} & = &  (S_2 \sin\alpha - S_1 \cos\alpha),  \\
S_{1} & = &  g_2 N_{13}(N_{12}-\tan\theta_W N_{11}),  \\
S_{2} & = &  g_2 N_{14}(N_{12}-\tan\theta_W N_{11})\,.
\label{eq:hchi_coupling}
\eea
In the above expressions, $h_1$  and $h_2$ denote the 125 GeV Higgs boson and 
the heavy Higgs boson mass eigenstates, respectively, $g_2$ denotes the $SU(2)$ 
gauge coupling,  and $\alpha$  denotes the mixing angle in the CP-even Higgs sector. 
Note that in the above equations, $N_{11}$ and $N_{12}$ denote the bino 
and wino composition of the lightest neutralino mass eigenstate, while $N_{13}^2$ and 
$N_{14}^2$ denote the respective down-type and up-type Higgsino fractions in the $\tilde{\chi}_1^0$ respectively. For higgsino-like $\tilde{\chi}_1^0$, the gaugino fraction is very small compared 
to the higgsino fraction, i.e., $|N_{13}|^2 + |N_{14}|^2 \gg |N_{11}|^2 + |N_{12}|^2$. Thus, the 
tree-level vertex involving the  CP-even Higgs bosons is suppressed by a rather small gaugino 
component of the mixing matrix (i.e., $N_{11}$ and $N_{12}$). As pointed out, the radiative corrections to 
these vertices play an important role in the spin-independent direct detection process;  
this will be illustrated in \ref{sec:results}. 

To treat the interaction Lagrangian at one-loop level,  the counterterm Lagrangian 
$\mathscr{L}_{CT}$ should be added to the tree-level Lagrangian $\mathscr{L}_{tree}$. The lagrangian, 
thus takes the following form : 
\beq
\mathscr{L} = \mathscr{L}_{Born} + \mathscr{L}_{CT}\, ,
\eeq
where $\mathscr{L}_{Born}$ is written using the renormalized fields, and $\mathscr{L}_{CT}$ 
involves the contributions from the relevant counterterms. The ``bare" and the renormalized  
neutralino mass eigenstates are related as follows 
\beq 
\tilde{\chi}_i^{0~bare} = \big(\delta_{ij} + \dfrac{1}{2} \delta {\rm Z}_{ij} P_L + \dfrac{1}{2} 
\delta {\rm Z}_{ij}^* P_R \big) \tilde{\chi}_j^{0~renormalized}, 
\eeq 
where the index $j$ has been summed over $j \in \{1,2,3,4\}$. The wave-function 
renormalization counterterms $\delta Z_{ij}$ are determined using the on-shell 
renormalization schemes \cite{Fritzsche:2002bi},  a comparison among different variants 
can be found in  ref.\cite{Chatterjee:2011wc}. Similarly, for the CP-even neutral Higgs
bosons the `bare" and the renormalized mass eigenstates are related as follows, 
\beq 
h_i^{bare} = \big(\delta_{ij} + \dfrac{1}{2} \delta {\rm Z}^{\rm H}_{ij} \big) h_{j}^{renormalized},
\eeq 
where the index $j$ has been summed over $j \in \{1,2\}$. The on-shell 
renormalization prescription is used to determine the wave-function renormalization 
counterterms $\delta Z^H_{ij}$.  With the above relations, the counterterm Lagrangian $
\mathscr{L}_{CT}$ relevant for the present discussion can be obtained as follows 
\begin{equation}
-\mathscr{L}_{CT}   \supset    \dfrac{1}{2}   h_1 \bar{\tilde{\chi}}_1^0 (\delta \mathscr{C}_1^R 
P_R + \delta \mathscr{C}_1^L P_L) \tilde{\chi}_1^0  + \dfrac{1}{2} h_2 \bar{\tilde{\chi}}_1^0 
(\delta \mathscr{C}_2^R P_R + \delta \mathscr{C}_2^L P_L) \tilde{\chi}_1^0 ,   \nonumber \\
\end{equation}
where the renormalized fields are used in the counterterm Lagrangian; we have 
dropped the respective superscript. 
At the one-loop level, the neutralinos, charginos, gauge bosons, and the Higgs bosons 
contribute to the $\tilde{\chi}_{1}^0, \tilde{\chi}_{1}^0, h_i$ vertices. Further, there can be 
sizable contributions from the third-generation (s)quarks, thanks to the sizable Yukawa 
couplings, as will be discussed in \ref{sec:results}. Note that for all our benchmark 
scenarios, as described in \ref{sec:results}, the lightest eigenvalue of the neutralino 
mass matrix $M^{\rm n}$ is positive.  Consequently, for the benchmarks presented 
in \ref{sec:results} (\ref{tab:bp1} and \ref{tab:bp2}),  $\mathscr{C}^{L}_i = 
\mathscr{C}^{R}_i$ and $\delta\mathscr{C}^{L}_i = \delta\mathscr{C}^{R}_i$ for $i \in \{1,2\}$.

\subsection{Constraints on the Parameter Space}
A small $\mu \ll |M_1|, M_2$, as discussed above, leads to a compressed spectrum with 
three closely spaced states $\tilde\chi^{0}_1, \tilde\chi^{0}_2, \tilde\chi^{\pm}_1$. The LHC 
sets stringent limits on chargino and neutralino masses, from pair productions of charginos 
and neutralinos and the subsequent decay of those to $\tilde\chi^{0}_1$ and SM particles. 
These limits are sensitive to the mass difference of the heavier chargino and neutralino 
states and the LSP ($\tilde\chi^{0}_1$). 
As for small mass splittings, the relevant bounds on the compressed spectrum can be found in 
ref.\cite{ATLAS:2019lng, ATLAS:2021moa,ATLAS:2022rme,CMS:2023mny}. For 300 (600) GeV
higgsino-like neutralinos, $\Delta m_1\lesssim 0.3~(0.2) $ GeV is disfavored \cite{CMS:2023mny}, 
and the constraints weaken for heavier mass. Searches targetting mass splittings around the 
electroweak scale may be found in ref.\cite{ATLAS:2021yqv, ATLAS:2019wgx, ATLAS:2022zwa,ATLAS:2022hbt,ATLAS:2021moa, CMS:2022sfi,CMS:2023xlp}, where decays of the 
heavier neutralinos into on-shell gauge bosons or Higgs bosons and the LSP, as well as 
their three-body decays have been considered. However, as $|M_1|,~M_2 \gg |\mu|$ in 
our context, these constraints are not very relevant to the present discussion. We have 
considered the following constraints on the spectrum for our benchmark scenarios 
presented in \ref{sec:results}.
\begin{itemize}
 \item We have constrained the lightest CP-even Higgs mass $m_h$ within the range : 
 $ 122 \leq m_{h} \text{ (GeV)} \leq 128$ \cite{Aad:2015zhl,Aad:2012tfa,
 Chatrchyan:2012xdj}. 
 Note that the experimental uncertainty is  about 0.25 GeV and the  
 uncertainty in the theoretical estimation of the Higgs mass is about $\pm 3$ GeV, 
 see e.g. \cite{Carena:2013qia} and references there. 
 \item The squarks and the sleptons masses have been assumed to be above 1.5 TeV, 
 and the gluino mass is kept above 2.2 TeV, respecting the constraints from the LHC.  
 
 \item In our scenario, with $|\mu| \ll |M_1|, M_{2}$, the low-lying higgsino-like states 
form a compressed spectrum. The $\mu$ parameter has been chosen such that 
LHC constraints on the compressed spectra are respected \cite{ATLAS:2021moa, 
ATLAS:2022rme, ATLAS:2019lng}. We have also used \texttt{SModelS} (version-2.3.0) 
\cite{Alguero:2021dig, Heisig:2018kfq, Dutta:2018ioj, Ambrogi:2017neo,Ambrogi:2018ujg,Sjostrand:2014zea,Sjostrand:2006za,Beenakker:1996ch,Buckley:2013jua} to check our benchmark scenarios.

 \item However, we have relaxed the constraints on the relic density of DM (i.e., $\Omega_{\rm DM} 
 \simeq 0.12$). As $\tilde{\chi}_1^0$ may not constitute all of the DM, the constraint from indirect 
 searches on the higgsino-like DM \cite{Fermi-LAT:2016uux, MAGIC:2016xys, Dessert} has also 
 been relaxed. 
\end{itemize}

\section{Direct detection of Dark Matter: Implications for a Higgsino-like LSP}
\label{sec:DD}
\subsection{Generalities of Direct Detection}
In this section, we describe the generalities of spin-independent direct detection and 
sketch the implications for the higgsino-like $\tilde{\chi}_1^0-$ nucleon scattering. 
  In the context of direct detection, the differential event-rate per unit time at 
a detector, as a function of the nuclear recoil energy $E_R$, is given by, 
\beq
\dfrac{d R}{dE_R} = n_T \dfrac{\rho_{\tilde{\chi}_1^0}}{m_{\tilde{\chi}_1^0}} \int_{v_{\rm min}}^{v_{\rm esc}} d^3 v f_E(\vec{v}) v \dfrac{d \sigma(v, E_R)}{d E_R}
\eeq
In the above equation, $\rho_{\tilde{\chi}_1^0}$ is the local density ($\simeq 0.3 {\rm ~GeV~ cm}^{-3}$), $n_T$ is the number of target nuclei in the detector and $\sigma(v, E_R)$ denotes 
the scattering cross-section with the nucleus. Further, $f_E(\vec{v})$ denotes the velocity 
distribution in the Earth's rest frame and $f_E(\vec{v})= f(\vec{v} + \vec{v}_E)$, where 
$f$ is the distribution function in the galactic rest frame and $\vec{v}_E$ is the velocity 
of Earth with respect to the galactic rest frame and $v_{\rm esc}$ is the escape velocity of 
our galaxy. Further, $v_{\rm min}^2 = \dfrac{m_T E_R}{2 M_{r}^2}$ is the minimum speed 
of the DM particle required to impart a recoil energy $E_R$,  where $m_T$ is the mass 
of the target nucleus, and $M_{r}$ is the reduced mass of  the DM-nucleus system. The 
cross-section with a nucleus (atomic number $A$ and charge $Z$) is given by, 
\beq
\dfrac{d \sigma}{d E_R}(v, E_R) =  \dfrac{m_T}{2 M_{r}^2 v^2} \sigma_0 F^2(q^2)
\eeq	
where $m_T$ is the mass of the target nucleus, $q^2 = 2 m_T E_R$ is the square of 
the momentum transfer, and $F$ stands for the form factor, which will be taken as
the Woods Saxon form factor \cite{Belanger:2008sj}. Further, $\sigma$ is the 
(spin-independent) DM-nucleus scattering cross-section. In the present context  
only spin-independent cross-section is relevant, which, at zero momentum transfer 
is given by $\sigma_0$. 
 
\subsection{Dark Matter-nucleon spin-independent  elastic scattering}		 
In the following, we briefly discuss the relevant parton level effective Lagrangian 
leading to the the spin-independent interaction \cite{Drees1},
\beq
\mathscr{L}_{\rm eff} \supset \lambda_q  \bar{\tilde{\chi}}^0_1\tilde{\chi}^0_1 \bar{q}q 
+ g_q \bar{\tilde{\chi}}^0_1 \gamma^{\mu} \partial^{\nu} \tilde{\chi}^0_1 (\bar{q} \gamma_{\mu} \partial_{\nu}q - \partial_{\mu}\bar{q} \gamma_{\nu} q) + \mathscr{L}_{\rm eff}^{g}
\label{eq:L_chi_q}
\eeq
In the above equation, the first term in the right-hand-side (RHS) receives contributions 
largely from the scattering processes mediated by the Higgs bosons. In particular, in the 
limit of no mixing in the squark sector, the contribution from the squark sector to this 
operator vanishes \cite{Drees1,Drees2}.  The next term captures the effect of squark-mediated $s-$channel scattering processes. Further, $\mathscr{L}_{\rm eff}^{g}$ denotes
the relevant effective interactions with gluons which contribute to the spin-independent 
neutralino nucleon scattering process \cite{Drees1, Drees2}.  

We now focus on the implications for a higgsino-like $\tilde{\chi}^0_1$, as we consider 
in the present context.  For such states, the higgsino fraction is much greater than the 
gaugino fraction  (i.e., $|N_{13}|,  |N_{14}| \gg |N_{11}|, |N_{12}|)$.  Therefore, the 
tree-level coupling involving two neutralinos and the Higgs bosons (see  \ref{eq:hchi_coupling}) 
are small, as these are suppressed by a factor of the gaugino component of the 
higgsino-like neutralino state. Note that, in the present context we consider 
$|M_1|, ~M_2 \lesssim 5 $ TeV and the higgsino mass parameter $|\mu| \lesssim 1$ TeV.  
Consequently, the gaugino fraction in the lightest neutralino is typically 
$\mathcal{O}(10^{-2})$. Thus, the tree-level Higgs boson exchange contributions, 
while small, can be significant and generally non-negligible. The tree-level contributions 
from the ($s-$channel) squark-mediated processes are suppressed by an additional factor of a 
rather small gaugino fraction in $\tilde{\chi}_1^0$ and/or an additional factor of 
Yukawa coupling for the first two generation (s)quarks as compared to the tree-level Higgs 
boson exchange processes. 
Further, we have considered the first two 
generations of squarks to be very heavy ($\gg \mathscr{O}(2)$) TeV for all our 
benchmark scenarios, as will be described in \ref{sec:results}. Therefore, 
contributions from the respective squark-mediated processes (and their contributions 
to the neutralino-gluon effective operators) remain sub-dominant in the present context. 
In the following, we first describe the Higgs exchange contribution, as the focus 
of the present study  is on the radiative corrections to the neutralino-Higgs boson 
vertices.

The effective parton-level interactions, as mentioned in (\ref{eq:L_chi_q}) 
leads to the following effective interaction Lagrangian with the  nucleon $N \in \{ n, p\}$,
where $n$ and $p$ stand for neutron and proton, respectively : 
\beq
\mathscr{L}_N^{\rm eff} \supset f_N  \bar{\tilde{\chi}}^0_1\tilde{\chi}^0_1 \bar{\psi}_N \psi_N,
\label{eq:L_chi_N}
\eeq
where $f_N$ denotes the effective coupling and $\psi_N$ denotes the field describing 
the nucleon $N$. The important contributions from the two CP-even neutral Higgs boson 
mediated processes in the spin-independent cross-section $ \sigma _{SI}$ comes 
from its contribution to the coefficient $\lambda_q$. The contribution  from the two 
CP-even neutral Higgs bosons $  \lambda_q^{H}$ is  given by, 
\beq
\lambda_q^{H} 
= \overset{2}{\underset{i=1}\Sigma}  \frac{C_{i} C_{i q}}{m_{h_i}^2}.
\label{eq:lam_chi_q}
\eeq
  In this 
expression, $C_{i} = \mathscr{C}_i^{L}$ as mentioned in \ref{eq:hchi_coupling} 
and $C_{i q}$ denotes the coupling of the same Higgs boson and quark($q$) 
\cite{Drees:2004jm}. The respective contribution to the spin-independent elastic 
scattering cross-section may be expressed in terms of their contribution to the 
effective interaction strength $f_N$ \cite{ELLIS1991,ELLIS1993,Drees1},
\begin{equation}
f_{N}^{(H)}= m_{N} \left( \overset{u,d,s} {\underset{q}{\Sigma}} f^N_{T_q}
\frac{\lambda^{H}_q}{m_q} + \frac{2}{27} \overset{c,b,t}{\underset{q}{\Sigma}} f^N_{T_G} \frac{\lambda^{H}_q}{m_q} + \frac{8\pi}{9\alpha_S}f^N_{T_G} m_N T_{\tilde{q}}\right); 
N \in \{p,n\}, 
\label{eq:higgs_nucleon}
\end{equation}
where \cite{SHIFMAN1978,Drees1}, 
\begin{equation}
f^N_{T_q} = \frac{1}{m_N} \langle N |m_q \bar{q}{q}| N\rangle,
~f^N_{T_G} = 1 - \overset{u,d,s}{\underset{q}{\Sigma}}f^N_{T_q}, 
~T_{\tilde{q}} = \dfrac{\alpha_S}{4 \pi}\dfrac{1}{24}  \overset{2} {\underset{i = 1} \Sigma} \dfrac{C_i}{m_{h_i}^2} {\underset{\tilde{q}_j}{\Sigma}} \dfrac{C^{i}_{\tilde{q}_j}}
{m_{\tilde{q}_j}^2} \, .
\end{equation}
In the above equation, $f^N_{T_q}$ denotes the contribution of the (light) quarks $q  \in \{u,d,s\}$ 
to the mass $m_N$ of the nucleon $N$.\footnote{We have used the scalar coefficients for the 
quark content in the nucleons as implemented in \texttt{micrOMEGAs} \cite{Belanger:2013oya}, 
where the mass ratios of the light quarks have been estimated using chiral perturbation theory, 
see, e.g., \cite{PDG}. For a discussion on hadronic uncertainties in the DM-nucleus 
scattering cross-section, see, e.g., \cite{Ellis:2008hf}. }

Further $\alpha_S = \dfrac{g_S^2}{4 \pi} $, $g_S$ 
denotes gauge coupling for the strong interaction; $m_{\tilde{q}_j}$ and $C^{i}_{\tilde{q}_j}$  
denote the mass of the $j$-th squark and its coupling  with the $i$-th (CP-even) neutral 
Higgs boson, respectively. The heavy quarks ($\{c,b,t\}$) contribute to $f_N$ through the 
loop-induced interactions with gluons. In \ref{eq:higgs_nucleon}, the first term 
includes a contribution from the effective neutralino-quark interactions; the second and the 
third term includes the contributions from the effective interaction with the gluon fraction. In 
particular, the second term (proportional to $f^N_{T_G} \frac{\lambda^{H}_q}{m_q}$)  and 
the third term (proportional to $f^N_{T_G} m_N T_{\tilde{q}}$) include  the relevant contributions 
from the heavy quarks and all the squarks to the Higgs bosons-gluon effective vertices 
respectively.

A brief discussion on various other important  contributions to the DM-nucleon 
scattering is in order. In addition to the contribution from the Higgs boson exchange 
processes, there are tree-level contributions to $f_N$ from squark exchange processes. 
As already mentioned, in the  present discussion we assume the (first two generations of) squarks to be very 
heavy. In such a scenario, the dominant contribution to the spin-independent neutralino-
nucleon interaction is mediated by the Higgs bosons. Further, the contribution to 
$f_{N}^{(H)}$ from the term proportional to $T_{\tilde{q}}$, which incorporates the 
squark contributions to the effective vertices involving Higgs bosons and gluons, is  
suppressed  for heavy squark masses. In the present context, the only the third generation 
squarks are relatively light, around $1.5$ TeV. Regarding other important radiative 
corrections, the supersymmetric-QCD corrections to the Higgs and the down-type 
quark vertices \cite{Djouadi:2000ck}; the one-loop corrections to the neutralino-gluon interactions 
originating from the triangle vertex corrections involving (s)quarks at the 
Higgs-gluon-gluon vertex, and also the  box-diagrams involving (s)quarks can be sizable 
\cite{Drees1}. These contributions have been implemented in the numerical 
package \texttt{micrOMEGAS} \cite{Belanger:2008sj} following 
ref.\cite{Drees1}.\footnote{Note that, for our benchmark scenarios, all the relevant 
tree-level processes, including the squark exchange processes, and the radiative 
corrections mentioned above, 
have also been considered; we have used $\texttt{micrOMEGAs}$ for the same. However, 
the percentage change in the scattering cross-section has been estimated for the vertex corrections to $C_i$. For the radiative corrections implemented, as described above, 
we have checked that the contributions are about 10\% to the  
spin-independent cross-section for the benchmark scenarios.}  Further, contributions from 
the box diagrams to the DM-quark scattering involving electroweak gauge bosons have 
been considered in the literature \cite{Hisano:2004pv, Hisano:2011cs, Hisano:2012wm}. 
In ref.\cite{Hisano:2012wm}, it has been shown that the tree-level Higgs boson 
exchange contribution to the (higgsino-like) neutralino-nucleon scattering dominates 
over these contributions when the gaugino mass parameters are less than 
$\mathcal{O}(5-10)$ TeV, as is relevant in the present context.

As is evident from the discussion above, the interaction rate is proportional to 
$f_N^2$, which involves the square of the LSP-Higgs bosons vertices $C_i$. As discussed 
above, several dominant one-loop contributions to the scattering process have been  
estimated and incorporated in the publicly available packages, e.g., \texttt{micrOMEGAs} 
\cite{Belanger:2004yn,Belanger:2008sj,Alguero:2022inz}. However, a detailed 
estimation of the one-loop corrections from the modification of the neutralino-Higgs 
boson vertices $C_i$ have not received adequate attention.\footnote{Certain subsets 
of the diagrams have been considered in ref. \cite{Drees:1996pk, Hisano:2004pv,
Hisano:2011cs, Hisano:2012wm} in the limit of a pure higgsino or higgsino-like 
neutralino DM; a full calculation of the vertex corrections, involving the respective 
counterterms, is not available in the literature to our knowledge.} In particular, 
for an almost pure higgsino-like LSP, which is often relevant for a natural 
supersymmetric spectrum, the small (but generally non-vanishing) gaugino 
fractions imply that the tree-level value of $C_i$ to be small. Therefore, 
radiative corrections to the same vertices can play a crucial role in the estimation 
of the cross-section. 
\begin{figure}[h!]
	\centering
	\includegraphics[scale = 0.5]{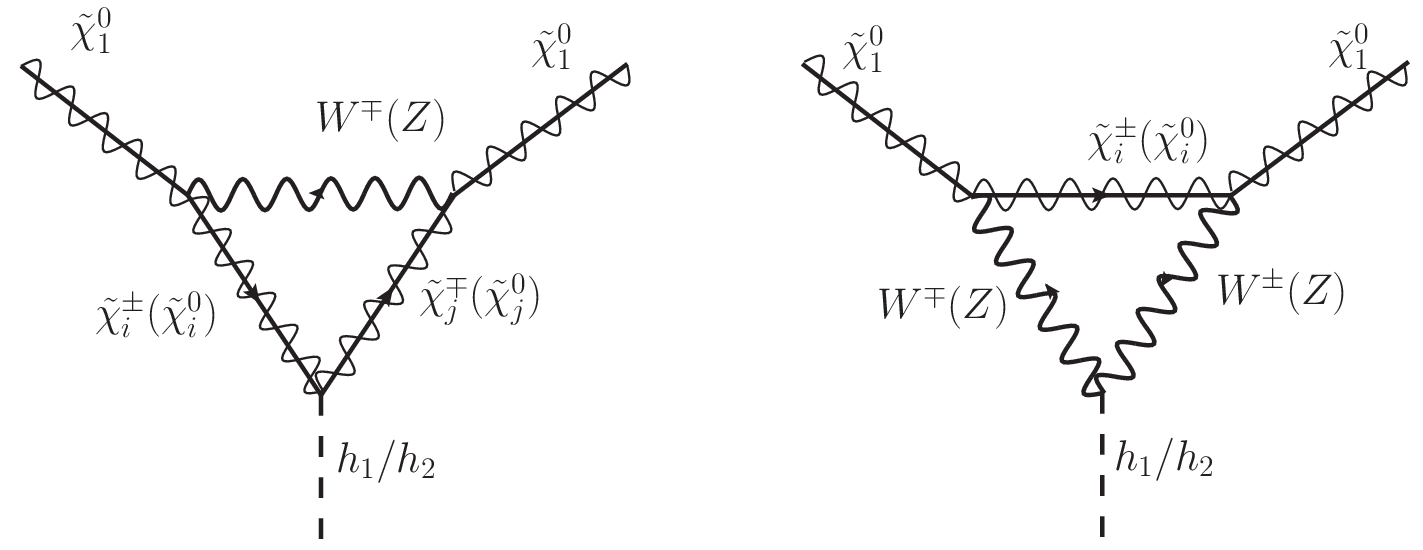}\\
	\includegraphics[scale = 0.5]{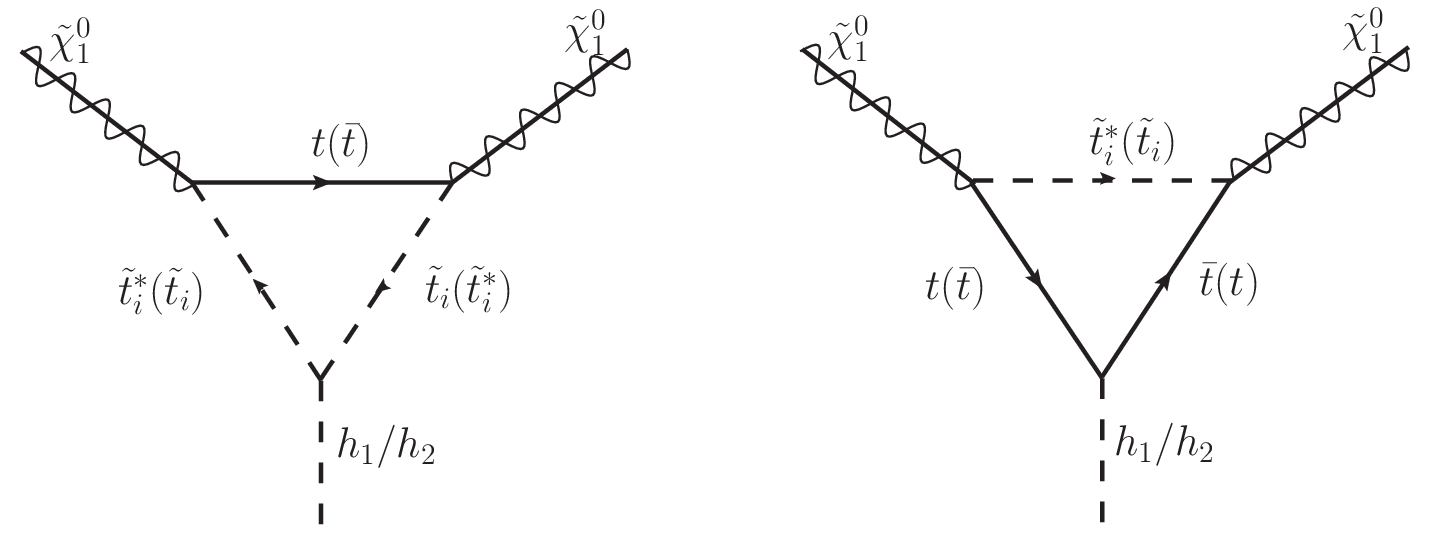}\\
 \caption{Some important contributions to the $\tilde{\chi}_1^0-\tilde{\chi}_1^0-h_i$ vertices at 
 one-loop level.}
	\label{fig:h_peng}
\end{figure}
In \ref{fig:h_peng}, some of the important diagrams contributing to the vertex correction 
have been depicted.  We consider all the triangle diagrams involving  
charginos, neutralinos, gauge bosons and Higgs bosons which contribute to the vertex 
corrections to the $\tilde{\chi}_1-\tilde{\chi}_1-h_i$ vertices. Further, as the Yukawa couplings 
for the third generation (s)quarks are large, contributions from the third generation (s)quarks 
have also been considerd. As the loop diagrams with two fermions and one boson are generally 
Ultra-Violet (UV) divergent, we have included the vertex counterterms, and ensured 
the UV finiteness of the overall contributions. Note that the wave function 
renormalization counterterms also include the effect of mixing of the tree-level fields 
(due to radiative corrections from the two-point functions) appearing in the external lines. 
The complete set of radiative contributions considered in this work have been described 
in Appendix B, and the counterterms have been mentioned in Appendix C.

\section{Results}
\label{sec:results}
In this section, we present the results highlighting the importance of the radiative 
corrections to the vertices involving neutralino and Higgs bosons, as discussed in  \ref{sec:DD}, to the (spin-independent) direct detection process in the context
of a higgsino-like $\tilde{\chi}_1^0$.  
\subsection{Implementation}
\begin{figure}[h!]
	\centering
	\includegraphics[width=0.8\textwidth]{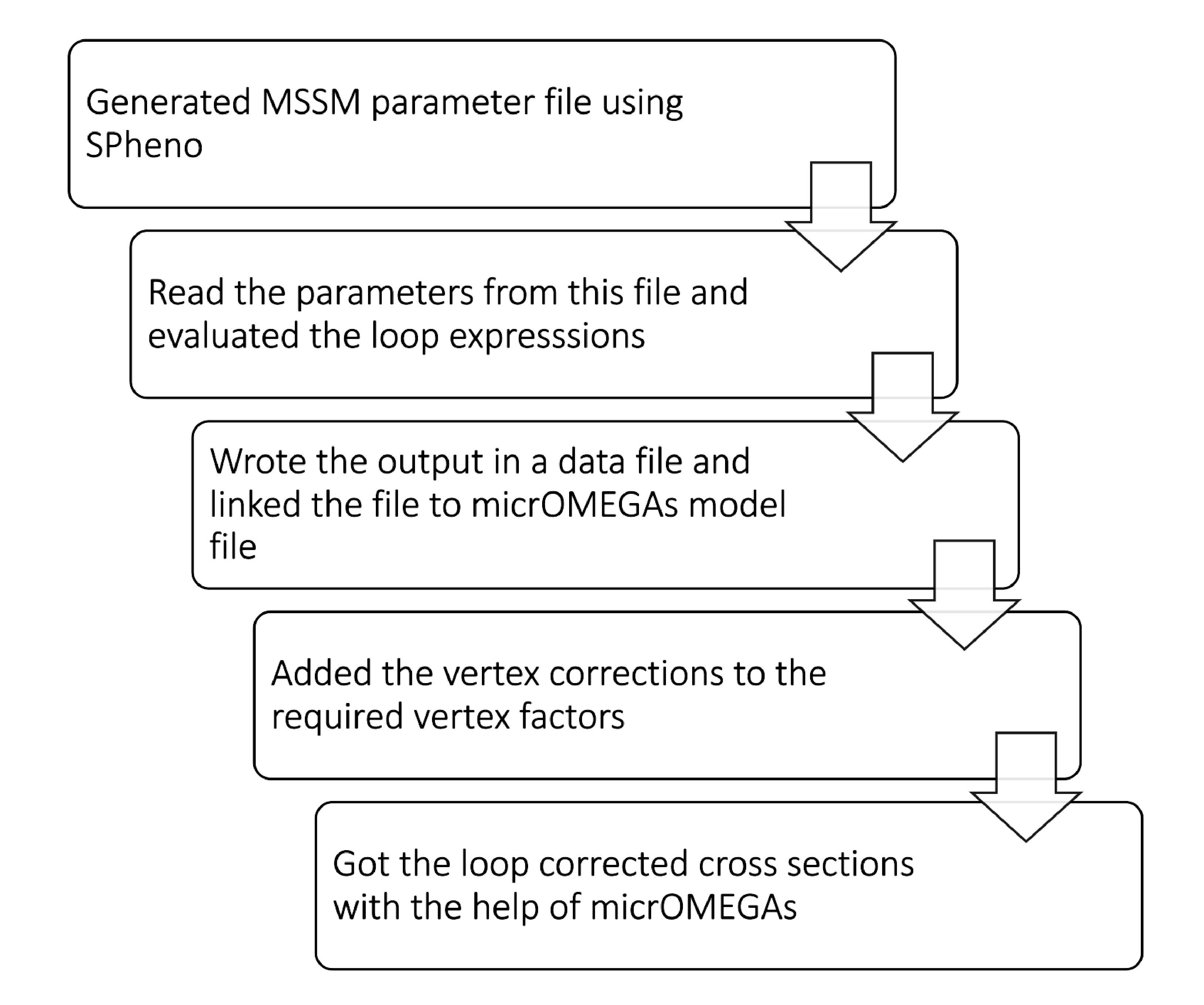}
	\caption{The flow-chart for implementation of the relevant corrections to the 
	neutralino-Higgs boson(s) vertices.}
	\label{fig:flowchart}
\end{figure}

We begin by describing the procedure to compute the radiative corrections.
The steps have been sketched in the flowchart shown in \ref{fig:flowchart}.
\begin{itemize}
    \item We  generated the benchmark scenarios using the spectrum generator 
     \texttt{SPheno} \cite{Porod:2003um} (version 4.0.4). The parameters are read from the output file and 
    the relevant radiative corrections are numerically evaluated using those parameters. 
    The input parameters $M_1$, $M_2$ and $\mu$ in the chargino-neutralino sector 
    are varied to obtain different benchmark scenarios, as presented in \ref{tab:bp1} and 
    \ref{tab:bp2}. The details of the benchmark scenarios will be discussed in the next subsection. 
    
    \item To evaluate the radiative  corrections to the $\tilde{\chi}_1^0-\tilde{\chi}_1^0-
    h_i$ vertices, we have used  the publicly available packages \texttt{FeynArts} (version 3.11)
    \cite{KUBLBECK1990165, HAHN2001418}, \texttt{FormCalc} (version 9.10) 
    \cite{HAHN1999153}, and \texttt{LoopTools} (version 2.15) \cite{HAHN1999153}. 
    In particular, the Feynman diagrams are evaluated using \texttt{FeynArts}, and the  
    vertex corrections are calculated using \texttt{FormCalc}. Further, the radiative 
    contributions are expressed in terms of the Passarino-Veltman integrals (briefly discussed in Appendix A) and numerically  
    evaluated for the benchmark scenarios using \texttt{FormCalc} and \texttt{LoopTools}. 
    Further, the UV finiteness of the radiatively corrected vertex factors (including the counter-term contributions) have been numerically checked using the packages mentioned above.
    Finally, the numerical results are stored in data files. To determine the relevant
    counterterms, we have used the on-shell renormalization scheme. We have used the 
    relevant counterterms implemented in \texttt{FormCalc}, as described in 
    ref. \cite{Fritzsche:2013fta}.
   \item To evaluate the direct detection cross-sections \texttt{micrOMEGAs} \cite{Belanger:2008sj,Belanger:2004yn, Belanger:2006is, Belanger:2010pz} (version 5.2.1)
    \cite{Alguero:2022inz} has been used. We generated the model files for \texttt{micrOMEGAs} 
    using \texttt{SARAH} (version 4.14.5) package \cite{STAUB20141773} on the Mathematica platform. 
    We modify the relevant vertices in the code to 
    include the radiatively corrected vertices. For each benchmark, then, the corrected vertices 
    are read from the output file, as described above, using a subroutine. Thus, the radiatively 
   corrected vertices are used to evaluate the spin-independent direct detection cross-section. 
   \end{itemize}
      
\subsection{Benchmark Scenarios} 
In this subsection, the benchmark scenarios have been discussed. 
The benchmark points have been described in \ref{tab:bp1} and \ref{tab:bp2}. 

As we focus on the higgsino-like $\tilde{\chi}_1^0$ DM, $|\mu| \ll |M_1|, M_2$ have 
been set for all the benchmark scenarios. The tree-level vertices $\mathscr{C}_i^{L/R}$, 
as described in \ref{eq:hchi_coupling}, are proportional to the product 
of the gaugino and higgsino components of $\tilde{\chi}_1^0$. Thus, the tree-level
spin-independent cross-section ($\sigma_{SI}$) is sensitive to the variation in the 
gaugino-higgsino mixing. With $|M_1|, ~M_2,$ and $|\mu|$ fixed, the gaugino-higgsino 
mixing is sensitive to the signs of $M_1$ and $\mu$.  Consequently, the tree-level spin-independent 
cross-sections and the relative contributions from the radiative corrections to the 
$\tilde{\chi}_1^0 - \tilde{\chi}_1^0 -h_i$ vertex factors can be very different even for 
very similar chargino-neutralino masses. In the benchmark scenarios, with $|\mu| \ll |M_1|, 
~M_2$; we have varied the sign of $\mu$ and $M_1$  to illustrate this variation. 
Further, the order of $M_1$ and $M_2$ have been altered to study the effect of the 
variation in the gaugino components. 
 
The benchmark points BP-1a to BP-6a, as shown in \ref{tab:bp1} reflect scenarios 
with $|\mu| =  300$ GeV. Setting $|M_1|, M_2 \gg |\mu|$ ensures that $\tilde{\chi}_1^0, 
\tilde{\chi}_2^0, \tilde{\chi}_1^{\pm}$ are closely spaced and are higgsino-like states. 
For the benchmark scenarios BP-1b to BP-6b, as shown in \ref{tab:bp2}, a heavier $|\mu| 
= 600$ GeV has been considered.  As discussed above, to illustrate the variation in the 
gaugino components of $\tilde{\chi}_1^0$ for very similar particle spectra, the sign of $\mu$ 
and the sign of $M_1$ have been varied. For BP-1a and 2a,  with $\mu = 300$ GeV and 
$\mu = -300$ GeV respectively, $M_1$ is set to  $-5$ TeV. For BP-3a and 4a,  with 
$\mu = 300$ GeV and $\mu = -300$ GeV respectively, $M_1$ is set to $5$ TeV. For all 
these benchmark scenarios, we fix $M_2 = 4$ TeV. For BP-5a and 6a, with $\mu = 300$ 
GeV and $M_2 = 5$ TeV, while $M_1$ assumes $-4$ TeV and $4$ TeV respectively. 
BP-1b to BP-6b resembles BP-1a to BP-6a respectively, only with $|\mu| = 600$ GeV. 
Note that for BP-1 to BP-4 (a and b), $|M_1| > M_2$, while for BP-5 and BP-6 (a and b) 
$|M_1| < M_2$. For all these benchmark scenarios $\tan\beta = 10$, the masses of 
the Higgs bosons and the third-generation squarks, which are also relevant for the present study, have been kept fixed. Further, constraints from LHC on such compressed spectra have been taken into account. 

\begin{table}
\begin{center}
\begin{tabular}{|c|c|c|c|c|c|c|}
 \hline
 Parameters &  BP1a  & BP2a & BP3a & BP4a & BP5a & BP6a \\
  \hline
$\mu$ (GeV) & 300 & -300 & 300 & -300 & 300 & 300 \\
  \hline
 $M_1$ (GeV) &  -5000 & -5000 & 5000 & 5000 & -4000 & 4000 \\
\hline
 $M_2$ (GeV) &  4000 &  4000 & 4000 & 4000 & 5000 & 5000 \\
\hline
 $m_{\tilde{\chi}^0_1}$ (GeV)  & 299.17  &    299.44  &  298.72 & 299.14 & 299.44 & 298.88  \\
 \hline
 $m_{\tilde{\chi}^0_2}$ (GeV)  & -300.44  &   -300.66  &  -300.74 & -301.11 & -300.29 & -300.66  \\
\hline
 $m_{\tilde{\chi}^0_3}$ (GeV)  & 4000  &   4000  &  4000 & 4000 & -4000 & 4000  \\
\hline
 $m_{\tilde{\chi}^0_4}$ (GeV)  & -5000  &  -5000  &  5000 & 5000 & 5000 & 5000  \\
\hline
 $m_{\tilde{\chi}^{\pm}_1}$ (GeV)  & 299.56  &  300.2  &  299.56 & 300.2 & 299.67 & 299.67  \\
\hline
 $m_{\tilde{\chi}^{\pm}_2}$ (GeV)  & 4000  &  4000  &  4000 & 4000 & 5000 & 5000  \\
 \hline
$m_{h_1}$  (GeV)  &   122.92  &  122.79  & 122.73  &  122.61  &   122.81  &   122.65   \\
 \hline
$m_{h_2}$  (GeV)  &  1386   &  1468  &  1407  &  1448  &  1425   &   1450   \\
 \hline
 \hline
 HF  & 0.9997  &    0.9998  &  0.9997 & 0.9998 & 0.9998 & 0.9998  \\
\hline
 $N_{11} (\times 10^{\textrm{-}3})$  & -$6.291$  &  -$5.145$  &  $7.087$ & $5.795$ & -$7.756$ & -$9.004$ \\
\hline
 $N_{12} (\times 10^{\textrm{-}2})$  & -$1.679 $ &  -$1.373$  &  -$1.677$ & -$1.372$ & -$1.322$ & $1.321$ \\
\hline
$N_{13}$ & 0.708 &  -0.707 & 0.708 & -0.708 & 0.708 & -0.708 \\
\hline
$N_{14}$ & -0.706 & -0.706 & -0.706 & -0.706 & -0.706 & 0.706 \\
\hline
\end{tabular} 

\caption{The benchmark scenarios with a higgsino-like $\tilde{\chi}^0_1$  have been tabulated for $|\mu|= 300$ GeV. HF stands for higgsino fraction. 
The fixed input parameters are: the mass of the pseudoscalar Higgs boson $m_{A} = 1.414$ TeV, and $\tan{\beta}=10$. The gluino mass parameter $M_3=3$ TeV. The trilinear coupling for two stops with the Higgs boson is set as $T_t=-3$ TeV. The soft-supersymmetry-breaking mass parameters for the left-type and the right-type stop and sbottom squarks are as follows: $m_{\tilde{Q}_L}=2.69$ TeV, $m_{\tilde{t}_R} = 2.06$ TeV and $m_{\tilde{b}_R}=2.50$ TeV. As for the physical masses, the charged 
Higgs boson mass $M_{H^{\pm}}= 1.416$ TeV, the CP-even Higgs mixing angle $ \alpha = \sin^{-1}
(-0.1)$.  For all the benchmarks, the third generation squark mass and mixing parameters are taken as:  the lightest stop mass $ {\rm m}_{\tilde{t}_1} = 2.05$ TeV, the heaviest stop mass 
${\rm m}_{\tilde{t}_2} = 2.71$ TeV, the lightest sbottom mass $ {\rm m}_{\tilde{b}_1} = 2.50$ TeV, the heaviest sbottom mass  ${\rm m}_{\tilde{b}_2}= 2.69$ TeV. }
\label{tab:bp1}
\end{center}
\end{table}

For all the benchmark scenarios, $\tilde{\chi}_1^0$ corresponds to a mass 
eigenstate with positive eigenvalue. In the benchmark scenarios with a negative 
$\mu$ parameter, i.e. BP-2a, BP-2b, BP-4a and BP-4b, $\tilde{\chi}_1^0$ is the 
symmetric higgsino-like state. For all other benchmarks (with positive $\mu$ parameter) 
$\tilde{\chi}_1^0$ is the antisymmetric higgsino-like state. Note that, irrespective of the 
sign of $M_1$, the gaugino components in $\tilde{\chi}_1^0$ are reduced substantially 
for negative $\mu$ (where $\tilde{\chi}_1^0$ is the symmetric state) as compared to 
positive $\mu$ (where $\tilde{\chi}_1^0$ is the symmetric state). This is evident from 
comparing the wino and the bino components ($N_{12},~N_{11}$ respectively) of 
$\tilde{\chi}_1^0$ in BP-1a(b) and BP-2a(b) respectively. In particular, the wino component 
is reduced by approximately 50\% and 25\% for benchmarks BP-1a(b) and BP-2a(b), respectively. 
The bino content, which contributes subdominantly, follows a similar trend, although 
by a smaller margin. As the tree-level $\tilde{\chi}_1^0-  \tilde{\chi}_1^0-h_i $ vertices 
are directly proportional to the gaugino fraction, the change in sign of the higgsino mass parameter $\mu$ 
leads to a significant change in the tree-level spin-independent 
direct detection cross-section. 

\begin{table}
\begin{center}
\begin{tabular}{|c|c|c|c|c|c|c|}
  \hline
 Parameters &  BP1b & BP2b & BP3b  & BP4b & BP5b & BP6b \\
 \hline
 $\mu$ (GeV) & 600  & -600 &  600  & -600 & 600  & 600  \\
 \hline
 $M_1$ (GeV) &  -5000 & -5000 & 5000 & 5000 & -4000 & 4000 \\
\hline
 $M_2$ (GeV) &  4000 & 4000 & 4000 &  4000 & 5000 & 5000 \\
\hline
 $m_{\tilde{\chi}^0_1}$ (GeV)  & 599.06   & 599.37 &  598.61   &  599.07  & 599.36  & 598.79  \\
  \hline
 $m_{\tilde{\chi}^0_2}$ (GeV)  & -600.39  & -600.59 & -600.7   &  -601.04 & -600.24 & -600.62  \\
\hline
 $m_{\tilde{\chi}^0_3}$ (GeV)  & 4000      & 4000  &  4000      &  4000     &   -4000    &  4000  \\
\hline
 $m_{\tilde{\chi}^0_4}$ (GeV)  & -5000     & -5000 &  -5000      &  5000     &   5000    & 5000  \\
\hline
 $m_{\tilde{\chi}^{\pm}_1}$ (GeV)  & 599.43  & 600.08 &  599.43  &  600.08 &   599.58 & 599.58  \\
\hline
 $m_{\tilde{\chi}^{\pm}_2}$ (GeV)  &    4000  & 4000 &  4000    &   4000    &    5000   & 5000  \\
 \hline
 $m_{h_1}$  (GeV)  &  122.94   &  122.68  &  122.75  & 122.51   &  122.83   &   122.65   \\
 \hline
 $m_{h_2}$  (GeV)  &   1347  &  1506  &  1390  &  1465  &  1423   &   1450   \\
 \hline
 \hline
 HF  & 0.9997  & 0.9998 &  0.9996   &  0.9997 & 0.9997 & 0.9997  \\
\hline
 $N_{11} ( \times 10^{\textrm{-}3})$  & $5.956$  & -$4.872$ &  -$7.575$  &  -$6.196$ & -$7.252$ & -$9.804$ \\
\hline
 $N_{12} (\times 10^{\textrm{-}2})$  & $1.827$ & -$1.495$ & $1.827$  &  $1.494$ & -$1.412$ & $1.412$ \\
\hline
$N_{13}$ & -0.707 & -0.707 & -0.708 & 0.708 & 0.707 & -0.708 \\
\hline
$N_{14}$ & 0.707 & -0.707 & 0.706   &  0.706 & -0.707 & 0.706 \\
\hline
\end{tabular} 
\caption{The benchmark scenarios with a higgsino-like $\tilde{\chi}^0_1$  have been tabulated for $|\mu|= 600$ GeV. HF stands for higgsino fraction. 
The fixed input parameters are: the mass of the pseudoscalar Higgs boson $m_{A} = 1.414$ TeV, and $\tan{\beta}=10$. The gluino mass parameter $M_3=3$ TeV. The trilinear coupling for two stops with the Higgs boson is set as $T_t=-3$ TeV. The soft-supersymmetry-breaking mass parameters for the left-type and the right-type stop and sbottom squarks are as follows: $m_{\tilde{Q}_L}=2.69$ TeV, $m_{\tilde{t}_R} = 2.06$ TeV and $m_{\tilde{b}_R}=2.50$ TeV.  As for the physical masses, the charged 
Higgs boson mass $M_{H^{\pm}}= 1.416$ TeV, the CP-even Higgs mixing angle $ \alpha = \sin^{-1}
(-0.1)$.  For all the benchmarks, the third generation squark mass and mixing parameters are taken as:  the lightest stop mass $ {\rm m}_{\tilde{t}_1} = 2.05$ TeV, the heaviest stop mass 
${\rm m}_{\tilde{t}_2} = 2.71$ TeV, the lightest sbottom mass $ {\rm m}_{\tilde{b}_1} = 2.50$ TeV, the heaviest sbottom mass  ${\rm m}_{\tilde{b}_2}= 2.69$ TeV. }
\label{tab:bp2} 
\end{center}
\end{table}

\subsection{Numerical Results and Discussion}

\begin{table}[H]
\begin{center}
\begin{tabular}{|c|c|c|c|c|}
 \hline
\multirow{3}{*}{BP}  & $\mathscr{C}_{1}^{L/R}$, $\mathscr{C}_{2}^{L/R}$ &    $ \Delta\mathscr{C}_{1}^{L/R}$ (\%)   &   $\Delta \mathscr{C}_{2}^{L/R}$ (\%) & $\sigma_{SI}$ [pb]   \\
  &  & Total  (SQ)  & Total  (SQ)  &  ($\Delta \sigma_{SI}$ \%)\\
  & &  (Loop, CT) & (Loop, CT) &  \\
 \hline
 \multirow{2}{*}{BP1a} &  $7.96 \times 10^{\textrm{-}3}$ & 19.74(-22.35)  & -13.96(-2.9) &  $4.13 \times 10^{\textrm{-}11}$ \\
 &  $4.68 \times 10^{\textrm{-}3}$ & (2.74, 17.0)& (-17.63, 3.67) & (41.7)  \\
 \hline
 \multirow{2}{*}{BP1b} & $8.64 \times 10^{\textrm{-}3}$ & 15.50(-26.62) & -14.29(-1.02) & $4.89 \times 10^{\textrm{-}11}$  \\
 & $5.24 \times 10^{\textrm{-}3}$ & (-1.52, 17.03) & (-18.0, 3.74) & (31.5) \\
 \hline
 \multirow{2}{*}{BP2a} & $6.12 \times 10^{\textrm{-}3}$ & 37.88(-29.52) & -19.87(-9.05) & {$2.29 \times 10^{\textrm{-}11}$} \\
  & -$4.36 \times 10^{\textrm{-}3}$ & (20.89, 16.99) & (-23.53, 3.66) & (96) \\
 \hline
 \multirow{2}{*}{BP2b}  & $6.63 \times 10^{\textrm{-}3}$ & 32.7(-35.78) & -21.22(-8.18) & {$2.71 \times 10^{\textrm{-}11}$}  \\
  & -$4.82 \times 10^{\textrm{-}3}$ & (15.73, 16.97) & (-25, 3.78) & (81) \\
\hline
 \multirow{2}{*}{BP3a} &  $1.13 \times 10^{\textrm{-}2}$ &  11.07(-18.05) & -6.98(-1.39)  & {$8.46 \times 10^{\textrm{-}11}$} \\
   & $7.77 \times 10^{\textrm{-}3}$ & (-7.1, 18.2) & (-11.83, 4.89) & (22.4) \\
 \hline
 \multirow{2}{*}{BP3b} &  $1.21 \times 10^{\textrm{-}2}$ & 9.14(-21.22) & -7.63(-0.26) & {$9.67 \times 10^{\textrm{-}11}$}  \\
    & $8.37 \times 10^{\textrm{-}3}$ & (-9.13, 18.27) & (-12.66, 5.03) & (18.25) \\
\hline
 \multirow{2}{*}{BP4a} &  $8.25 \times 10^{\textrm{-}3}$ &   21.11(-24.33) & -12.36(-6.99)  & {$4.13 \times 10^{\textrm{-}11}$} \\
  & -$7.33 \times 10^{\textrm{-}3}$ & (2.81, 18.3) & (-17.31, 4.95) & (49.75) \\
 \hline
 \multirow{2}{*}{BP4b} &  $8.82 \times 10^{\textrm{-}3}$ &  19.21(-28.89) & -13.74(-6.57)  &  {$4.77 \times 10^{\textrm{-}11}$}  \\
  & -$7.83 \times 10^{\textrm{-}3}$ & (0.87, 18.34) & (-18.83, 5.09) & (45) \\
\hline
  \multirow{2}{*}{BP5a} &  $6.24 \times 10^{\textrm{-}3}$ & 38.95(-25.5) & -16.6(-2.68)  & {$2.53 \times 10^{\textrm{-}11}$} \\
    & $3.06 \times 10^{\textrm{-}3}$ & (22.47, 16.48) & (-20.35, 3.75) & (89.6) \\
 \hline
  \multirow{2}{*}{BP5b} & $6.74 \times 10^{\textrm{-}3}$ &  32.88(-31.58) & -15.77(-0.26)  & {$2.97 \times 10^{\textrm{-}11}$}  \\
     & $3.49 \times 10^{\textrm{-}3}$ & (16.27, 16.61) & (-19.72, 3.95) & (73.8) \\
\hline
  \multirow{2}{*}{BP6a} & $1.05 \times 10^{\textrm{-}2}$ &  17.0(-17.81) & -5.32(-0.65)  &  {$7.26 \times 10^{\textrm{-}11}$} \\
   & $6.94 \times 10^{\textrm{-}3}$ & (-1.42, 18.42) & (-10.42, 5.10) & (35.8)  \\
 \hline
  \multirow{2}{*}{BP6b} &   $1.11 \times 10^{\textrm{-}2}$ & 15.41(-21.43) & -5.44(-0.74)  & {$8.13 \times 10^{\textrm{-}11}$}  \\
 & $7.43 \times 10^{\textrm{-}3}$ & (-3.20, 18.61) & (-10.8, 5.36) & (32.2) \\
  \hline
\end{tabular}
\caption{\label{tab:result} $\Delta\mathscr{C}_{i}^{L/R}$ ($\%$) denotes the percentage 
correction to the $\tilde{\chi}_1^0-\tilde{\chi}_1^0-h_i$ vertices $\mathscr{C}_{i}^{L/R}$. 
$\sigma_{SI}$  denotes spin-independent cross-section (with proton) including the radiative corrections and $\Delta\sigma_{SI}$  ($\%$) denotes the percentage contribution to the same from the radiative corrections under consideration. In the third and the fourth column title, 
``Total" refers to total percentage correction to $\mathscr{C}_{i}^{L/R}$, ``CT" refers 
to the percentage contribution from the counter-term vertex, ``Loop" denotes the percentage 
contribution from the one-loop diagrams, and ``SQ" denotes the percentage contribution from the 
third-generation quarks and squarks running in the loops. }
\end{center}
\end{table} 

In this section, we discuss the numerical results.  The radiative corrections to 
the $\tilde{\chi}_1^0-\tilde{\chi}_1^0-h_i$ vertices for the benchmark scenarios, as described 
in \ref{tab:bp1} and \ref{tab:bp2}, have been computed and have been presented in  \ref{tab:result}. In 
\ref{tab:result}, the one-loop corrected $\tilde{\chi}_1^0-\tilde{\chi}_1^0-h_i$ vertices ($\mathscr{C}_{i}^{L/R}
$) for the respective benchmark scenarios (as mentioned in the first column) have been presented in the 
second column. In the third and the fourth column the percentage contribution from the radiative corrections 
to the $\tilde{\chi}_1^0$-proton spin-independent scattering cross-sections $\Delta \mathscr{C}_{i}^{L/R} = \dfrac{\mathscr{C}_{i}^{L/R}- \mathscr{C}_{i ~{\rm tree}}^{L/R}}{\mathscr{C}_{i ~{\rm tree}}^{L/R}} \times 100\%$ have been described for $i=1$ and $i=2$ respectively. Note that in the present scenario, the results 
are similar for $\tilde{\chi}_1^0$-neutron spin-independent scattering cross-sections. For estimating 
the radiative corrections, contributions from the loops involving all the neutralinos and charginos, gauge 
bosons, Higgs bosons, and third-generation (s)quarks have been considered. Individual contributions from 
all the loops, counterterms, and also the third generation (s)quarks to the respective vertices have been 
mentioned. Finally, the radiatively corrected $\tilde{\chi}_1^0-{\rm nucleon}$ cross-section 
and the percentage contribution to the same $\Delta \sigma_{SI}  = \dfrac{\sigma_{SI} - 
\sigma_{SI ~{\rm tree}} }{\sigma_{SI ~{\rm tree}}} \times 100\%$ have been presented in 
the  fifth column. In the above discussion, the subscript ``tree" denotes the respective quantities without 
including the radiative corrections considered in this article. As discussed in the previous section, we have 
used \texttt{FeynArts}, \texttt{FormCalc}, and \texttt{LoopTools} for the numerical evaluation of the radiative 
contributions and the relevant counterterms.  

As described in \ref{tab:bp1} and \ref{tab:bp2}, for all the benchmark scenarios 
$\tilde{\chi}_1^0$ is dominantly higgsino-like. The higgsino fraction ($\textrm {HF} = |N_{13}|^2+ |N_{14}|^2$) is above $99\%$. The radiative corrections to the $\tilde{\chi}_1^0-\tilde{\chi}_1^0-h_1$ vertex $
\mathscr{C}_1^{L/R}$ contributes dominantly to the spin-independent cross-section $\sigma_{SI}$. The 
contribution to the spin-independent cross-section $\sigma_{SI}$ from the heavy Higgs boson $h_2$ is only 
about $\lesssim 3 \%$ for all the benchmark scenarios. This is because $m_{h_2}\gg m_{h_1}$ (about ten 
times) in the present context.  Therefore, its contribution to the $\tilde{\chi}^0_1$- nucleon coupling $
\lambda_q^H$ $( \propto \frac{1}{m_{h_2}^2})$ is suppressed, as can be inferred from \ref{eq:lam_chi_q}. 
Thus, for all the benchmark scenarios, the percentage corrections to the cross-section $\sigma_{SI}$  are 
approximately twice that of the percentage corrections to the $\tilde{\chi}_1^0-\tilde{\chi}_1^0-h_1$ vertex 
factor $\mathscr{C}_1^{L/R}$.

The radiative corrections to $\tilde{\chi}_1^0-\tilde{\chi}_1^0-h_1/h_2$ vertices are significant for all the 
benchmark scenarios and vary between approximately 9\%-40\% for the light Higgs boson vertex and 
between approximately 5\%-21\% for the vertex involving the heavy Higgs boson. Comparing the first eight 
benchmarks (BP-1a to BP-4b), the percentage change in the $\tilde{\chi}_1^0-\tilde{\chi}_1^0-h_1$  vertices 
are significant for the benchmarks with negative $\mu$ (BP-2a, BP-2b and BP-4a, BP-4b), as compared to 
their counterparts with positive $\mu$ (BP-1a, BP-1b and BP-3a, BP-3b). Let us consider BP-1a(b) and 
BP-2a(b). While BP-1a(b) and BP-2a(b) only differ by the sign of $\mu$, thus, the percentage contribution 
to $\mathscr{C}_1^{L/R}$ from the radiative correction for BP-2a(b) is significantly higher as compared to 
BP-1a(b). This is largely because of a substantial reduction in the tree-level vertex factor for BP-2a and 
BP-2b, while the radiative corrections are also marginally higher. Note that, for positive (negative) $\mu$, $
\tilde{\chi}_1^0$ is the symmetric (anti-symmetric) higgsino-like state. A similar argument explains the 
larger percentage corrections in the context of BP-4a(b), as compared to BP-3a(b). It follows from 
\ref{eq:hchi_coupling}, that for the symmetric states, the respective tree-level vertex suffers from 
cancellation between two terms proportional to $N_{13}$ and $N_{14}$ respectively. In all the benchmark 
scenarios, the dominant loop contributions to $\mathscr{C}_1^{L/R}$ come from the triangle loops involving 
two vector bosons and one neutralino/chargino. Further, the third generation (s)quarks contribute 
significantly thanks to the large Yukawa couplings. The contributions from the loops involving, in 
particular, two quarks and one squark tend to negate the contributions from the loops involving the vector 
bosons and one neutralino/ chargino. In BP-5a(b) and BP-6a(b), the difference in the percentage 
contribution to $\mathscr{C}_1^{L/R}$ is largely attributed to the cancellation from the (s)quark loop. Further, contributions from the vertex counterterms are substantial. 
In particular, we find sizable contributions from the terms proportional to the diagonal 
and off-diagonal wave-function renormalization counterterms. 
 
The vertex counterterms are evaluated following the implementation in \texttt{FormCalc}  
\cite{Fritzsche:2013fta}. The details have been discussed in Appendix C. On-shell renormalization 
schemes have been adopted for the neutralino-chargino sector \cite{Fritzsche:2002bi}. In 
particular, for BP1a to BP-4b, two chargino masses and the heaviest neutralino mass (CCN[4]) 
have been used as on-shell input masses. For BP-5a, BP-5b, BP-6a, and BP-6b, two chargino 
masses and the third neutralino mass (CCN[3]) have been used as on-shell input masses. This 
ensures that there is always a bino-like neutralino among the input masses \cite{Baro2009,Chatterjee:2011wc}. The respective contributions from the counterterms have been 
shown in \ref{tab:result}. Note that we have used tree-level masses for all the neutralinos 
and charginos, including $\tilde{\chi}_1^0$ for the eastimation of the spin-independent scattering 
cross-section. This ensures that the percentage corrections to the cross-section reflects only 
the contributions from the vertex corrections, which we intend to illustrate.

As the spin-independent cross-section of $\tilde{\chi}_1^0$ with the nucleons 
(protons and neutrons) receive dominant contributions from the light Higgs boson-mediated 
processes, the percentage corrections to the cross-sections are about twice 
the respective percentage corrections to the $\tilde{\chi}_1^0-\tilde{\chi}_1^0-h_1$ vertex. 
These cross-sections can be enhanced by up to about 100\% for the benchmark scenarios. 
This highlights the importance of these corrections in the present context. 

Note that, as mentioned in the \ref{sec:intro} and elaborated further in \ref{sec:DD}, 
certain important loop corrections 
to the Higgs bosons-nucleon interactions, which contribute to the effective 
neutralino-nucleon effective operators (see \ref{eq:L_chi_N} and 
\ref{eq:L_chi_q}), have been already included in \texttt{micrOMEGAs}. Thus, the 
cross-sections computed using the one-loop corrections to $\tilde{\chi}_1^0-
\tilde{\chi}_1^0-h_1/h_2$ vertices also effectively include certain two-loop 
contributions. These corrections are also included in the cross-sections 
with which we have compared the final results after including the vertex corrections. 
Thus, the percentage corrections to the cross-sections, as mentioned in the 
\ref{tab:result}, solely come from the corrections to the 
$\tilde{\chi}_1^0-\tilde{\chi}_1^0-h_1/h_2$ vertices. 

\begin{figure}[H]
	\centering
	\includegraphics[scale=0.305]{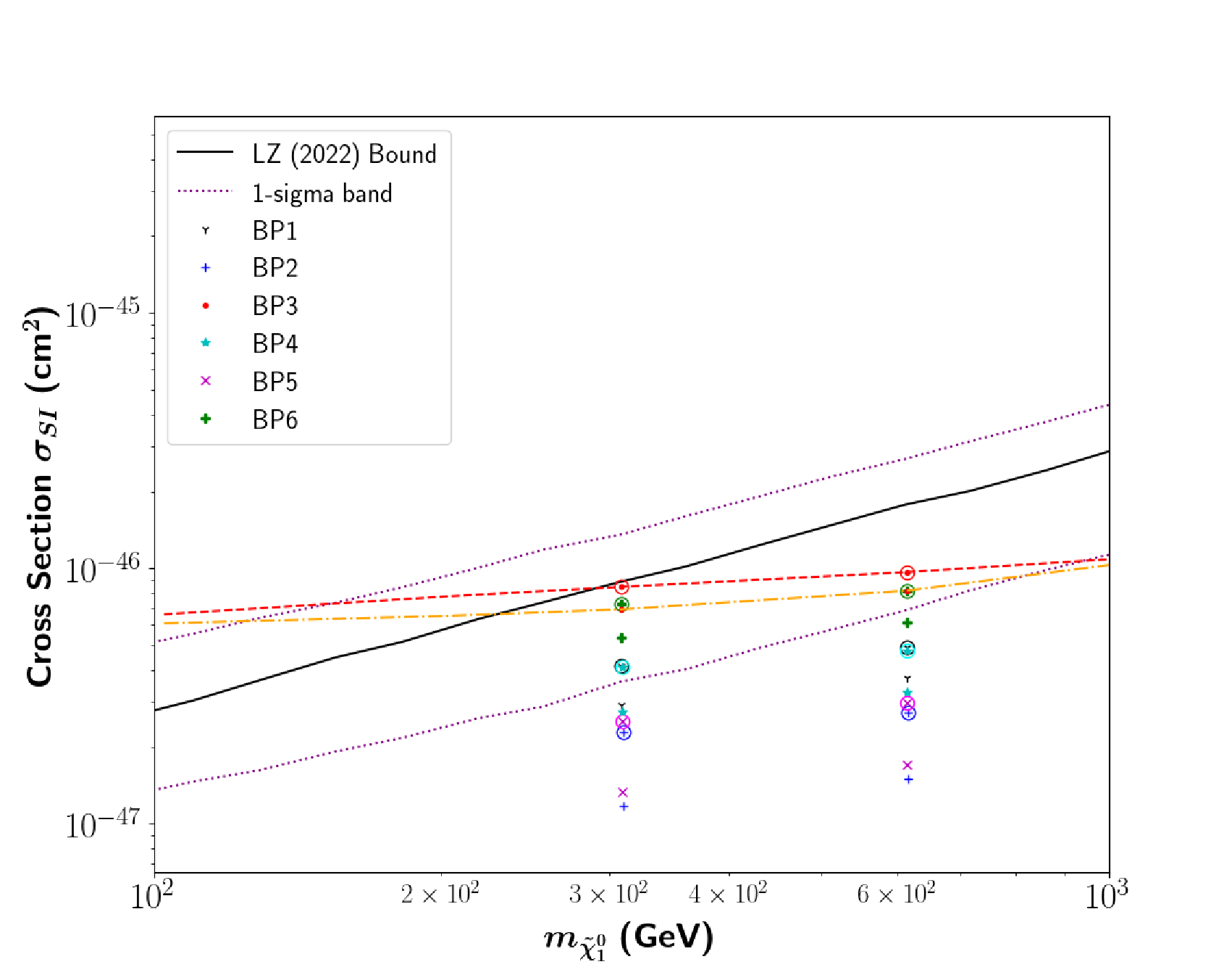}
	\includegraphics[scale=0.305]{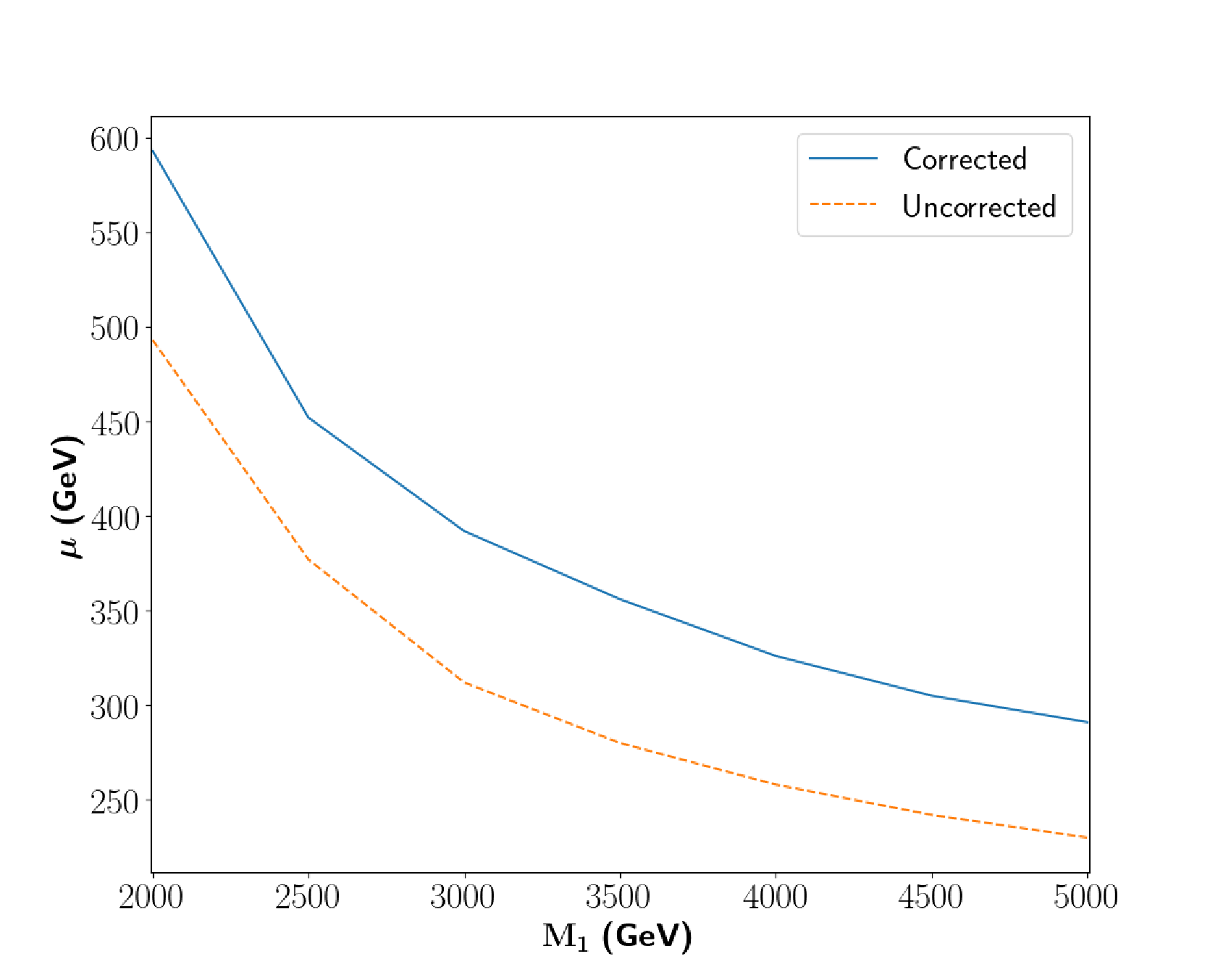}\\
	(a) \hspace{8cm} (b)
	\caption{Panel (a) shows the comparison of the shift of various benchmark points (Table \ref{tab:result}) before and after adding the vertex corrections ($\mathscr{C}_{i}^{L/R}$) with the direct detection bound of LUX-ZEPLIN (LZ) experiment \cite{LZ:2022lsv}. The circled points depict the corrected cross-sections ($\sigma_{SI}$) and the uncircled ones are without the corrections ($\sigma_{SI \rm ~tree}$). Panel (b) shows the shift in the $\mu$ parameter for different values of $M_1$ as constrained by LZ (2022) \cite{LZ:2022lsv} after adding the vertex corrections ($\mathscr{C}_{i}^{L/R}$). The change in the constraint on the $\mu$ parameter (for $\mu \ll M_1$,$M_2$) corresponding to the cross-section after adding the vertex corrections $\mathscr{C}_{i}^{L/R}$ is shown by the solid line, the dashed line represents the case without the corrections ($\mathscr{C}_{i~\mathrm{tree}}^{L/R}$). Here, $M_2$ is taken as 4 TeV and the other parameters are assumed to be the same as mentioned in Table \ref{tab:bp1}.}
	\label{fig:DDplot}
\end{figure}

Assuming that  $\tilde{\chi}_1^0$ constitutes the entirety of DM, we have further considered the 
implications of these large corrections for the viability of sub-TeV higgsino-like DM in light of 
stringent limits from the direct detection experiments. We consider the DM-nucleon 
(proton) cross-section limits from the LUX-ZEPLIN (LZ) experiment \cite{LZ:2022lsv} and 
compare the status of the benchmark scenarios after including the radiative corrections as shown in Fig.\ref{fig:DDplot}(a). 
We find that, thanks to  the radiative corrections, benchmark point BP-1a is pushed 
above the lower limit of the 1$\sigma$ sensitivity band (dotted line), and BP-1b is pushed close to the 
1$\sigma$ band. Benchmark points BP-2a and BP-2b are pushed close to the 1$\sigma$ 
band while lying below it. The benchmark BP-3a falls on the exclusion line (solid line) and is close to being 
ruled out after the corrections are added, and BP-3b is also close to the exclusion limit. As for 
benchmark BP4a, it is pushed above the 1$\sigma$ lower band, and BP-4b is pushed close to 
it. BP-5a and BP-5b are also pushed closer to the 1$\sigma$ band of the exclusion region;
Finally, BP6a and BP6b benchmarks are pushed above the 1$\sigma$ band and close 
to the exclusion limit when the corrections are added. Although, to 
estimate the overall impact on the scattering cross-section all the radiative corrections 
need to be considered together, the above discussion aims to demonstrate the 
relative importance of the vertex corrections, in comparison with the same 
cross-section evaluated using the tree-level vertices $C_i$. \footnote{
Note that by changing the stop-stop-Higgs boson soft-supersymmetry-breaking trilinear 
term $T_t$ to -4 TeV, the light Higgs mass $m_{h_1}$, as computed by \texttt{SPheno}, 
becomes about 125 GeV. We have checked that using $m_{h_1} \simeq 125$  GeV, 
with the above modifications to the stop sector parameters, does not affect 
the vertex corrections and the percentage corrections to the direct detection 
cross-section appreciably. For most of the benchmark scenarios, which assume 
$m_{h_1} \simeq 123$ GeV, using the parameters as mentioned above lead to 
variations in the percentage correction to the neutralino-proton cross-section 
($\Delta \sigma_{SI}$) by less than $\sim$3\%. Further, note that while using $m_{h_1} 
\simeq 125$ GeV, keeping all the other parameters as the benchmark scenarios, 
does not change in the vertex corrections appreciably, and thus, the percentage
change in the spin-independent cross-sections ($\Delta \sigma_{SI}$) are also well 
below a percent.}
To demonstrate the significance of the radiative corrections on constraining the 
Higgsino mass parameter 
$\mu$ in the present context, we further vary the $\mu$ parameter  keeping all the other 
relevant parameters the same as BP-3a(or b). The cross-sections with the radiatively 
corrected $\tilde{\chi}_1^0$-$\tilde{\chi}_1^0$-$h_1/h_2$ vertices ($\mathscr{C}_{i}^{L/R}$) 
and the respective tree-level vertices ($\mathscr{C}_{i~\mathrm{tree}}^{L/R}$) have been 
used to obtain the dashed red line and the dot-dashed orange line 
respectively in Fig.\ref{fig:DDplot}(a). As demonstrated in the figure, 
the dashed red line intersects the  90\% confidence limit from the LZ experiment \cite{LZ:2022lsv} 
for a heavier $m_{\tilde{\chi}_1^0}$, as compared to the dot-dashed orange line. As in the present 
context $m_{\tilde{\chi}_1^0} \simeq |\mu|$ (as $|\mu| \ll |M_1|, M_2$), 
therefore, the constraint on the $\mu$ parameter is improved. This is further illustrated in 
Fig.\ref{fig:DDplot}(b) with positive $M_1$ and $\mu$ $(\mu \ll M_1, M_2)$. 
In this figure, the constraint on $\mu$ parameter is shown to vary with respect to $M_1$. 
We have assumed, as in the benchmark scenarios, $\tan {\beta}=10,~M_2=4$ TeV and 
$M_A=1.414$ TeV; the other parameters are also kept the same as mentioned in Table \ref{tab:bp1} 
and \ref{tab:bp2}. 
As shown in Fig.\ref{fig:DDplot}(b), for $M_1 = 2$ TeV, $\mu \lesssim 493$ GeV (as shown by the dashed line) 
is excluded by the direct detection experiment LZ, when tree-level $\tilde{\chi}_1^0$-$\tilde{\chi}_1^0$-$h_1/h_2$ 
vertices are used to estimate the respective cross-sections. While estimating the cross-section using the 
radiatively corrected vertices ($\mathscr{C}_{i}^{L/R}$), the constraint shifts to $\mu \lesssim 593$ GeV 
(as shown by the solid line), a shift of 100 GeV. Likewise, the bound on $\mu$ shifts from 230 GeV to 291 GeV 
for $M_1=5$ TeV. Thus, the constraint on the $\mu$ parameter space (with $\mu \ll M_1$,$M_2$) 
becomes more stringent by about 60-100 GeV, as illustrated in this figure.
{\footnote{The cases for other combinations of signs of $M_1$ and $\mu$ are not shown as 
their cross-sections lie below the LZ bounds in the parameter space of our interest.}}

\section{Conclusion}
\label{sec:Conclusion}
 
Light higgsino-like $\tilde{\chi}_1^0$ fits well within the framework of natural supersymmetry. 
In this article, we have considered higgsino-like $\tilde{\chi}_1^0$ DM within R-parity conserving 
MSSM and have studied the importance of a class of radiative corrections to the $\tilde{\chi}_1^0-
\tilde{\chi}_1^0-h_1/h_2$ vertices in the context of spin-independent direct detection. The 
tree-level couplings between $\tilde{\chi}_1^0$ and the CP-even neutral Higgs bosons 
($h_1,~h_2$), in such a scenario, are suppressed by small gaugino-higgsino mixing. However, 
as demonstrated in this article, the radiative contributions to these vertices (including 
the respective counterterms) from the loops 
involving the charginos, neutralinos, gauge bosons, and Higgs bosons can have significant 
implications for direct detection. Further, third-generation (s)quark contributions are significant 
and tend to cancel the former to some extent in the parameter region considered in this article. 
For the benchmark scenarios presented,  the radiatively corrected vertices can be enhanced 
by about 40\% compared to the respective tree-level vertices. The spin-independent 
cross-section of $\tilde{\chi}_1^0$ with the nucleons (protons and neutrons), 
which receives a significant contribution from the CP-even neutral Higgs boson mediated processes through the respective effective operators, thus, can be enhanced by  
about 100\% in certain benchmark scenarios. We further illustrate that 
the corrections are sensitive to the sign of $\mu$ and the choice of the gaugino mass 
parameters $M_1$ and $M_2$, even though $|\mu| \ll |M_1|, M_2$. 
Note that,  the  ``tree-level" cross-section in such scenarios is quite sensitive to 
the small gaugino admixture in the $\tilde{\chi}_1^0$. Thus, generally, the constraint 
on the mass of sub-TeV higgsino-like $\tilde{\chi}_1^0$, after including these corrections, 
is sensitive to the sign of $\mu$ and the choice of the gaugino mass parameters 
$M_1$ and $M_2$. As mentioned in the Introduction, in the sub-TeV mass region, the thermal relic 
abundance of a higgsino-like $\tilde{\chi}_1^0$ LSP is inadequate to fulfill the 
required relic abundance of DM ($\Omega_{\rm DM} h^2 = 0.12$ \cite{Planck:2018vyg}). 
Thus, assuming only thermal production of $\tilde{\chi}_1^0$ will lead to a 
dilution of the direct detection constraints on $\tilde{\chi}_1^0$, in proportion 
to the relative abundance of $\tilde{\chi}_1^0$. However, considering the 
possibility of non-thermal production of $\tilde{\chi}_1^0$ in the early Universe, 
there is a possibility that $\tilde{\chi}_1^0$ constitutes the entire DM. In any scenario, 
the result demonstrates the significance of  the complete vertex corrections to the 
$\tilde{\chi}_1^0-
\tilde{\chi}_1^0-h_1/h_2$ vertices in the spin-independent
scattering cross-section of a higgsino-like $\tilde{\chi}_1^0$ DM.

\section*{Acknowledgement}
The authors thank B. De for his contributions in the early stages of this work. We 
acknowledge useful discussions with S. Heinemeyer, A. Pukhov, and C. Schappacher. AC acknowledges 
the hospitality at IoP, Bhubaneswar, during the meeting IMHEP-19 and IMHEP-22 which facilitated this work. AC 
and SAP also acknowledge the hospitality at IoP, Bhubaneswar, during a visit. SB acknowledges 
the local hospitality at SNIoE, Greater Noida, during the meeting at WPAC-2023 where 
this work was finalised.

 \newpage
\renewcommand{\theequation}{A.\arabic{equation}}
\setcounter{equation}{0}
\section*{Appendix A}
\begin{figure}[H]
	\centering
	\includegraphics[scale=0.5]{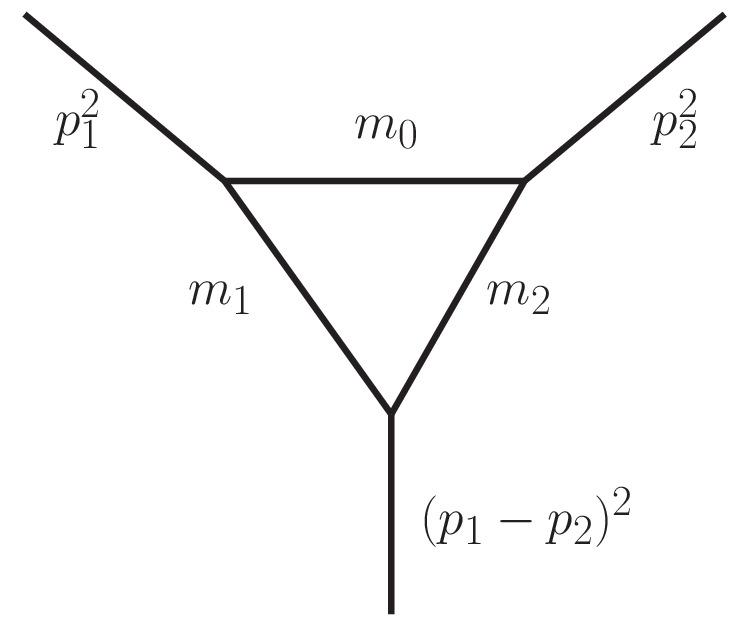}
	\caption{The above figure shows the mass and momentum convention for the Passarino-Veltman functions.}
	\label{fig:Cfunction}
\end{figure}
In this appendix, we summarize the Passarino-Veltman functions \cite{PASSARINO1979151}, 
which appear in the radiative corrections, as described in Appendix B. We follow the convention of refs. \cite{Patel:2015tea,vanOldenborgh:1990yc}.  
The Passarino-Veltman C functions have the following form:
\beq
\begin{gathered}
\mathbf{C}_0\left(p_1^2,\left(p_1-p_2\right)^2, p_2^2, m_0, m_1, m_2\right)=-\int_0^1 \mathrm{~d} x \int_0^{1-x} \mathrm{~d} y\left[x^2 p_1^2+y^2 p_2^2+x y 2 p_1 p_2 \right. \\
 \left.-x\left(p_1^2-m_1^2+m_0^2\right)-y\left(p_2^2-m_2^2+m_0^2\right)+m_0^2-i \epsilon\right]^{-1}. 
\end{gathered} 
\eeq
 We have used the following abbreviation:
\beq
\begin{gathered}
\quad\langle\cdots\rangle_q:=\frac{(2 \pi \tilde{\mu})^{4-D}}{i \pi^2} \int \mathrm{d}^{D}q \cdots \, ,
\end{gathered}
\eeq
where $\tilde{\mu}$ denotes a parameter with dimension of mass. Further, 
\beq
\begin{aligned}
\mathbf{C}_\mu\left(p_1^2,\left(p_1-p_2\right)^2, p_2^2, m_0, m_1, m_2\right) & =\left\langle\frac{q_\mu}{\left(q^2-m_0^2\right)\left[\left(q+p_1\right)^2-m_1^2\right]\left[\left(q+p_2\right)^2-m_2^2\right]}\right\rangle_q \\
& =p_{1, \mu} \mathbf{C}_1+p_{2, \mu} \mathbf{C}_2 .
\end{aligned}\\
\eeq
Contraction with $p_1^\mu$, then, gives 
\beq
\begin{split}
p_1^2 \mathbf{C}_1+p_1 p_2 \mathbf{C}_2 &=   \left\langle\frac{\frac{1}{2}\left[\left(q+p_1\right)^2-m_1^2\right]-\frac{1}{2}\left(q^2-m_0^2\right)-\frac{1}{2}\left(p_1^2-m_1^2+m_0^2\right)}{\left(q^2-m_0^2\right)\left[\left(q+p_1\right)^2-m_1^2\right]\left[\left(q+p_2\right)^2-m_2^2\right]}\right\rangle_q \nonumber \\
&=  \frac{1}{2} \mathbf{B}_0\left(p_2^2, m_0, m_2\right)-\frac{1}{2} \mathbf{B}_0\left(\left(p_1-p_2\right)^2, m_1, m_2\right)-\frac{1}{2}\left(p_1^2-m_1^2+m_0^2\right) \mathbf{C}_0\, ,  
\end{split}
\eeq
where the momenta and masses are as shown in \ref{fig:Cfunction}. 
Further, 
\beq
\begin{aligned}
\left(\begin{array}{c}
\mathbf{C}_1 \\
\mathbf{C}_2
\end{array}\right)= & \left(\begin{array}{cc}
p_1^2 & p_1 p_2 \\
p_1 p_2 & p_2^2
\end{array}\right)^{-1} \cdot\left(\begin{array}{c}
\frac{1}{2} \mathbf{B}_0\left(p_2^2, m_0, m_2\right)-\frac{1}{2} \mathbf{B}_0\left(\left(p_1-p_2\right)^2, m_1, m_2\right)-\frac{1}{2} f_1 \mathbf{C}_0 \\
\frac{1}{2} \mathbf{B}_0\left(p_1^2, m_0, m_1\right)-\frac{1}{2} \mathbf{B}_0\left(\left(p_1-p_2\right)^2, m_1, m_2\right)-\frac{1}{2} f_2 \mathbf{C}_0
\end{array}\right)
\end{aligned}
\eeq
where $f_i=p_i^2-m_i^2+m_0^2$,  for $i \in \{1,2\}$.
In the expressions above, $\mathbf{B}_0$ is given by
\begin{equation}
\mathbf{B}_0\left(p_1, m_0, m_1\right) =\Delta-\int_0^1 \mathrm{~d} x \log \left[\frac{x^2 p_1^2-x\left(p_1^2-m_1^2+m_0^2\right)+m_0^2-i \epsilon}{\mu^2}\right]+\mathscr{O}(D-4) \, , 
\end{equation}
where $\Delta := \frac{2}{4-d}-\gamma_E+\log 4 \pi$, $\gamma_E$ is the Euler-Mascheroni constant and $d$ stands for space-time dimension. 

\renewcommand{\theequation}{B.\arabic{equation}}
\setcounter{equation}{0} 
\section*{Appendix B}
In this Appendix, we discuss the radiative corrections to the $\tilde{\chi}^0_1-\tilde{\chi}^0_1-h_i$ vertices originating from the triangle diagrams. In particular, generic expressions for 
contributions from scalar bosons, vector bosons, and fermions running in the loops have been 
provided. In the following discussion, $F$ and $F'$ denote fermions, $S$ and $S'$ are used 
for scalar bosons, and $V$ denotes vector bosons. Further, $q^2$ denotes the square of the 
momentum transferred from the incident $\tilde{\chi}^0_1$ to the quarks in the nucleons, and 
$d$ stands for space-time dimension. Here, $G$ and $G^{\pm}$ refer to the neutral and 
charged Goldstone Bosons respectively. We have evaluated the expressions using 
\texttt{Package-X} (version 2.1.1) \cite{Patel:2015tea}, and have also checked some of 
these expressions by explicit calculations. Feynman gauge has been used for the calculation. 
The vertices may be found in ref. \cite{Drees:2004jm}.

\begin{figure}[H]
	\centering
	\includegraphics[width=0.8\linewidth]{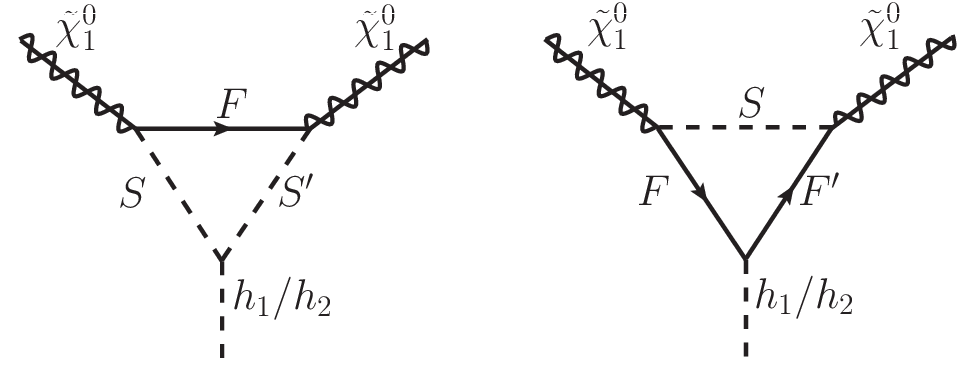}\\
	(a)~~~~~~~~~~~~~~~~~~~~~~~~~~~~~~~~~~~~~~~~~~~~~~~~~~~~~~~~~~~(b)
	\caption{Topology 1(a) and 1(b)}
	\label{fig:topology1}
\end{figure}

{\bf Topology-(1a):} \\
The respective Feynman diagram is shown in Fig. \ref{fig:topology1}(a).
\bea
i\delta\Gamma^{(a)} =& -\dfrac{i}{16\pi^2}\Big[P_L\Bigl\{\xi_{LL}m_F\mathbf{C}_0 - \xi_{LR}m_{\tilde{\chi}_1^0}\mathbf{C}_1 - \xi_{RL}m_{\tilde{\chi}_1^0}\mathbf{C}_2\Bigr\}+ P_R\Bigl\{\xi_{RR}m_F\mathbf{C}_0-\xi_{RL}m_{\tilde{\chi}_1^0}\mathbf{C}_1 \nonumber\\
&-\xi_{LR}m_{\tilde{\chi}_1^0}\mathbf{C}_2\Bigr\}\Big]~,
\eea

where $\mathbf{C}_i=\mathbf{C}_i\big(m_{\tilde{\chi}_1^0}^2,q^2,m_{\tilde{\chi}_1^0}^2;m_F,m_S,m_S^\prime\big)$,
and 
\bea
\xi_{LL} = \lambda_{h_iSS^\prime} \mathscr{G}_{\tilde{\chi}_1^0FS^\prime}^{L}\mathscr{G}_{\tilde{\chi}_1^0FS}^{L}~,\qquad\qquad \xi_{LR} = \lambda_{h_iSS^\prime} \mathscr{G}_{\tilde{\chi}_1^0FS^\prime}^{L}\mathscr{G}_{\tilde{\chi}_1^0FS}^{R}~,\\
\xi_{RL} = \lambda_{h_iSS^\prime} \mathscr{G}_{\tilde{\chi}_1^0FS^\prime}^{R}\mathscr{G}_{\tilde{\chi}_1^0FS}^{L}~,\qquad\qquad \xi_{RR} = \lambda_{h_iSS^\prime} \mathscr{G}_{\tilde{\chi}_1^0FS^\prime}^{R}\mathscr{G}_{\tilde{\chi}_1^0FS}^{R}~.
\eea
\\
(1) $h_i=h_1/h_2$, $F=\tilde{\chi}_{\ell}^0$, and $S=S^\prime=h_1$.
\begin{align}
\xi_{LL}= \lambda_{h_ih_1h_1}\mathcal{G}^{L}_{\tilde{\chi}_1^0\tilde{\chi}_{\ell}^0h_1}\mathcal{G}^{R*}_{\tilde{\chi}_1^0\tilde{\chi}_{\ell}^0h_1}~,\qquad\qquad\xi_{LR}= \lambda_{h_ih_1h_1}\mathcal{G}^{L}_{\tilde{\chi}_1^0\tilde{\chi}_{\ell}^0h_1}\mathcal{G}^{L*}_{\tilde{\chi}_1^0\tilde{\chi}_{\ell}^0h_1}~,\\
\xi_{RL}= \lambda_{h_ih_1h_1}\mathcal{G}^{R}_{\tilde{\chi}_1^0\tilde{\chi}_{\ell}^0h_1}\mathcal{G}^{R*}_{\tilde{\chi}_1^0\tilde{\chi}_{\ell}^0h_1}~,\qquad\qquad \xi_{RR}= \lambda_{h_ih_1h_1}\mathcal{G}^{R}_{\tilde{\chi}_1^0\tilde{\chi}_{\ell}^0h_1}\mathcal{G}^{L*}_{\tilde{\chi}_1^0\tilde{\chi}_{\ell}^0h_1}~,
\end{align}
where $\lambda_{h_ih_ih_i} = -3\dfrac{g_2M_Z}{2c_W}B_{h_i},\quad {\rm with}\,\, B_{h_i}= \left\{
\begin{array}{ll}
c_{2\alpha} s_{\beta+\alpha}; & h_i=h_1 \\
c_{2\alpha} c_{\beta+\alpha}; & h_i=h_2 \\
\end{array} 
\right.$,\\
\\
 $\mathcal{G}^{L}_{\tilde{\chi}_1^0\tilde{\chi}_{\ell}^0h_{i}} = \left\{
\begin{array}{ll}
g_2\big(Q_{\ell 1}^{\prime\prime*}s_\alpha+S_{\ell 1}^{\prime\prime*}c_\alpha\big); & h_i=h_1 \\
g_2\big(-Q_{\ell 1}^{\prime\prime*}c_\alpha+S_{\ell 1}^{\prime\prime*}s_\alpha\big); & h_i=h_2 \\
\end{array} 
\right.~,$\,\,
$\mathcal{G}^{R}_{\tilde{\chi}_1^0\tilde{\chi}_{\ell}^0h_{i}} = \left\{
\begin{array}{ll}
g_2\big(Q_{1\ell}^{\prime\prime}s_\alpha+S_{1\ell}^{\prime\prime}c_\alpha\big); & h_i=h_1 \\
g_2\big(-Q_{1\ell}^{\prime\prime}c_\alpha+S_{1\ell}^{\prime\prime}s_\alpha\big); & h_i=h_2 \\
\end{array} 
\right..$\\ 
\vspace{0.5cm}\newline
(2) $h_i=h_1/h_2$, $F=\tilde{\chi}_{\ell}^0$, and $S=h_1$, $S^\prime=h_2$ or $S=h_2$, $S^\prime=h_1$. 
\begin{align}
\xi_{LL}= \lambda_{h_ih_1h_2}\mathcal{G}^{L}_{\tilde{\chi}_1^0\tilde{\chi}_{\ell}^0h_2}\mathcal{G}^{R*}_{\tilde{\chi}_1^0\tilde{\chi}_{\ell}^0h_1}~,\qquad\qquad\xi_{LR}= \lambda_{h_ih_1h_2}\mathcal{G}^{L}_{\tilde{\chi}_1^0\tilde{\chi}_{\ell}^0 h_2}\mathcal{G}^{L*}_{\tilde{\chi}_1^0\tilde{\chi}_{\ell}^0h_1}~,\\
\xi_{RL}= \lambda_{h_ih_1h_2}\mathcal{G}^{R}_{\tilde{\chi}_1^0\tilde{\chi}_{\ell}^0h_2}\mathcal{G}^{R*}_{\tilde{\chi}_1^0\tilde{\chi}_{\ell}^0h_1}~,\qquad\qquad \xi_{RR}= \lambda_{h_ih_1h_2}\mathcal{G}^{R}_{\tilde{\chi}_1^0\tilde{\chi}_{\ell}^0h_2}\mathcal{G}^{L*}_{\tilde{\chi}_1^0\tilde{\chi}_{\ell}^0h_1}~,
\end{align}
where $\lambda_{h_ih_1h_2}=\dfrac{g_2M_Z}{2c_W}C_{h_i}$, with $C_{h_i} = \left\{
\begin{array}{ll}
-2s_{2\alpha} s_{\beta+\alpha}+c_{\beta+\alpha}c_{2\alpha}; & h_i=h_1 \\
\,\,\,\,\,2s_{2\alpha}c_{\beta+\alpha}+s_{\beta+\alpha}c_{2\alpha}; & h_i=h_2 \\
\end{array} 
\right..$\\
\vspace{0.5cm}\newline
(3) $h_i=h_1/h_2$, $F=\tilde{\chi}_{\ell}^0$, and $S=S^\prime=h_2$.
\begin{align}
\xi_{LL}= \lambda_{h_ih_2h_2}\mathcal{G}^{L}_{\tilde{\chi}_1^0\tilde{\chi}_{\ell}^0h_2}\mathcal{G}^{R*}_{\tilde{\chi}_1^0\tilde{\chi}_{\ell}^0h_2}~,\qquad\qquad\xi_{LR}= \lambda_{h_ih_2h_2}\mathcal{G}^{L}_{\tilde{\chi}_1^0\tilde{\chi}_{\ell}^0h_2}\mathcal{G}^{L*}_{\tilde{\chi}_1^0\tilde{\chi}_{\ell}^0h_2}~,\\
\xi_{RL}= \lambda_{h_ih_2h_2}\mathcal{G}^{R}_{\tilde{\chi}_1^0\tilde{\chi}_{\ell}^0h_2}\mathcal{G}^{R*}_{\tilde{\chi}_1^0\tilde{\chi}_{\ell}^0 h_2}~,\qquad\qquad \xi_{RR}= \lambda_{h_ih_2h_2}\mathcal{G}^{R}_{\tilde{\chi}_1^0\tilde{\chi}_{\ell}^0 h_2}\mathcal{G}^{L*}_{\tilde{\chi}_1^0\tilde{\chi}_{\ell}^0 h_2}~,
\end{align}
\vspace{0.5cm}\newline
(4) $h_i=h_1/h_2$, $F=\tilde{\chi}_{\ell}^0$, and $S=S^\prime=A$.
\begin{align}
\xi_{LL}= \lambda_{h_iAA}\mathcal{G}^{L}_{\tilde{\chi}_1^0\tilde{\chi}_{\ell}^0A}\mathcal{G}^{R*}_{\tilde{\chi}_1^0\tilde{\chi}_{\ell}^0A}~,\qquad\qquad\xi_{LR}= \lambda_{h_iAA}\mathcal{G}^{L}_{\tilde{\chi}_1^0\tilde{\chi}_{\ell}^0A}\mathcal{G}^{L*}_{\tilde{\chi}_1^0\tilde{\chi}_{\ell}^0A}~,\\
\xi_{RL}= \lambda_{h_iAA}\mathcal{G}^{R}_{\tilde{\chi}_1^0\tilde{\chi}_{\ell}^0A}\mathcal{G}^{R*}_{\tilde{\chi}_1^0\tilde{\chi}_{\ell}^0A}~,\qquad\qquad \xi_{RR}= \lambda_{h_iAA}\mathcal{G}^{R}_{\tilde{\chi}_1^0\tilde{\chi}_{\ell}^0A}\mathcal{G}^{L*}_{\tilde{\chi}_1^0\tilde{\chi}_{\ell}^0A}~,
\end{align}
where $\lambda_{h_iAA}=-\dfrac{g_2M_Z}{2c_W} c_{2\beta}\,D_{h_i}$, with $D_{h_i} = \left\{
\begin{array}{ll}
\,\,\,\,\,s_{\beta+\alpha}; & h_i=h_1 \\
-c_{\beta+\alpha}; & h_i=h_2 \\
\end{array} 
\right.,$\\

$\mathcal{G}^{L}_{\tilde{\chi}_1^0\tilde{\chi}_{\ell}^0A}=i\big(Q_{\ell 1}^{\prime\prime*}s_\beta-S_{\ell 1}^{\prime\prime*}c_\beta\big)$, and \,\,$\mathcal{G}^{R}_{\tilde{\chi}_1^0\tilde{\chi}_{\ell}^0A}=i\big(-Q_{1\ell}^{\prime\prime}s_\beta+S_{1\ell}^{\prime\prime}c_\beta\big)$.\\
\vspace{0.5cm}\newline
(5) $h_i=h_1/h_2$, $F=\tilde{\chi}_{\ell}^0$, and $S=A$, $S^\prime=G$ or $S=G$, $S^\prime=A$.
\begin{align}
\xi_{LL}= \lambda_{h_iAG}\mathcal{G}^{L}_{\tilde{\chi}_1^0\tilde{\chi}_{\ell}^0G}\mathcal{G}^{R*}_{\tilde{\chi}_1^0\tilde{\chi}_{\ell}^0A}~,\qquad\qquad\xi_{LR}= \lambda_{h_iAG}\mathcal{G}^{L}_{\tilde{\chi}_1^0\tilde{\chi}_{\ell}^0G}\mathcal{G}^{L*}_{\tilde{\chi}_1^0\tilde{\chi}_{\ell}^0A}~,\\
\xi_{RL}= \lambda_{h_iAG}\mathcal{G}^{R}_{\tilde{\chi}_1^0\tilde{\chi}_{\ell}^0G}\mathcal{G}^{R*}_{\tilde{\chi}_1^0\tilde{\chi}_{\ell}^0A}~,\qquad\qquad \xi_{RR}= \lambda_{h_iAG}\mathcal{G}^{R}_{\tilde{\chi}_1^0\tilde{\chi}_{\ell}^0G}\mathcal{G}^{L*}_{\tilde{\chi}_1^0\tilde{\chi}_{\ell}^0A}~,
\end{align}
where $\lambda_{h_iAG} = -\dfrac{g_2M_Z}{2c_W}s_{2\beta}\, D_{h_i}$,\,\, 
$\mathcal{G}^{L}_{\tilde{\chi}_1^0\tilde{\chi}_{\ell}^0G} = ig_2\big(-Q_{\ell 1}^{\prime\prime*}c_\beta - S_{\ell 1}^{\prime\prime*}s_\beta\big)$, and $\mathcal{G}^{R}_{\tilde{\chi}_1^0\tilde{\chi}_{\ell}^0G} = ig_2\big(Q_{1\ell}^{\prime\prime}c_\beta + S_{1\ell}^{\prime\prime}s_\beta\big)$.\\
\vspace{0.5cm}\newline
(6) $h_i=h_1/h_2$, $F=\tilde{\chi}_{\ell}^0$, and $S=S^\prime=G$.
\begin{align}
\xi_{LL}= \lambda_{h_iGG}\mathcal{G}^{L}_{\tilde{\chi}_1^0\tilde{\chi}_{\ell}^0G}\mathcal{G}^{R*}_{\tilde{\chi}_1^0\tilde{\chi}_{\ell}^0G}~,\qquad\qquad\xi_{LR}= \lambda_{h_iGG}\mathcal{G}^{L}_{\tilde{\chi}_1^0\tilde{\chi}_{\ell}^0G}\mathcal{G}^{L*}_{\tilde{\chi}_1^0\tilde{\chi}_{\ell}^0G}~,\\
\xi_{RL}= \lambda_{h_iGG}\mathcal{G}^{R}_{\tilde{\chi}_1^0\tilde{\chi}_{\ell}^0G}\mathcal{G}^{R*}_{\tilde{\chi}_1^0\tilde{\chi}_{\ell}^0G}~,\qquad\qquad \xi_{RR}= \lambda_{h_iGG}\mathcal{G}^{R}_{\tilde{\chi}_1^0\tilde{\chi}_{\ell}^0G}\mathcal{G}^{L*}_{\tilde{\chi}_1^0\tilde{\chi}_{\ell}^0G}~,
\end{align}
where $\lambda_{h_iGG}=-\dfrac{g_2M_Z}{2c_W}c_{2\beta}\,D^\prime_{h_i}$, with $D^\prime_{h_i} = \left\{
\begin{array}{ll}
-s_{\beta+\alpha}; & h_i=h_1 \\
\,\,\,\,\,c_{\beta+\alpha}; & h_i=h_2 \\
\end{array} 
\right.,$\\
\vspace{0.5cm}\newline
(7) $h_i=h_1/h_2$, $F=\tilde{\chi}_{\ell}^{\pm}$, and $S=S^\prime=H^{\pm}$.
\begin{align}
\xi_{LL}= \lambda_{h_iH^\pm H^\pm}\mathcal{G}^{L}_{\tilde{\chi}_1^0\tilde{\chi}_{\ell}^\pm H^\pm}\mathcal{G}^{R*}_{\tilde{\chi}_1^0\tilde{\chi}_{\ell}^\pm H^\pm}~,\qquad\qquad\xi_{LR}= \lambda_{h_iH^\pm H^\pm}\mathcal{G}^{L}_{\tilde{\chi}_1^0\tilde{\chi}_{\ell}^\pm H^\pm}\mathcal{G}^{L*}_{\tilde{\chi}_1^0\tilde{\chi}_{\ell}^\pm H^\pm}~,\\
\xi_{RL}= \lambda_{h_iH^\pm H^\pm}\mathcal{G}^{R}_{\tilde{\chi}_1^0\tilde{\chi}_{\ell}^\pm H^\pm}\mathcal{G}^{R*}_{\tilde{\chi}_1^0\tilde{\chi}_{\ell}^\pm H^\pm}~,\qquad\qquad \xi_{RR}= \lambda_{h_iH^\pm H^\pm}\mathcal{G}^{R}_{\tilde{\chi}_1^0\tilde{\chi}_{\ell}^\pm H^\pm}\mathcal{G}^{L*}_{\tilde{\chi}_1^0\tilde{\chi}_{\ell}^\pm H^\pm}~,
\end{align}
where, $\lambda_{h_iH^\pm H^\pm} = -g_2 A_{h_i}$, with $A_{h_i} = \left\{
\begin{array}{ll}
M_Ws_{\beta-\alpha}+\dfrac{M_Z}{2c_W}c_{2\beta} s_{\beta+\alpha}; & h_i=h \\
M_W c_{\beta-\alpha}-\dfrac{M_Z}{2c_W}c_{2\beta}c_{\beta+\alpha}; & h_i=H \\
\end{array} 
\right.,$ \\ $\mathcal{G}^{L}_{\tilde{\chi}_1^0\tilde{\chi}_{\ell}^\pm H^\pm}=-g_2 Q^{\prime L}_{1\ell}$, and $\mathcal{G}^{R}_{\tilde{\chi}_1^0\tilde{\chi}_{\ell}^\pm H^\pm} = -g_2 Q^{\prime R}_{1\ell}~.$\\
\vspace{0.5cm}\newline
(8) $h_i=h_1/h_2$, $F=\tilde{\chi}_{\ell}^{\pm}$, and $S=H^\pm$, $S^\prime=G^{\pm}$ or $S=G^\pm$, $S^\prime=H^{\pm}$.
\begin{align}
\xi_{LL}= \lambda_{h_iH^\pm G^\pm}\mathcal{G}^{L}_{\tilde{\chi}_1^0\tilde{\chi}_{\ell}^\pm G^\pm}\mathcal{G}^{R*}_{\tilde{\chi}_1^0\tilde{\chi}_{\ell}^\pm H^\pm}~,\qquad\qquad\xi_{LR}= \lambda_{h_iH^\pm G^\pm}\mathcal{G}^{L}_{\tilde{\chi}_1^0\tilde{\chi}_{\ell}^\pm G^\pm}\mathcal{G}^{L*}_{\tilde{\chi}_1^0\tilde{\chi}_{\ell}^\pm H^\pm}~,\\
\xi_{RL}= \lambda_{h_iH^\pm G^\pm}\mathcal{G}^{R}_{\tilde{\chi}_1^0\tilde{\chi}_{\ell}^\pm G^\pm}\mathcal{G}^{R*}_{\tilde{\chi}_1^0\tilde{\chi}_{\ell}^\pm H^\pm}~,\qquad\qquad \xi_{RR}= \lambda_{h_iH^\pm G^\pm}\mathcal{G}^{R}_{\tilde{\chi}_1^0\tilde{\chi}_{\ell}^\pm G^\pm}\mathcal{G}^{L*}_{\tilde{\chi}_1^0\tilde{\chi}_{\ell}^\pm H^\pm}~,
\end{align}
where $\lambda_{h_iH^\pm G^\pm} = -\frac{g_2M_W}{2}A^{\prime}_{h_i}$, with $A^\prime_{h_i} = \left\{
\begin{array}{ll}
\,\,\,\,\,\dfrac{s_{2\beta} s_{\beta+\alpha}}{c_W^2}-c_{\beta-\alpha}; & h_i=h_1 \\
-\dfrac{s_{2\beta} c_{\beta+\alpha}}{c_W^2}-s_{\beta-\alpha}; & h_i=h_2 \\
\end{array} 
\right.,$\\
 $\mathcal{G}^{L}_{\tilde{\chi}_1^0\tilde{\chi}_{\ell}^\pm G^\pm} = -g_2 t_\beta Q_{1\ell}^{\prime L}$, and $\mathcal{G}^{R}_{\tilde{\chi}_1^0\tilde{\chi}_{\ell}^\pm G^\pm} = \dfrac{g_2}{t_\beta} Q_{1\ell}^{\prime R}$.\\ 
\vspace{0.5cm}\newline
(9) $h_i=h_1/h_2$, $F=\tilde{\chi}_{\ell}^{\pm}$, and $S=S^\prime=G^{\pm}$.
\begin{align}
\xi_{LL}= \lambda_{h_iG^\pm G^\pm}\mathcal{G}^{L}_{\tilde{\chi}_1^0\tilde{\chi}_{\ell}^\pm G^\pm}\mathcal{G}^{R*}_{\tilde{\chi}_1^0\tilde{\chi}_{\ell}^\pm G^\pm}~,\qquad\qquad\xi_{LR}= \lambda_{h_iG^\pm G^\pm}\mathcal{G}^{L}_{\tilde{\chi}_1^0\tilde{\chi}_{\ell}^\pm G^\pm}\mathcal{G}^{L*}_{\tilde{\chi}_1^0\tilde{\chi}_{\ell}^\pm G^\pm}~,\\
\xi_{RL}= \lambda_{h_iG^\pm G^\pm}\mathcal{G}^{R}_{\tilde{\chi}_1^0\tilde{\chi}_{\ell}^\pm G^\pm}\mathcal{G}^{R*}_{\tilde{\chi}_1^0\tilde{\chi}_{\ell}^\pm G^\pm}~,\qquad\qquad \xi_{RR}= \lambda_{h_iG^\pm G^\pm}\mathcal{G}^{R}_{\tilde{\chi}_1^0\tilde{\chi}_{\ell}^\pm G^\pm}\mathcal{G}^{L*}_{\tilde{\chi}_1^0\tilde{\chi}_{\ell}^\pm G^\pm}~,
\end{align} 
where $\lambda_{h_iG^\pm G^\pm} = -\dfrac{g_2M_Z}{2c_W} c_{2\beta} \,D^\prime_{h_i}$.\\
\vspace{0.5cm}\newline
	(10) $h_i=h_1/h_2$, $F=q_i$, $S=\tilde{q}_t$, $S^\prime=\tilde{q}_s$
	\begin{align}
	\xi_{LL} = \lambda_{h_i\tilde{q}_t \tilde{q}_s} \mathcal{G}_{\tilde{\chi}_1^0q_i \tilde{q}_s}^{L}\mathcal{G}_{\tilde{\chi}_1^0q_i\tilde{q}_t}^{R*}~,\qquad\qquad \xi_{LR} = \lambda_{h_i\tilde{q}_t \tilde{q}_s} \mathcal{G}_{\tilde{\chi}_1^0q_i \tilde{q}_s}^{L}\mathcal{G}_{\tilde{\chi}_1^0q_i\tilde{q}_t}^{L*}~,\\
	\xi_{RL} = \lambda_{h_i\tilde{q}_t \tilde{q}_s} \mathcal{G}_{\tilde{\chi}_1^0q_i \tilde{q}_s}^{R}\mathcal{G}_{\tilde{\chi}_1^0q_i\tilde{q}_t}^{R*}~,\qquad\qquad \xi_{RR} = \lambda_{h_i\tilde{q}_t \tilde{q}_s} \mathcal{G}_{\tilde{\chi}_1^0q_i \tilde{q}_s}^{R}\mathcal{G}_{\tilde{\chi}_1^0q_i\tilde{q}_t}^{L*}~,
	\end{align}
	where $\lambda_{h_i\tilde{q}_t \tilde{q}_s}= C\big[h_i, \tilde{q}_t, \tilde{q}_s\big]$ are defined as 
	\begin{align*}
	C\big[h_1, \tilde{u}_t,\tilde{u}_s\big]&= c_A\big[\tilde{u}_t, \tilde{u}_s\big]c_\alpha -c_\mu \big[\tilde{u}_t, \tilde{u}_s\big]s_\alpha+c_g\big[\tilde{u}_t,\tilde{u}_s\big]s_{\alpha+\beta}~,\\	
	C\big[h_1, \tilde{d}_t,\tilde{d}_s\big]&= -c_A\big[\tilde{d}_t, \tilde{d}_s\big]s_\alpha +c_\mu \big[\tilde{d}_t, \tilde{d}_s\big]c_\alpha+c_g\big[\tilde{d}_t,\tilde{d}_s\big]s_{\alpha+\beta}~,\\	
	C\big[h_2,\tilde{u}_t,\tilde{u}_s\big]&=c_A\big[\tilde{u}_t, \tilde{u}_s\big]s_\alpha +c_\mu \big[\tilde{u}_t, \tilde{u}_s\big]c_\alpha-c_g\big[\tilde{u}_t,\tilde{u}_s\big]c_{\alpha+\beta}~,\\
	C\big[h_2,\tilde{d}_t,\tilde{d}_s\big]&=c_A\big[\tilde{d}_t, \tilde{d}_s\big]c_\alpha +c_\mu \big[\tilde{d}_t, \tilde{d}_s\big]s_\alpha-c_g\big[\tilde{d}_t,\tilde{d}_s\big]c_{\alpha+\beta}~,
	\end{align*}
	
	with
	\begin{align*}
	c_A\big[\tilde{u}_t,\tilde{u}_s\big] =& \frac{g_2}{M_W s_\beta} \biggl\{\frac{1}{2}\bigg[W_{it}^{\tilde{u}*}W_{j+3\,\,s}^{\tilde{u}}\big(\mathbf{m_u}A^{u*}\big)_{ij} + W_{j+3\,\,t}^{\tilde{u}*}W_{is}^{\tilde{u}}\big(\mathbf{m_u}^*A^{u}\big)_{ij}\bigg]\biggr\}\nonumber\\
	&- m^2_{u_k}\big[U_{ik}^{u_L}U_{jk}^{u_L*}W_{it}^{\tilde{u}*}W_{js}^{\tilde{u}} + U_{ik}^{u_R}U_{jk}^{u_R*}W_{i+3\,\,t}^{\tilde{u}*}W_{j+3\,\,s}^{\tilde{u}}\big]~,
\\
	c_A\big[\tilde{d}_t,\tilde{d}_s\big] =& \frac{g_2}{M_W c_\beta} \biggl\{\frac{1}{2}\bigg[W_{it}^{\tilde{d}*}W_{j+3\,\,s}^{\tilde{d}}\big(\mathbf{m_d}A^{d*}\big)_{ij} + W_{j+3\,\,t}^{\tilde{d}*}W_{is}^{\tilde{d}}\big(\mathbf{m_d}^*A^{d}\big)_{ij}\bigg]\biggr\}\nonumber\\
	&- m^2_{d_k}\big[U_{ik}^{d_L}U_{jk}^{d_L*}W_{it}^{\tilde{d}*}W_{js}^{\tilde{d}} + U_{ik}^{d_R}U_{jk}^{d_R*}W_{i+3\,\,t}^{\tilde{d}*}W_{j+3\,\,s}^{\tilde{d}}\big]~,
	\\
	c_\mu\big[\tilde{u}_t,\tilde{u}_s\big] =& \frac{g_2}{2M_W s_\beta} m_{u_k}\big[\mu U_{ik}^{u_L}U_{jk}^{u_R*}W_{it}^{\tilde{u}*}W_{j+3\,\,s}^{\tilde{u}} + \mu^* U_{ik}^{u_L*}U_{jk}^{u_R}W_{j+3\,\,t}^{\tilde{u}*}W_{is}^{\tilde{u}}\big]~,\\
	c_\mu\big[\tilde{d}_t,\tilde{d}_s\big] =& \frac{g_2}{2M_W c_\beta} m_{d_k}\big[\mu U_{ik}^{d_L}U_{jk}^{d_R*}W_{it}^{\tilde{d}*}W_{j+3\,\,s}^{\tilde{d}} + \mu^* U_{ik}^{d_L*}U_{jk}^{d_R}W_{j+3\,\,t}^{\tilde{d}*}W_{is}^{\tilde{d}}\big]~,
	\\
	c_g\big[\tilde{u}_t, \tilde{u}_s\big] =& \frac{g_2 M_W}{2} \bigg[W_{it}^{\tilde{u}*}W_{is}^{\tilde{u}}\bigg(1-\frac{1}{3}t_W^2\bigg) + \frac{4}{3}W_{i+3\,\,t}^{\tilde{u}*}W_{i+3\,\,s}^{\tilde{u}}t_W^2\bigg]~,\\
	c_g\big[\tilde{d}_t, \tilde{d}_s\big] =& -\frac{g_2 M_W}{2} \bigg[W_{it}^{\tilde{d}*}W_{is}^{\tilde{d}}\bigg(1+\frac{1}{3}t_W^2\bigg) + \frac{2}{3}W_{i+3\,\,t}^{\tilde{d}*}W_{i+3\,\,s}^{\tilde{d}}t_W^2\bigg]~.
	\end{align*}
In the present context, only the third-generation quarks and squarks have been included, and no flavor mixing in the squark sector has been assumed. 
Further, the third-generation squarks are kept lighter than the first two generations. 
Thus, in the subscripts only $t, s \in \{1,2\}$ are relevant; further $i,j,k=3$. 
The same treatment follows for all the subsequent diagrams, including the third 
generation (s)quark loops. \\
\vspace{1cm}

{\bf Topology-(1b):}\\
The respective Feynman diagram is shown in Fig. \ref{fig:topology1}(b).
\begin{center}
\begin{align}
i\delta\Gamma^{(b)} = &-\frac{i}{16\pi^2}\Big[P_L\Bigl\{\zeta_{LRL}\mathbf{B}_0+\zeta_{LLL} m_F m_{F^\prime}\mathbf{C}_0+\zeta_{LLR}m_{\tilde{\chi}_1^0}m_{F^\prime}\big(\mathbf{C}_0 + \mathbf{C}_1\big) + \zeta_{LRL}m_S^2\mathbf{C}_0  \nonumber\\
& + \zeta_{LRL}m_{\tilde{\chi}_1^0}^2 \big(2\mathbf{C}_0+3\mathbf{C}_1+ 3\mathbf{C}_2\big) - \zeta_{LRL} q^2\big(\mathbf{C}_0 + \mathbf{C}_1 + \mathbf{C}_2\big) - \zeta_{LRL} \big(2m_{\tilde{\chi}_1^0}^2-q^2\big). \nonumber\\
&\big(\mathbf{C}_0+ \mathbf{C}_1 + \mathbf{C}_2\big)+ \zeta_{LRR}m_{\tilde{\chi}_1^0}m_F\mathbf{C}_1+\zeta_{RLL}m_{\tilde{\chi}_1^0}m_F\big(\mathbf{C}_0+\mathbf{C}_2\big)\nonumber\\
&+\zeta_{RLR}m_{\tilde{\chi}_1^0}^2\big(\mathbf{C}_0+\mathbf{C}_1 +\mathbf{C}_2\big)+\zeta_{RRL}m_{\tilde{\chi}_1^0}m_{F^\prime}\mathbf{C}_2\Bigr\}\nonumber\\
&+P_R\Bigl\{\zeta_{RLR}\mathbf{B}_0 + \zeta_{LLR}m_{\tilde{\chi}_1^0}m_{F^\prime}\mathbf{C}_2 + \zeta_{LRL}m_{\tilde{\chi}_1^0}^2\big(\mathbf{C}_0+\mathbf{C}_1+\mathbf{C}_2\big)+\zeta_{LRR}m_{\tilde{\chi}_1^0}m_F\big(\mathbf{C}_0 \nonumber\\
&+ \mathbf{C}_2\big)+ \zeta_{RLL}m_{\tilde{\chi}_1^0}m_F\mathbf{C}_1 + \zeta_{RLR}m_S^2\mathbf{C}_0 + \zeta_{RLR}m_{\tilde{\chi}_1^0}^2\big(2\mathbf{C}_0 + 3\mathbf{C}_1 + 3\mathbf{C}_2\big) - \zeta_{RLR}q^2. \nonumber\\
&\big(\mathbf{C}_0 + \mathbf{C}_1 + \mathbf{C}_2\big)- \zeta_{RLR}\big(2m_{\tilde{\chi}_1^0}^2-q^2\big)\big(\mathbf{C}_0 + \mathbf{C}_1 + \mathbf{C}_2\big) + \zeta_{RRL}m_{\tilde{\chi}_1^0}m_{F^\prime}\big(\mathbf{C}_0 + \mathbf{C}_1\big) \nonumber\\
& + \zeta_{RRR}m_F m_{F^\prime}\mathbf{C}_0\Bigr\}\Big]~,
\end{align}
\end{center}

where $\mathbf{B}_0 = \mathbf{B}_0\big(q^2 ; m_F, m_{F^\prime}\big)$, $\mathbf{C}_i = \mathbf{C}_i\big(m_{\tilde{\chi}_1^0}^2, q^2, m_{\tilde{\chi}_1^0}^2 ; m_S,m_F,m_{F^\prime}\big)$ and 
\begin{center}
\begin{align}
\zeta_{LLL} = \mathscr{G}_{\tilde{\chi}_1^0 F^\prime S}^L \mathscr{G}_{FF^\prime h_i}^L \mathscr{G}_{\tilde{\chi}_1^0 FS}^L~, \qquad\qquad \zeta_{LLR} = \mathscr{G}_{\tilde{\chi}_1^0 F^\prime S}^L \mathscr{G}_{FF^\prime h_i}^L \mathscr{G}_{\tilde{\chi}_1^0 FS}^R~,\\
\zeta_{LRL} = \mathscr{G}_{\tilde{\chi}_1^0 F^\prime S}^L \mathscr{G}_{FF^\prime h_i}^R \mathscr{G}_{\tilde{\chi}_1^0 FS}^L~, \qquad\qquad \zeta_{LRR} = \mathscr{G}_{\tilde{\chi}_1^0 F^\prime S}^L \mathscr{G}_{FF^\prime h_i}^R \mathscr{G}_{\tilde{\chi}_1^0 FS}^R~,\\
\zeta_{RLL} = \mathscr{G}_{\tilde{\chi}_1^0 F^\prime S}^R \mathscr{G}_{FF^\prime h_i}^L \mathscr{G}_{\tilde{\chi}_1^0 FS}^L~, \qquad\qquad \zeta_{RLR} = \mathscr{G}_{\tilde{\chi}_1^0 F^\prime S}^R \mathscr{G}_{FF^\prime h_i}^L \mathscr{G}_{\tilde{\chi}_1^0 FS}^R~,\\
\zeta_{RRL} = \mathscr{G}_{\tilde{\chi}_1^0 F^\prime S}^R \mathscr{G}_{FF^\prime h_i}^R \mathscr{G}_{\tilde{\chi}_1^0 FS}^L~, \qquad\qquad \zeta_{RRR} = \mathscr{G}_{\tilde{\chi}_1^0 F^\prime S}^R \mathscr{G}_{FF^\prime h_i}^R \mathscr{G}_{\tilde{\chi}_1^0 FS}^R~.
\end{align}
\end{center}
 \vspace{0.5cm}
(1) $h_i=h_1/h_2$, $S = h_1/h_2$, $F=\tilde{\chi}_\ell^0$, $F^\prime=\tilde{\chi}_n^0$.
\begin{align}
\zeta_{LLL} = \mathcal{G}_{\tilde{\chi}_1^0 \tilde{\chi}_n^0 h_i}^L \mathcal{G}_{\tilde{\chi}_\ell^0\tilde{\chi}_n^0 h_i}^L \mathcal{G}_{\tilde{\chi}_1^0 \tilde{\chi}_\ell^0 h_i}^{R*}~, \qquad\qquad \zeta_{LLR} = \mathcal{G}_{\tilde{\chi}_1^0 \tilde{\chi}_n^0 h_i}^L \mathcal{G}_{\tilde{\chi}_\ell^0\tilde{\chi}_n^0 h_i}^L \mathcal{G}_{\tilde{\chi}_1^0 \tilde{\chi}_\ell^0 h_i}^{L*}~,\\
\zeta_{LRL} = \mathcal{G}_{\tilde{\chi}_1^0 \tilde{\chi}_n^0 h_i}^L \mathcal{G}_{\tilde{\chi}_\ell^0\tilde{\chi}_n^0 h_i}^R \mathcal{G}_{\tilde{\chi}_1^0 \tilde{\chi}_\ell^0 h_i}^{R*}~, \qquad\qquad \zeta_{LRR} = \mathcal{G}_{\tilde{\chi}_1^0 \tilde{\chi}_n^0 h_i}^L \mathcal{G}_{\tilde{\chi}_\ell^0\tilde{\chi}_n^0 h_i}^R \mathcal{G}_{\tilde{\chi}_1^0 \tilde{\chi}_\ell^0 h_i}^{L*}~,\\
\zeta_{RLL} = \mathcal{G}_{\tilde{\chi}_1^0 \tilde{\chi}_n^0 h_i}^R \mathcal{G}_{\tilde{\chi}_\ell^0\tilde{\chi}_n^0 h_i}^L \mathcal{G}_{\tilde{\chi}_1^0 \tilde{\chi}_\ell^0 h_i}^{R*}~, \qquad\qquad \zeta_{RLR} = \mathcal{G}_{\tilde{\chi}_1^0 \tilde{\chi}_n^0 h_i}^R \mathcal{G}_{\tilde{\chi}_\ell^0\tilde{\chi}_n^0 h_i}^L \mathcal{G}_{\tilde{\chi}_1^0 \tilde{\chi}_\ell^0 h_i}^{L*}~,\\
\zeta_{RRL} = \mathcal{G}_{\tilde{\chi}_1^0 \tilde{\chi}_n^0 h_i}^R \mathcal{G}_{\tilde{\chi}_\ell^0\tilde{\chi}_n^0 h_i}^R \mathcal{G}_{\tilde{\chi}_1^0 \tilde{\chi}_\ell^0 h_i}^{R*}~, \qquad\qquad \zeta_{RRR} = \mathcal{G}_{\tilde{\chi}_1^0 \tilde{\chi}_n^0 h_i}^R \mathcal{G}_{\tilde{\chi}_\ell^0\tilde{\chi}_n^0 h_i}^R \mathcal{G}_{\tilde{\chi}_1^0 \tilde{\chi}_\ell^0 h_i}^{L*}~,
\end{align}
where $\mathcal{G}_{\tilde{\chi}_\ell^0\tilde{\chi}_n^0 h_i}^L = \left\{
\begin{array}{ll}
\,\,\,\,\, g_2\big(Q_{\ell n}^{\prime\prime*}\,s_\alpha+S_{\ell n}^{\prime\prime*}\,c_\alpha\big); & h_i=h_1 \\
-g_2\big(Q_{\ell n}^{\prime\prime*}\,c_\alpha-S_{\ell n}^{\prime\prime*}\,s_\alpha\big); & h_i=h_2 \\
\end{array} 
\right.$~,\\
\, \indent and\, 
 $\mathcal{G}_{\tilde{\chi}_\ell^0\tilde{\chi}_n^0 h_i}^R = \left\{
\begin{array}{ll}
\,\,\,\,\, g_2\big(Q_{n\ell}^{\prime\prime}\,s_\alpha+S_{n\ell}^{\prime\prime}\,c_\alpha\big); & h_i=h_1 \\
-g_2\big(Q_{n\ell}^{\prime\prime}\,c_\alpha-S_{n\ell}^{\prime\prime}\,s_\alpha\big); & h_i=h_2 \\
\end{array} 
\right..$\\
\vspace{0.5cm}\newline
(2) $h_i=h_1/h_2$, $S = A$, $F=\tilde{\chi}_\ell^0$, $F^\prime=\tilde{\chi}_n^0$.
\begin{align}
\zeta_{LLL} = \mathcal{G}_{\tilde{\chi}_1^0 \tilde{\chi}_n^0 A}^L \mathcal{G}_{\tilde{\chi}_\ell^0\tilde{\chi}_n^0 h_i}^L \mathcal{G}_{\tilde{\chi}_1^0 \tilde{\chi}_\ell^0 A}^{R*}~, \qquad\qquad \zeta_{LLR} = \mathcal{G}_{\tilde{\chi}_1^0 \tilde{\chi}_n^0 A}^L \mathcal{G}_{\tilde{\chi}_\ell^0\tilde{\chi}_n^0 h_i}^L \mathcal{G}_{\tilde{\chi}_1^0 \tilde{\chi}_\ell^0 A}^{L*}~,\\
\zeta_{LRL} = \mathcal{G}_{\tilde{\chi}_1^0 \tilde{\chi}_n^0 A}^L \mathcal{G}_{\tilde{\chi}_\ell^0\tilde{\chi}_n^0 h_i}^R \mathcal{G}_{\tilde{\chi}_1^0 \tilde{\chi}_\ell^0 A}^{R*}~, \qquad\qquad \zeta_{LRR} = \mathcal{G}_{\tilde{\chi}_1^0 \tilde{\chi}_n^0 A}^L \mathcal{G}_{\tilde{\chi}_\ell^0\tilde{\chi}_n^0 h_i}^R \mathcal{G}_{\tilde{\chi}_1^0 \tilde{\chi}_\ell^0 A}^{L*}~,\\
\zeta_{RLL} = \mathcal{G}_{\tilde{\chi}_1^0 \tilde{\chi}_n^0 A}^R \mathcal{G}_{\tilde{\chi}_\ell^0\tilde{\chi}_n^0 h_i}^L \mathcal{G}_{\tilde{\chi}_1^0 \tilde{\chi}_\ell^0 A}^{R*}~, \qquad\qquad \zeta_{RLR} = \mathcal{G}_{\tilde{\chi}_1^0 \tilde{\chi}_n^0 A}^R \mathcal{G}_{\tilde{\chi}_\ell^0\tilde{\chi}_n^0 h_i}^L \mathcal{G}_{\tilde{\chi}_1^0 \tilde{\chi}_\ell^0 A}^{L*}~,\\
\zeta_{RRL} = \mathcal{G}_{\tilde{\chi}_1^0 \tilde{\chi}_n^0 A}^R \mathcal{G}_{\tilde{\chi}_\ell^0\tilde{\chi}_n^0 h_i}^R \mathcal{G}_{\tilde{\chi}_1^0 \tilde{\chi}_\ell^0 A}^{R*}~, \qquad\qquad \zeta_{RRR} = \mathcal{G}_{\tilde{\chi}_1^0 \tilde{\chi}_n^0 A}^R \mathcal{G}_{\tilde{\chi}_\ell^0\tilde{\chi}_n^0 h_i}^R \mathcal{G}_{\tilde{\chi}_1^0 \tilde{\chi}_\ell^0 A}^{L*}~,
\end{align}
where $\mathcal{G}_{\tilde{\chi}_1^0 \tilde{\chi}_n^0 A}^L = i\big(Q_{n1}^{\prime\prime*}s_\beta - S_{n1}^{\prime\prime*}c_\beta\big)$\, and \, $\mathcal{G}_{\tilde{\chi}_1^0 \tilde{\chi}_n^0 A}^R = i\big(-Q_{1n}^{\prime\prime}s_\beta + S_{1n}^{\prime\prime}c_\beta\big)$.\\
\vspace{0.5cm}\newline
(3) $h_i=h_1/h_2$, $S = G$, $F=\tilde{\chi}_\ell^0$, $F^\prime=\tilde{\chi}_n^0$.
\begin{align}
\zeta_{LLL} = \mathcal{G}_{\tilde{\chi}_1^0 \tilde{\chi}_n^0 G}^L \mathcal{G}_{\tilde{\chi}_\ell^0\tilde{\chi}_n^0 h_i}^L \mathcal{G}_{\tilde{\chi}_1^0 \tilde{\chi}_\ell^0 G}^{R*}~, \qquad\qquad \zeta_{LLR} = \mathcal{G}_{\tilde{\chi}_1^0 \tilde{\chi}_n^0 G}^L \mathcal{G}_{\tilde{\chi}_\ell^0\tilde{\chi}_n^0 h_i}^L \mathcal{G}_{\tilde{\chi}_1^0 \tilde{\chi}_\ell^0 G}^{L*}~,\\
\zeta_{LRL} = \mathcal{G}_{\tilde{\chi}_1^0 \tilde{\chi}_n^0 G}^L \mathcal{G}_{\tilde{\chi}_\ell^0\tilde{\chi}_n^0 h_i}^R \mathcal{G}_{\tilde{\chi}_1^0 \tilde{\chi}_\ell^0 G}^{R*}~, \qquad\qquad \zeta_{LRR} = \mathcal{G}_{\tilde{\chi}_1^0 \tilde{\chi}_n^0 G}^L \mathcal{G}_{\tilde{\chi}_\ell^0\tilde{\chi}_n^0 h_i}^R \mathcal{G}_{\tilde{\chi}_1^0 \tilde{\chi}_\ell^0 G}^{L*}~,\\
\zeta_{RLL} = \mathcal{G}_{\tilde{\chi}_1^0 \tilde{\chi}_n^0 G}^R \mathcal{G}_{\tilde{\chi}_\ell^0\tilde{\chi}_n^0 h_i}^L \mathcal{G}_{\tilde{\chi}_1^0 \tilde{\chi}_\ell^0 G}^{R*}~, \qquad\qquad \zeta_{RLR} = \mathcal{G}_{\tilde{\chi}_1^0 \tilde{\chi}_n^0 G}^R \mathcal{G}_{\tilde{\chi}_\ell^0\tilde{\chi}_n^0 h_i}^L \mathcal{G}_{\tilde{\chi}_1^0 \tilde{\chi}_\ell^0 G}^{L*}~,\\
\zeta_{RRL} = \mathcal{G}_{\tilde{\chi}_1^0 \tilde{\chi}_n^0 G}^R \mathcal{G}_{\tilde{\chi}_\ell^0\tilde{\chi}_n^0 h_i}^R \mathcal{G}_{\tilde{\chi}_1^0 \tilde{\chi}_\ell^0 G}^{R*}~, \qquad\qquad \zeta_{RRR} = \mathcal{G}_{\tilde{\chi}_1^0 \tilde{\chi}_n^0 G}^R \mathcal{G}_{\tilde{\chi}_\ell^0\tilde{\chi}_n^0 h_i}^R \mathcal{G}_{\tilde{\chi}_1^0 \tilde{\chi}_\ell^0 G}^{L*}~.
\end{align}
\vspace{0.5cm}\newline
(4) $h_i=h_1/h_2$, $S = H^\pm$, $F=\tilde{\chi}_\ell^\pm$, $F^\prime=\tilde{\chi}_n^\pm$.
\begin{align}
\zeta_{LLL} = \mathcal{G}_{\tilde{\chi}_1^0 \tilde{\chi}_n^\pm H^\pm}^L \mathcal{G}_{\tilde{\chi}_\ell^\pm\tilde{\chi}_n^\pm h_i}^L \mathcal{G}_{\tilde{\chi}_1^0 \tilde{\chi}_\ell^\pm H^\pm}^{R*}~, \qquad\qquad \zeta_{LLR} = \mathcal{G}_{\tilde{\chi}_1^0 \tilde{\chi}_n^\pm H^\pm}^L \mathcal{G}_{\tilde{\chi}_\ell^\pm\tilde{\chi}_n^\pm h_i}^L \mathcal{G}_{\tilde{\chi}_1^0 \tilde{\chi}_\ell^\pm H^\pm}^{L*}~,\\
\zeta_{LRL} = \mathcal{G}_{\tilde{\chi}_1^0 \tilde{\chi}_n^\pm H^\pm}^L \mathcal{G}_{\tilde{\chi}_\ell^\pm\tilde{\chi}_n^\pm h_i}^R \mathcal{G}_{\tilde{\chi}_1^0 \tilde{\chi}_\ell^\pm H^\pm}^{R*}~, \qquad\qquad \zeta_{LRR} = \mathcal{G}_{\tilde{\chi}_1^0 \tilde{\chi}_n^\pm H^\pm}^L \mathcal{G}_{\tilde{\chi}_\ell^\pm\tilde{\chi}_n^\pm h_i}^R \mathcal{G}_{\tilde{\chi}_1^0 \tilde{\chi}_\ell^\pm H^\pm}^{L*}~,\\
\zeta_{RLL} = \mathcal{G}_{\tilde{\chi}_1^0 \tilde{\chi}_n^\pm H^\pm}^R \mathcal{G}_{\tilde{\chi}_\ell^\pm\tilde{\chi}_n^\pm h_i}^L \mathcal{G}_{\tilde{\chi}_1^0 \tilde{\chi}_\ell^\pm H^\pm}^{R*}~, \qquad\qquad \zeta_{RLR} = \mathcal{G}_{\tilde{\chi}_1^0 \tilde{\chi}_n^\pm H^\pm}^R \mathcal{G}_{\tilde{\chi}_\ell^\pm\tilde{\chi}_n^\pm h_i}^L \mathcal{G}_{\tilde{\chi}_1^0 \tilde{\chi}_\ell^\pm H^\pm}^{L*}~,\\
\zeta_{RRL} = \mathcal{G}_{\tilde{\chi}_1^0 \tilde{\chi}_n^\pm H^\pm}^R \mathcal{G}_{\tilde{\chi}_\ell^\pm\tilde{\chi}_n^\pm h_i}^R \mathcal{G}_{\tilde{\chi}_1^0 \tilde{\chi}_\ell^\pm H^\pm}^{R*}~, \qquad\qquad \zeta_{RRR} = \mathcal{G}_{\tilde{\chi}_1^0 \tilde{\chi}_n^\pm H^\pm}^R \mathcal{G}_{\tilde{\chi}_\ell^\pm\tilde{\chi}_n^\pm h_i}^R \mathcal{G}_{\tilde{\chi}_1^0 \tilde{\chi}_\ell^\pm H^\pm}^{L*}~.
\end{align}
\vspace{0.5cm}\newline
(5) $h_i=h_1/h_2$, $S = G^\pm$, $F=\tilde{\chi}_\ell^\pm$, $F^\prime=\tilde{\chi}_n^\pm$.
\begin{align}
\zeta_{LLL} = \mathcal{G}_{\tilde{\chi}_1^0 \tilde{\chi}_n^\pm G^\pm}^L \mathcal{G}_{\tilde{\chi}_\ell^\pm\tilde{\chi}_n^\pm h_i}^L \mathcal{G}_{\tilde{\chi}_1^0 \tilde{\chi}_\ell^\pm G^\pm}^{R*}~, \qquad\qquad \zeta_{LLR} = \mathcal{G}_{\tilde{\chi}_1^0 \tilde{\chi}_n^\pm G^\pm}^L \mathcal{G}_{\tilde{\chi}_\ell^\pm\tilde{\chi}_n^\pm h_i}^L \mathcal{G}_{\tilde{\chi}_1^0 \tilde{\chi}_\ell^\pm G^\pm}^{L*}~,\\
\zeta_{LRL} = \mathcal{G}_{\tilde{\chi}_1^0 \tilde{\chi}_n^\pm G^\pm}^L \mathcal{G}_{\tilde{\chi}_\ell^\pm\tilde{\chi}_n^\pm h_i}^R \mathcal{G}_{\tilde{\chi}_1^0 \tilde{\chi}_\ell^\pm G^\pm}^{R*}~, \qquad\qquad \zeta_{LRR} = \mathcal{G}_{\tilde{\chi}_1^0 \tilde{\chi}_n^\pm G^\pm}^L \mathcal{G}_{\tilde{\chi}_\ell^\pm\tilde{\chi}_n^\pm h_i}^R \mathcal{G}_{\tilde{\chi}_1^0 \tilde{\chi}_\ell^\pm G^\pm}^{L*}~,\\
\zeta_{RLL} = \mathcal{G}_{\tilde{\chi}_1^0 \tilde{\chi}_n^\pm G^\pm}^R \mathcal{G}_{\tilde{\chi}_\ell^\pm\tilde{\chi}_n^\pm h_i}^L \mathcal{G}_{\tilde{\chi}_1^0 \tilde{\chi}_\ell^\pm G^\pm}^{R*}~, \qquad\qquad \zeta_{RLR} = \mathcal{G}_{\tilde{\chi}_1^0 \tilde{\chi}_n^\pm G^\pm}^R \mathcal{G}_{\tilde{\chi}_\ell^\pm\tilde{\chi}_n^\pm h_i}^L \mathcal{G}_{\tilde{\chi}_1^0 \tilde{\chi}_\ell^\pm G^\pm}^{L*}~,\\
\zeta_{RRL} = \mathcal{G}_{\tilde{\chi}_1^0 \tilde{\chi}_n^\pm G^\pm}^R \mathcal{G}_{\tilde{\chi}_\ell^\pm\tilde{\chi}_n^\pm h_i}^R \mathcal{G}_{\tilde{\chi}_1^0 \tilde{\chi}_\ell^\pm G^\pm}^{R*}~, \qquad\qquad \zeta_{RRR} = \mathcal{G}_{\tilde{\chi}_1^0 \tilde{\chi}_n^\pm G^\pm}^R \mathcal{G}_{\tilde{\chi}_\ell^\pm\tilde{\chi}_n^\pm h_i}^R \mathcal{G}_{\tilde{\chi}_1^0 \tilde{\chi}_\ell^\pm G^\pm}^{L*}~.
\end{align}
\vspace{0.5cm}\newline
	(6) $h_i=h_1/h_2$, $F=F^\prime=q_i$, $S=\tilde{q}_s$.
	\begin{align}
	\zeta_{LLL} = \mathcal{G}_{\tilde{\chi}_1^0 q_i \tilde{q}_s}^L \mathcal{G}_{q_i q_i h_i}^L \mathcal{G}_{\tilde{\chi}_1^0 q_i\tilde{q}_s}^{R*}~, \qquad\qquad \zeta_{LLR} = \mathcal{G}_{\tilde{\chi}_1^0 q_i \tilde{q}_s}^L \mathcal{G}_{q_iq_i h_i}^L \mathcal{G}_{\tilde{\chi}_1^0 q_i\tilde{q}_s}^{L*}~,\\
	\zeta_{LRL} = \mathcal{G}_{\tilde{\chi}_1^0 q_i \tilde{q}_s}^L \mathcal{G}_{q_iq_i h_i}^R \mathcal{G}_{\tilde{\chi}_1^0 q_i\tilde{q}_s}^{R*}~, \qquad\qquad \zeta_{LRR} = \mathcal{G}_{\tilde{\chi}_1^0 q_i \tilde{q}_s}^L \mathcal{G}_{q_iq_i h_i}^R \mathcal{G}_{\tilde{\chi}_1^0 q_i\tilde{q}_s}^{L*}~,\\
	\zeta_{RLL} = \mathcal{G}_{\tilde{\chi}_1^0 q_i \tilde{q}_s}^R \mathcal{G}_{q_iq_i h_i}^L \mathcal{G}_{\tilde{\chi}_1^0 q_i\tilde{q}_s}^{R*}~, \qquad\qquad \zeta_{RLR} = \mathcal{G}_{\tilde{\chi}_1^0 q_i \tilde{q}_s}^R \mathcal{G}_{q_iq_i h_i}^L \mathcal{G}_{\tilde{\chi}_1^0 q_i\tilde{q}_s}^{L*}~,\\
	\zeta_{RRL} = \mathcal{G}_{\tilde{\chi}_1^0q_i \tilde{q}_s}^R \mathcal{G}_{q_iq_i h_i}^R \mathcal{G}_{\tilde{\chi}_1^0 q_i\tilde{q}_s}^{R*}~, \qquad\qquad \zeta_{RRR} = \mathcal{G}_{\tilde{\chi}_1^0 q_i \tilde{q}_s}^R \mathcal{G}_{q_iq_i h_i}^R \mathcal{G}_{\tilde{\chi}_1^0 q_i\tilde{q}_s}^{L*}~,
	\end{align}
	
	where $\mathcal{G}_{q_i q_i h_i}^L= \mathcal{G}_{q_i q_i h_i}^R=-g_2\dfrac{m_{q_i}}{2M_W}\mathcal{X}_{q_i q_i h_i}$,  $\mathcal{G}_{\tilde{\chi}_1^0 q_i \tilde{q}_s}^L=G_{is1}^{q_L}$ and $\mathcal{G}_{\tilde{\chi}_1^0 q_i \tilde{q}_s}^R=G_{is1}^{q_R}$; $q=u,d$; $i= 3$; $s=1, 2 $; with\\
	
    $\mathcal{X}_{u_i u_i h_i} = \left\{
	\begin{array}{ll}
	\dfrac{c_\alpha}{s_\beta}; & h_i=h_1 \\
	\dfrac{s_\alpha}{s_\beta}; & h_i=h_2 \\
	\end{array} 
	\right.~,$\qquad\quad $\mathcal{X}_{d_i d_i h_i} = \left\{
	\begin{array}{ll}
	\dfrac{-s_\alpha}{c_\beta}; & h_i=h_1 \\
	\dfrac{c_\alpha}{c_\beta}; & h_i=h_2 \\
	\end{array} 
	\right.~,$	
    \begin{align*}
    \mathrm{G}_{is1}^{u_L}&= -\sqrt{2}g_2 \Bigg(\frac{1}{2}N^{*}_{12}+\frac{1}{6}\tan\theta_W N^{*}_{11}\Bigg) W_{js}^{\tilde{u}*}U_{ji}^{u_L}-\frac{g_2}{\sqrt{2}M_W\sin\beta}m_{u_i}N^{*}_{14} W_{j+3}^{\tilde{u}*}U_{ji}^{u_R}~,\\
    \mathrm{G}_{is1}^{u_R}&=\frac{2\sqrt{2}}{3}g_2\tan\theta_W N_{11} W_{j+3}^{\tilde{u}*}U_{ji}^{u_R}-\frac{g_2}{\sqrt{2}M_W\sin\beta}m_{u_i}N_{14} W_{js}^{\tilde{u}*}U_{ji}^{u_L}~,\\
    \mathrm{G}_{is1}^{d_L}&= \sqrt{2}g_2 \Bigg(\frac{1}{2}N^{*}_{12}-\frac{1}{6}\tan\theta_W N^{*}_{11}\Bigg) W_{js}^{\tilde{d}*}U_{ji}^{d_L}-\frac{g_2}{\sqrt{2}M_W\cos\beta}m_{d_i}N^{*}_{13} W_{j+3}^{\tilde{d}*}U_{ji}^{d_R}~,\\
    \mathrm{G}_{is1}^{d_R}&=-\frac{\sqrt{2}}{3}g_2\tan\theta_W N_{11} W_{j+3}^{\tilde{d}*}U_{ji}^{d_R}-\frac{g_2}{\sqrt{2}M_W\cos\beta}m_{d_i}N_{13} W_{js}^{\tilde{d}*}U_{ji}^{d_L}~.
    \end{align*}	
For third-generation quarks and squarks: $i,j=3$ and $s \in \{1,2\}$.\\
\vspace{1cm}

\begin{figure}[H]
	\centering
	\includegraphics[width=0.8\linewidth]{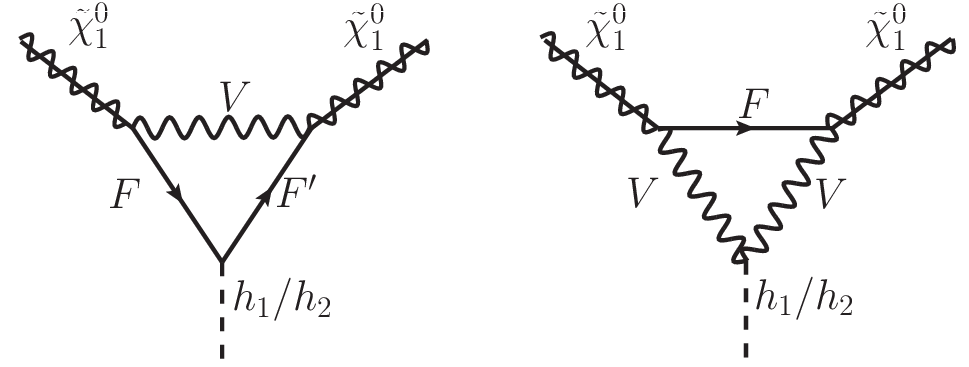}\\
		(a)~~~~~~~~~~~~~~~~~~~~~~~~~~~~~~~~~~~~~~~~~~~~~~~~~~~~~~~~~~~(b)
	\caption{Topology 2(a) and 2(b)}
	\label{fig:topology2}
\end{figure}

{\bf Topology-(2a):}\\
The respective Feynman diagram is shown in Fig. \ref{fig:topology2}(a).
\begin{center}
\begin{align}
i\delta\Gamma^{(c)} =& \frac{i}{16\pi^2}\Big[P_L\Bigl\{\Lambda_{LLL}m_{\tilde{\chi}_1^0} m_{F^\prime}\big(2-d\big)\mathbf{C}_2 + \Lambda_{LRL}m_{\tilde{\chi}_1^0} m_F \big(2-d\big)\big(\mathbf{C}_0+\mathbf{C}_2\big) \nonumber\\
&+ \Lambda_{LRR}m_{\tilde{\chi}_1^0}^2\big(d-4\big)\big(\mathbf{C}_0+\mathbf{C}_1+\mathbf{C}_2\big) + \Lambda_{RLL}\bigl\{d\mathbf{B}_0 + \big(4m_{\tilde{\chi}_1^0}^2 + m_V^2d - 2q^2\big)\mathbf{C}_0 \nonumber\\
&+ \big(4m_{\tilde{\chi}_1^0}^2 + m_{\tilde{\chi}_1^0}^2d - 2q^2\big)\big(\mathbf{C}_1+\mathbf{C}_2\big)\bigr\}+ \Lambda_{RLR}m_{\tilde{\chi}_1^0}m_F\big(2-d\big)\mathbf{C}_1 \nonumber\\
&+ \Lambda_{RRL}m_Fm_{F^\prime}d\mathbf{C}_0 +\Lambda_{RRR}m_{\tilde{\chi}_1^0}m_{F^\prime}\big(2-d\big)\big(\mathbf{C}_0+\mathbf{C}_1\big)\Bigr\} \nonumber\\
&+ P_R\Bigl\{\Lambda_{LLL}m_{\tilde{\chi}_1^0}m_{F^\prime}\big(\mathbf{C}_0+\mathbf{C}_1\big)+\Lambda_{LLR}m_F m_{F^\prime}d\mathbf{C}_0 + \Lambda_{LRL}m_{\tilde{\chi}_1^0}m_F \big(2-d\big)\mathbf{C}_1 \nonumber\\
&+ \Lambda_{LRR}\bigl\{d\mathbf{B}_0 + \big(4m_{\tilde{\chi}_1^0}^2 + m_V^2d -2q^2\big)\mathbf{C}_0 + \big(4m_{\tilde{\chi}_1^0}^2 + m_{\tilde{\chi}_1^0}^2d -2q^2\big)\big(\mathbf{C}_1+\mathbf{C}_2\big)\bigr\} \nonumber\\
&+ \Lambda_{RLL}m_{\tilde{\chi}_1^0}^2\big(d-4\big)\big(\mathbf{C}_0+\mathbf{C}_1+\mathbf{C}_2\big) + \Lambda_{RLR}m_{\tilde{\chi}_1^0}m_F\big(2-d\big)\big(\mathbf{C}_0+\mathbf{C}_2\big) \nonumber\\
&+ \Lambda_{RRR}m_{\tilde{\chi}_1^0}m_{F^\prime}\big(2-d\big)\mathbf{C}_2\Bigr\}\Big]~,
\end{align}
\end{center}

where $\mathbf{B}_0=\mathbf{B}_0\big(q^2, m_F, m_{F^\prime}\big)$, $\mathbf{C}_i = \mathbf{C}_i\big(m_{\tilde{\chi}_1^0}^2, q^2, m_{\tilde{\chi}_1^0}^2 ; m_V, m_F, m_{F^\prime}\big)$ and 
\begin{center}
\begin{align}
\Lambda_{LLL} = \mathscr{G}_{\tilde{\chi}_1^0 F^\prime V}^L \mathscr{G}_{F F^\prime h_i}^L \mathscr{G}_{\tilde{\chi}_1^0 F V}^L~,\qquad\qquad \Lambda_{LLR} = \mathscr{G}_{\tilde{\chi}_1^0 F^\prime V}^L \mathscr{G}_{F F^\prime h_i}^L \mathscr{G}_{\tilde{\chi}_1^0 F V}^R~,\\
\Lambda_{LRL} = \mathscr{G}_{\tilde{\chi}_1^0 F^\prime V}^L \mathscr{G}_{F F^\prime h_i}^R \mathscr{G}_{\tilde{\chi}_1^0 F V}^L~,\qquad\qquad \Lambda_{LRR} = \mathscr{G}_{\tilde{\chi}_1^0 F^\prime V}^L \mathscr{G}_{F F^\prime h_i}^R \mathscr{G}_{\tilde{\chi}_1^0 F V}^R~,\\
\Lambda_{RLL} = \mathscr{G}_{\tilde{\chi}_1^0 F^\prime V}^R \mathscr{G}_{F F^\prime h_i}^L \mathscr{G}_{\tilde{\chi}_1^0 F V}^L~,\qquad\qquad \Lambda_{RLR} = \mathscr{G}_{\tilde{\chi}_1^0 F^\prime V}^R \mathscr{G}_{F F^\prime h_i}^L \mathscr{G}_{\tilde{\chi}_1^0 F V}^R~,\\
\Lambda_{RRL} = \mathscr{G}_{\tilde{\chi}_1^0 F^\prime V}^R \mathscr{G}_{F F^\prime h_i}^R \mathscr{G}_{\tilde{\chi}_1^0 F V}^L~,\qquad\qquad \Lambda_{RRR} = \mathscr{G}_{\tilde{\chi}_1^0 F^\prime V}^R \mathscr{G}_{F F^\prime h_i}^R \mathscr{G}_{\tilde{\chi}_1^0 F V}^R~.
\end{align}
\end{center}
\vspace{0.5cm}
(1) $h_i=h_1/h_2$, $F=\tilde{\chi}_\ell^0$, $F^\prime=\tilde{\chi}_n^0$, $V=Z$.
\begin{align}
\Lambda_{LLL} = \mathcal{G}_{\tilde{\chi}_1^0 \tilde{\chi}_n^0 Z}^L \mathcal{G}_{\tilde{\chi}_\ell^0 \tilde{\chi}_n^0 h_i}^L \mathcal{G}_{\tilde{\chi}_1^0 \tilde{\chi}_\ell^0 Z}^{L*}~,\qquad\qquad \Lambda_{LLR} = -\mathcal{G}_{\tilde{\chi}_1^0 \tilde{\chi}_n^0 Z}^L \mathcal{G}_{\tilde{\chi}_\ell^0 \tilde{\chi}_n^0 h_i}^L \mathcal{G}_{\tilde{\chi}_1^0 \tilde{\chi}_\ell^0 Z}^L~,\\
\Lambda_{LRL} = \mathcal{G}_{\tilde{\chi}_1^0 \tilde{\chi}_n^0 Z}^L \mathcal{G}_{\tilde{\chi}_\ell^0 \tilde{\chi}_n^0 h_i}^R \mathcal{G}_{\tilde{\chi}_1^0 \tilde{\chi}_\ell^0 Z}^{L*}~,\qquad\qquad \Lambda_{LRR} = -\mathcal{G}_{\tilde{\chi}_1^0 \tilde{\chi}_n^0 Z}^L \mathcal{G}_{\tilde{\chi}_\ell^0 \tilde{\chi}_n^0 h_i}^R \mathcal{G}_{\tilde{\chi}_1^0 \tilde{\chi}_\ell^0 Z}^L~,\\
\Lambda_{RLL} = \mathcal{G}_{\tilde{\chi}_1^0 \tilde{\chi}_n^0 Z}^R \mathcal{G}_{\tilde{\chi}_\ell^0 \tilde{\chi}_n^0 h_i}^L \mathcal{G}_{\tilde{\chi}_1^0 \tilde{\chi}_\ell^0 Z}^{L*}~,\qquad\qquad \Lambda_{RLR} = -\mathcal{G}_{\tilde{\chi}_1^0 \tilde{\chi}_n^0 Z}^R \mathcal{G}_{\tilde{\chi}_\ell^0 \tilde{\chi}_n^0 h_i}^L \mathcal{G}_{\tilde{\chi}_1^0 \tilde{\chi}_\ell^0 Z}^L~,\\
\Lambda_{RRL} = \mathcal{G}_{\tilde{\chi}_1^0 \tilde{\chi}_n^0 Z}^R \mathcal{G}_{\tilde{\chi}_\ell^0 \tilde{\chi}_n^0 h_i}^R \mathcal{G}_{\tilde{\chi}_1^0 \tilde{\chi}_\ell^0 Z}^{L*}~,\qquad\qquad \Lambda_{RRR} = -\mathcal{G}_{\tilde{\chi}_1^0 \tilde{\chi}_n^0 Z}^R \mathcal{G}_{\tilde{\chi}_\ell^0 \tilde{\chi}_n^0 h_i}^R \mathcal{G}_{\tilde{\chi}_1^0 \tilde{\chi}_\ell^0 Z}^L~,
\end{align}
where $\mathcal{G}_{\tilde{\chi}_\ell^0 \tilde{\chi}_n^0 Z}^L = \dfrac{g_2}{c_W}N_{\ell n}^L$\, and \, $\mathcal{G}_{\tilde{\chi}_\ell^0 \tilde{\chi}_n^0 Z}^R = \dfrac{g_2}{c_W}N_{\ell n}^R$.
\vspace{0.5cm}\newline
(2) $h_i=h_1/h_2$, $F=\tilde{\chi}_\ell^\pm$, $F^\prime=\tilde{\chi}_n^\pm$, $V=W^\pm$.
 \begin{align}
\Lambda_{LLL} = \mathcal{G}_{\tilde{\chi}_1^0 \tilde{\chi}_n^\pm W^\pm}^L \mathcal{G}_{\tilde{\chi}_\ell^\pm \tilde{\chi}_n^\pm h_i}^L \mathcal{G}_{\tilde{\chi}_1^0 \tilde{\chi}_\ell^\pm W^\pm}^{R*}~,\qquad\qquad \Lambda_{LLR} = \mathcal{G}_{\tilde{\chi}_1^0 \tilde{\chi}_n^\pm W^\pm}^L \mathcal{G}_{\tilde{\chi}_\ell^\pm \tilde{\chi}_n^\pm h_i}^L \mathcal{G}_{\tilde{\chi}_1^0 \tilde{\chi}_\ell^\pm W^\pm}^{L*}~,\\
\Lambda_{LRL} = \mathcal{G}_{\tilde{\chi}_1^0 \tilde{\chi}_n^\pm W^\pm}^L \mathcal{G}_{\tilde{\chi}_\ell^\pm \tilde{\chi}_n^\pm h_i}^R \mathcal{G}_{\tilde{\chi}_1^0 \tilde{\chi}_\ell^\pm W^\pm}^{R*}~,\qquad\qquad \Lambda_{LRR} = \mathcal{G}_{\tilde{\chi}_1^0 \tilde{\chi}_n^\pm W^\pm}^L \mathcal{G}_{\tilde{\chi}_\ell^\pm \tilde{\chi}_n^\pm h_i}^R \mathcal{G}_{\tilde{\chi}_1^0 \tilde{\chi}_\ell^\pm W^\pm}^{L*}~,\\
\Lambda_{RLL} = \mathcal{G}_{\tilde{\chi}_1^0 \tilde{\chi}_n^\pm W^\pm}^R \mathcal{G}_{\tilde{\chi}_\ell^\pm \tilde{\chi}_n^\pm h_i}^L \mathcal{G}_{\tilde{\chi}_1^0 \tilde{\chi}_\ell^\pm W^\pm}^{R*}~,\qquad\qquad \Lambda_{RLR} = \mathcal{G}_{\tilde{\chi}_1^0 \tilde{\chi}_n^\pm W^\pm}^R \mathcal{G}_{\tilde{\chi}_\ell^\pm \tilde{\chi}_n^\pm h_i}^L \mathcal{G}_{\tilde{\chi}_1^0 \tilde{\chi}_\ell^\pm W^\pm}^{L*}~,\\
\Lambda_{RRL} = \mathcal{G}_{\tilde{\chi}_1^0 \tilde{\chi}_n^\pm W^\pm}^R \mathcal{G}_{\tilde{\chi}_\ell^\pm \tilde{\chi}_n^\pm h_i}^R \mathcal{G}_{\tilde{\chi}_1^0 \tilde{\chi}_\ell^\pm W^\pm}^{R*}~,\qquad\qquad \Lambda_{RRR} = \mathcal{G}_{\tilde{\chi}_1^0 \tilde{\chi}_n^\pm W^\pm}^R \mathcal{G}_{\tilde{\chi}_\ell^\pm \tilde{\chi}_n^\pm h_i}^R \mathcal{G}_{\tilde{\chi}_1^0 \tilde{\chi}_\ell^\pm W^\pm}^{L*}~,
\end{align}
where $\mathcal{G}_{\tilde{\chi}_\ell^0 \tilde{\chi}_n^\pm W^\pm}^L = g_2 C_{\ell n}^L$\, and \, $\mathcal{G}_{\tilde{\chi}_\ell^0 \tilde{\chi}_n^\pm W^\pm}^R = g_2 C_{\ell n}^R$.\\
\vspace{1cm}

{\bf Topology-(2b):}\\
The respective Feynman diagram is shown in Fig. \ref{fig:topology2}(b).
\begin{center}
\begin{align}
i\delta\Gamma^{(d)} =& -\frac{i}{16\pi^2}\Big[P_L\Bigl\{\eta_{LL}m_{\tilde{\chi}_1^0}\big(d-2\big)\mathbf{C}_2 + \eta_{RL}m_F d \mathbf{C}_0+\eta_{RR}m_{\tilde{\chi}_1^0}\big(d-2\big)\mathbf{C}_1\Bigr\}\nonumber\\
&+P_R\Bigl\{\eta_{LL}m_{\tilde{\chi}_1^0}\big(d-2\big)\mathbf{C}_1 + \eta_{LR}m_F d \mathbf{C}_0+\eta_{RR}m_{\tilde{\chi}_1^0}\big(d-2\big)\mathbf{C}_2\Bigr\}\Big]~,
\end{align}
\end{center}
where $\mathbf{C}_i = \mathbf{C}_i\big(m_{\tilde{\chi}_1^0}^2, q^2, m_{\tilde{\chi}_1^0}^2 ; m_F, m_V, m_{V}\big)$ and 
\begin{center}
\begin{align}
\eta_{LL} = \mathscr{G}_{VVh_i}\mathscr{G}_{\tilde{\chi}_1^0 F V}^L \mathscr{G}_{\tilde{\chi}_1^0 F V}^L~, \qquad\qquad \eta_{LR} = \mathscr{G}_{VVh_i}\mathscr{G}_{\tilde{\chi}_1^0 F V}^L \mathscr{G}_{\tilde{\chi}_1^0 F V}^R~,\\
\eta_{RL} = \mathscr{G}_{VVh_i}\mathscr{G}_{\tilde{\chi}_1^0 F V}^R \mathscr{G}_{\tilde{\chi}_1^0 F V}^L~, \qquad\qquad \eta_{RR} = \mathscr{G}_{VVh_i}\mathscr{G}_{\tilde{\chi}_1^0 F V}^R \mathscr{G}_{\tilde{\chi}_1^0 F V}^R~.
\end{align}
\end{center}
\vspace{0.5cm}
(1) $h_i=h_1/h_2$, $F=\tilde{\chi}_\ell^0$, $V=Z$.
\begin{align}
\eta_{LL} = \mathcal{G}_{ZZh_i}\mathcal{G}_{\tilde{\chi}_1^0 \tilde{\chi}_\ell^0 Z}^L \mathcal{G}_{\tilde{\chi}_1^0 \tilde{\chi}_\ell^0 Z}^{L*}~, \qquad\qquad \eta_{LR} = -\mathcal{G}_{ZZh_i}\mathcal{G}_{\tilde{\chi}_1^0 \tilde{\chi}_\ell^0 Z}^L \mathcal{G}_{\tilde{\chi}_1^0 \tilde{\chi}_\ell^0 Z}^L~,\\
\eta_{RL} = \mathcal{G}_{ZZh_i}\mathcal{G}_{\tilde{\chi}_1^0 \tilde{\chi}_\ell^0 Z}^R \mathcal{G}_{\tilde{\chi}_1^0 \tilde{\chi}_\ell^0 Z}^{L*}~, \qquad\qquad \eta_{RR} = -\mathcal{G}_{ZZh_i}\mathcal{G}_{\tilde{\chi}_1^0 \tilde{\chi}_\ell^0 Z}^R \mathcal{G}_{\tilde{\chi}_1^0 \tilde{\chi}_\ell^0 Z}^L~,
\end{align}
where $\mathcal{G}_{ZZh_i} = g_2M_Z g^{\mu\nu}Y_{h_i}$,\, with\, $Y_{h_i} = \left\{
\begin{array}{ll}
\dfrac{s_{\beta-\alpha}}{c_W} ; & h_i=h_1 \\
\dfrac{c_{\beta-\alpha}}{c_W}; & h_i=h_2 \\
\end{array} 
\right..$ \\
\vspace{0.5cm}\newline
(2) $h_i=h_1/h_2$, $F=\tilde{\chi}_\ell^\pm$, $V=W^\pm$.
\begin{align}
\eta_{LL} = \mathcal{G}_{W^\pm W^\pm h_i}\mathcal{G}_{\tilde{\chi}_1^0 \tilde{\chi}_\ell^\pm W^\pm}^L \mathcal{G}_{\tilde{\chi}_1^0 \tilde{\chi}_\ell^\pm W^\pm}^{R*}~, \qquad\qquad \eta_{LR} = \mathcal{G}_{W^\pm W^\pm h_i}\mathcal{G}_{\tilde{\chi}_1^0 \tilde{\chi}_\ell^\pm W^\pm}^L \mathcal{G}_{\tilde{\chi}_1^0 \tilde{\chi}_\ell^\pm W^\pm}^{L*}~,\\
\eta_{RL} = \mathcal{G}_{W^\pm W^\pm h_i}\mathcal{G}_{\tilde{\chi}_1^0 \tilde{\chi}_\ell^\pm W^\pm}^R \mathcal{G}_{\tilde{\chi}_1^0 \tilde{\chi}_\ell^\pm W^\pm}^{R*}~, \qquad\qquad \eta_{RR} = \mathcal{G}_{W^\pm W^\pm h_i}\mathcal{G}_{\tilde{\chi}_1^0 \tilde{\chi}_\ell^\pm W^\pm}^R \mathcal{G}_{\tilde{\chi}_1^0 \tilde{\chi}_\ell^\pm W^\pm}^{L*}~,
\end{align}
where $\mathcal{G}_{W^\pm W^\pm h_i} = g_2M_W g^{\mu\nu}Y^\prime_{h_i}$,\, with\, $Y^\prime_{h_i} = \left\{
\begin{array}{ll}
s_{\beta-\alpha} ; & h_i=h_1 \\
c_{\beta-\alpha}; & h_i=h_2 \\
\end{array} 
\right..$ \\

\begin{figure}[H]
	\centering
	\includegraphics[width=0.8\linewidth]{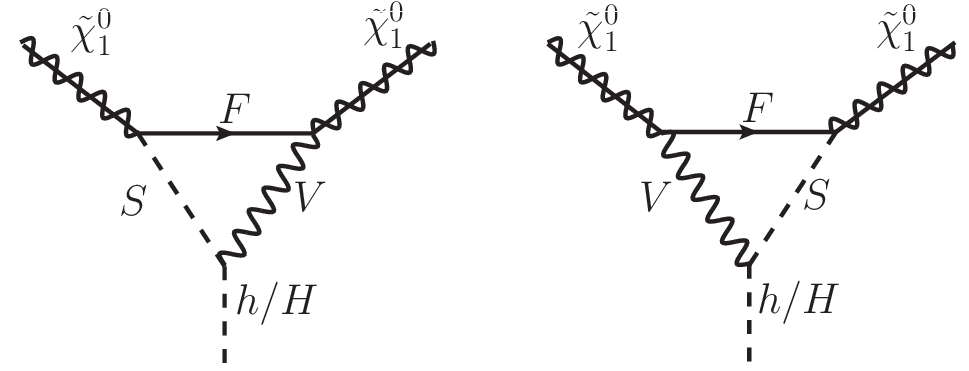}\\
(a)~~~~~~~~~~~~~~~~~~~~~~~~~~~~~~~~~~~~~~~~~~~~~~~~~~~~~~~~~~~(b)
	\caption{Topology 3(a) and 3(b)}
	\label{fig:topology3}
\end{figure}

{\bf Topology-(3a):}\\
The respective Feynman diagram is shown in Fig. \ref{fig:topology3}(a).
\begin{center}
\begin{align}
i\delta\Gamma^{(e)} =& \frac{i}{16\pi^2}\Big[P_L\Bigl\{\psi_{LL}m_{\tilde{\chi}_1^0}m_F\big(\mathbf{C}_2-\mathbf{C}_0\big) + \psi_{LR}m_{\tilde{\chi}_1^0}^2 \big(\mathbf{C}_1+2\mathbf{C}_2\big) + \psi_{RL}\bigl\{-d\mathbf{C}_{00} \nonumber\\
&- m_{\tilde{\chi}_1^0}^2\big(\mathbf{C}_{22}+2\mathbf{C}_{12}+\mathbf{C}_{11}+2\mathbf{C}_1\big)+q^2\mathbf{C}_{12} + \big(2q^2-3m_{\tilde{\chi}_1^0}^2\big)\mathbf{C}_2\bigr\} \nonumber\\
&+ \psi_{RR}m_{\tilde{\chi}_1^0}m_F\big(\mathbf{C}_1+2\mathbf{C}_0\big)\Bigr\} + P_R\Bigl\{\psi_{LL}m_{\tilde{\chi}_1^0}m_F\big(\mathbf{C}_1+2\mathbf{C}_0\big)\nonumber\\
&+\psi_{LR}\bigl\{-d\mathbf{C}_{00} - m_{\tilde{\chi}_1^0}^2\big(\mathbf{C}_{22}+2\mathbf{C}_{12}+\mathbf{C}_{11}+2\mathbf{C}_1\big)+q^2\mathbf{C}_{12} + \big(2q^2-3m_{\tilde{\chi}_1^0}^2\big)\mathbf{C}_2\bigr\}\nonumber\\
& + \psi_{RL}m_{\tilde{\chi}_1^0}^2\big(\mathbf{C}_1+2\mathbf{C}_2\big) + \psi_{RR}m_{\tilde{\chi}_1^0}m_F\big(\mathbf{C}_2-\mathbf{C}_0\big)\Bigr\}\Big]~,
\end{align}
\end{center}
where $\mathbf{C}_i = \mathbf{C}_i\big(m_{\tilde{\chi}_1^0}^2, q^2, m_{\tilde{\chi}_1^0}^2 ; m_F, m_S, m_V\big)$, $\mathbf{C}_{ij} = \mathbf{C}_{ij}\big(m_{\tilde{\chi}_1^0}^2, q^2, m_{\tilde{\chi}_1^0}^2 ; m_F, m_S, m_V\big)$ and 
\begin{center}
\begin{align}
\psi_{LL} = \mathscr{G}_{h_i SV}\mathscr{G}_{\tilde{\chi}_1^0FV}^L \mathscr{G}_{\tilde{\chi}_1^0FS}^L~,\qquad\qquad \psi_{LR} = \mathscr{G}_{h_i SV}\mathscr{G}_{\tilde{\chi}_1^0FV}^L \mathscr{G}_{\tilde{\chi}_1^0FS}^R~, \\
\psi_{RL} = \mathscr{G}_{h_i SV}\mathscr{G}_{\tilde{\chi}_1^0FV}^R \mathscr{G}_{\tilde{\chi}_1^0FS}^L~,\qquad\qquad \psi_{RR} = \mathscr{G}_{h_i SV}\mathscr{G}_{\tilde{\chi}_1^0FV}^R \mathscr{G}_{\tilde{\chi}_1^0FS}^R~.
\end{align}
\end{center}
\vspace{0.5cm}
(1) $h_i=h_1/h_2$, $F=\tilde{\chi}_\ell^0$, $S=A$, $V=Z$.
\begin{align}
\psi_{LL} = \mathcal{G}_{h_i AZ}\mathcal{G}_{\tilde{\chi}_1^0\tilde{\chi}_\ell^0 Z}^L \mathcal{G}_{\tilde{\chi}_1^0\tilde{\chi}_\ell^0 A}^{R*}~,\qquad\qquad \psi_{LR} = \mathcal{G}_{h_i AZ}\mathcal{G}_{\tilde{\chi}_1^0\tilde{\chi}_\ell^0 Z}^L \mathcal{G}_{\tilde{\chi}_1^0\tilde{\chi}_\ell^0 A}^{L*}~, \\
\psi_{RL} = \mathcal{G}_{h_i AZ}\mathcal{G}_{\tilde{\chi}_1^0\tilde{\chi}_\ell^0 Z}^R \mathcal{G}_{\tilde{\chi}_1^0\tilde{\chi}_\ell^0 A}^{R*}~,\qquad\qquad \psi_{RR} = \mathcal{G}_{h_i AZ}\mathcal{G}_{\tilde{\chi}_1^0\tilde{\chi}_\ell^0 Z}^R \mathcal{G}_{\tilde{\chi}_1^0\tilde{\chi}_\ell^0 A}^{L*}~,
\end{align}
where $\mathcal{G}_{h_i AZ} = \dfrac{g_2}{2c_W}Y^{\prime\prime}_{h_i}$,\, with\, $Y^{\prime\prime}_{h_i} = \left\{
\begin{array}{ll}
\,\,\,\,c_{\beta-\alpha} ; & h_i=h_1 \\
-s_{\beta-\alpha}; & h_i=h_2 \\
\end{array} 
\right..$\\
\vspace{0.5cm}\newline
(2) $h_i=h_1/h_2$, $F=\tilde{\chi}_\ell^0$, $S=G$, $V=Z$.
\begin{align}
\psi_{LL} = \mathcal{G}_{h_i GZ}\mathcal{G}_{\tilde{\chi}_1^0\tilde{\chi}_\ell^0 Z}^L \mathcal{G}_{\tilde{\chi}_1^0\tilde{\chi}_\ell^0 G}^{R*}~,\qquad\qquad \psi_{LR} = \mathcal{G}_{h_i GZ}\mathcal{G}_{\tilde{\chi}_1^0\tilde{\chi}_\ell^0 Z}^L \mathcal{G}_{\tilde{\chi}_1^0\tilde{\chi}_\ell^0 G}^{L*}~, \\
\psi_{RL} = \mathcal{G}_{h_i GZ}\mathcal{G}_{\tilde{\chi}_1^0\tilde{\chi}_\ell^0 Z}^R \mathcal{G}_{\tilde{\chi}_1^0\tilde{\chi}_\ell^0 G}^{R*}~,\qquad\qquad \psi_{RR} = \mathcal{G}_{h_i GZ}\mathcal{G}_{\tilde{\chi}_1^0\tilde{\chi}_\ell^0 Z}^R \mathcal{G}_{\tilde{\chi}_1^0\tilde{\chi}_\ell^0 G}^{L*}~,
\end{align}
where $\mathcal{G}_{h_i GZ} = \left\{
\begin{array}{ll}
\dfrac{g_2}{2c_W}s_{\beta-\alpha} ; & h_i=h_1 \\
\dfrac{g_2}{2c_W}c_{\beta-\alpha}; & h_i=h_2 \\
\end{array} 
\right..$\\
\vspace{0.5cm}\newline
(3) $h_i=h_1/h_2$, $F=\tilde{\chi}_\ell^\pm$, $S=H^\pm$, $V=W^\pm$.
\begin{align}
\psi_{LL} = \mathcal{G}_{h_i H^\pm W^\pm}\mathcal{G}_{\tilde{\chi}_1^0\tilde{\chi}_\ell^\pm W^\pm}^L \mathcal{G}_{\tilde{\chi}_1^0\tilde{\chi}_\ell^\pm H^\pm
}^{R*}~,\qquad\qquad \psi_{LR} = \mathcal{G}_{h_i H^\pm W^\pm}\mathcal{G}_{\tilde{\chi}_1^0\tilde{\chi}_\ell^\pm W^\pm}^L \mathcal{G}_{\tilde{\chi}_1^0\tilde{\chi}_\ell^\pm H^\pm}^{L*}~, \\
\psi_{RL} = \mathcal{G}_{h_i H^\pm W^\pm}\mathcal{G}_{\tilde{\chi}_1^0\tilde{\chi}_\ell^\pm W^\pm}^R \mathcal{G}_{\tilde{\chi}_1^0\tilde{\chi}_\ell^\pm H^\pm}^{R*}~,\qquad\qquad \psi_{RR} = \mathcal{G}_{h_i H^\pm W^\pm}\mathcal{G}_{\tilde{\chi}_1^0\tilde{\chi}_\ell^\pm W^\pm}^R \mathcal{G}_{\tilde{\chi}_1^0\tilde{\chi}_\ell^\pm H^\pm}^{L*}~,
\end{align}
where $\mathcal{G}_{h_i H^\pm W^\pm} = \dfrac{g_2}{2}Y^{\prime\prime}_{h_i}$.\\
\vspace{0.5cm}\newline
(4) $h_i=h_1/h_2$, $F=\tilde{\chi}_\ell^\pm$, $S=G^\pm$, $V=W^\pm$.
\begin{align}
\psi_{LL} = \mathcal{G}_{h_i G^\pm W^\pm}\mathcal{G}_{\tilde{\chi}_1^0\tilde{\chi}_\ell^\pm W^\pm}^L \mathcal{G}_{\tilde{\chi}_1^0\tilde{\chi}_\ell^\pm G^\pm
}^{R*}~,\qquad\qquad \psi_{LR} = \mathcal{G}_{h_i G^\pm W^\pm}\mathcal{G}_{\tilde{\chi}_1^0\tilde{\chi}_\ell^\pm W^\pm}^L \mathcal{G}_{\tilde{\chi}_1^0\tilde{\chi}_\ell^\pm G^\pm}^{L*}~, \\
\psi_{RL} = \mathcal{G}_{h_i G^\pm W^\pm}\mathcal{G}_{\tilde{\chi}_1^0\tilde{\chi}_\ell^\pm W^\pm}^R \mathcal{G}_{\tilde{\chi}_1^0\tilde{\chi}_\ell^\pm G^\pm}^{R*}~,\qquad\qquad \psi_{RR} = \mathcal{G}_{h_i G^\pm W^\pm}\mathcal{G}_{\tilde{\chi}_1^0\tilde{\chi}_\ell^\pm W^\pm}^R \mathcal{G}_{\tilde{\chi}_1^0\tilde{\chi}_\ell^\pm G^\pm}^{L*}~,
\end{align}
where $\mathcal{G}_{h_i G^\pm W^\pm} =  \left\{
\begin{array}{ll}
-\dfrac{g_2}{2}s_{\beta-\alpha} ; & h_i=h_1 \\
-\dfrac{g_2}{2}c_{\beta-\alpha}; & h_i=h_2 \\
\end{array} 
\right..$\\
\vspace{1cm}

{\bf Topology-(3b):}\\
The respective Feynman diagram is shown in Fig. \ref{fig:topology3}(b).
\begin{center}
\begin{align}
i\delta\Gamma^{(f)} =& \frac{i}{16\pi^2}\Big[P_L\Bigl\{\Xi_{LL}\bigl\{d\mathbf{C}_{00} + m_{\tilde{\chi}_1^0}^2\big(\mathbf{C}_{22}+2\mathbf{C}_{12}+\mathbf{C}_{11}+2\mathbf{C}_2+3\mathbf{C}_1\big)-q^2\big(\mathbf{C}_{12}\nonumber\\
&+2\mathbf{C}_1\big)\bigr\}+\Xi_{LR} m_{\tilde{\chi}_1^0}m_F\big(\mathbf{C}_0-\mathbf{C}_1\big)-\Xi_{RL}m_{\tilde{\chi}_1^0} m_F\big(\mathbf{C}_2+2\mathbf{C}_0\big)-\Xi_{RR} m_{\tilde{\chi}_1^0}^2\big(\mathbf{C}_2\nonumber\\
&+2\mathbf{C}_1\big)\Bigr\} + P_R\Bigl\{-\Xi_{LL}m_{\tilde{\chi}_1^0}^2 \big(\mathbf{C}_2+2\mathbf{C}_1\big)-\Xi_{LR}m_{\tilde{\chi}_1^0}m_F\big(\mathbf{C}_2+2\mathbf{C}_0\big) \nonumber\\
&+ \Xi_{RL} m_{\tilde{\chi}_1^0}m_F\big(\mathbf{C}_0-\mathbf{C}_1\big)+ \Xi_{RR}\bigl\{d\mathbf{C}_{00} + m_{\tilde{\chi}_1^0}^2\big(\mathbf{C}_{22} \nonumber\\
&+2\mathbf{C}_{12}+\mathbf{C}_{11}+2\mathbf{C}_2+3\mathbf{C}_1\big)-q^2\big(\mathbf{C}_{12}+2\mathbf{C}_1\big)\bigr\}\Bigr\}\Big],
\end{align}
\end{center}
where $\mathbf{C}_i = \mathbf{C}_i\big(m_{\tilde{\chi}_1^0}^2, q^2, m_{\tilde{\chi}_1^0}^2 ; m_F, m_V, m_S\big)$, $\mathbf{C}_{ij} = \mathbf{C}_{ij}\big(m_{\tilde{\chi}_1^0}^2, q^2, m_{\tilde{\chi}_1^0}^2 ; m_F, m_V, m_S\big)$ and 
\begin{center}
\begin{align}
\Xi_{LL} = \mathscr{G}_{h_i SV}\mathscr{G}_{\tilde{\chi}_1^0FS}^L \mathscr{G}_{\tilde{\chi}_1^0FV}^L, \qquad\qquad \Xi_{LR} = \mathscr{G}_{h_i SV}\mathscr{G}_{\tilde{\chi}_1^0FS}^L \mathscr{G}_{\tilde{\chi}_1^0FV}^R,\\
\Xi_{RL} = \mathscr{G}_{h_i SV}\mathscr{G}_{\tilde{\chi}_1^0FS}^R \mathscr{G}_{\tilde{\chi}_1^0FV}^L, \qquad\qquad \Xi_{RR} = \mathscr{G}_{h_i SV}\mathscr{G}_{\tilde{\chi}_1^0FS}^R \mathscr{G}_{\tilde{\chi}_1^0FV}^R.
\end{align}
\end{center}
\vspace{0.5cm}
(1) $h_i=h_1/h_2$, $F=\tilde{\chi}_\ell^0$, $S=A$, $V=Z$.
\begin{align}
\Xi_{LL} = \mathcal{G}_{h_i AZ}\mathcal{G}_{\tilde{\chi}_1^0 \tilde{\chi}_\ell^0 A}^L \mathcal{G}_{\tilde{\chi}_1^0 \tilde{\chi}_\ell^0 Z}^{L*}~, \qquad\qquad \Xi_{LR} = -\mathcal{G}_{h_i AZ}\mathcal{G}_{\tilde{\chi}_1^0 \tilde{\chi}_\ell^0 A}^L \mathcal{G}_{\tilde{\chi}_1^0 \tilde{\chi}_\ell^0 Z}^{L}~,\\
\Xi_{RL} = \mathcal{G}_{h_i AZ}\mathcal{G}_{\tilde{\chi}_1^0 \tilde{\chi}_\ell^0 A}^R \mathcal{G}_{\tilde{\chi}_1^0 \tilde{\chi}_\ell^0 Z}^{L*}~, \qquad\qquad \Xi_{RR} = -\mathcal{G}_{h_i AZ}\mathcal{G}_{\tilde{\chi}_1^0 \tilde{\chi}_\ell^0 A}^R \mathcal{G}_{\tilde{\chi}_1^0 \tilde{\chi}_\ell^0 Z}^{L}~.
\end{align}
\vspace{0.5cm}\newline
(2) $h_i=h_1/h_2$, $F=\tilde{\chi}_\ell^0$, $S=G$, $V=Z$.
\begin{align}
\Xi_{LL} = \mathcal{G}_{h_i GZ}\mathcal{G}_{\tilde{\chi}_1^0 \tilde{\chi}_\ell^0 G}^L \mathcal{G}_{\tilde{\chi}_1^0 \tilde{\chi}_\ell^0 Z}^{L*}~, \qquad\qquad \Xi_{LR} = -\mathcal{G}_{h_i GZ}\mathcal{G}_{\tilde{\chi}_1^0 \tilde{\chi}_\ell^0 G}^L \mathcal{G}_{\tilde{\chi}_1^0 \tilde{\chi}_\ell^0 Z}^{L}~,\\
\Xi_{RL} = \mathcal{G}_{h_i GZ}\mathcal{G}_{\tilde{\chi}_1^0 \tilde{\chi}_\ell^0 G}^R \mathcal{G}_{\tilde{\chi}_1^0 \tilde{\chi}_\ell^0 Z}^{L*}~, \qquad\qquad \Xi_{RR} = -\mathcal{G}_{h_i GZ}\mathcal{G}_{\tilde{\chi}_1^0 \tilde{\chi}_\ell^0 G}^R \mathcal{G}_{\tilde{\chi}_1^0 \tilde{\chi}_\ell^0 Z}^{L}~.
\end{align}
\vspace{0.5cm}\newline
(3) $h_i=h_1/h_2$, $F=\tilde{\chi}_\ell^\pm$, $S=H^\pm$, $V=W^\pm$.
\begin{align}
\Xi_{LL} = \mathcal{G}_{h_i H^\pm W^\pm}\mathcal{G}_{\tilde{\chi}_1^0\tilde{\chi}_\ell^\pm H^\pm}^L \mathcal{G}_{\tilde{\chi}_1^0\tilde{\chi}_\ell^\pm W^\pm}^{R*}~, \qquad\qquad \Xi_{LR} = \mathcal{G}_{h_i H^\pm W^\pm}\mathcal{G}_{\tilde{\chi}_1^0\tilde{\chi}_\ell^\pm H^\pm}^L \mathcal{G}_{\tilde{\chi}_1^0\tilde{\chi}_\ell^\pm W^\pm}^{L*}~,\\
\Xi_{RL} = \mathcal{G}_{h_i H^\pm W^\pm}\mathcal{G}_{\tilde{\chi}_1^0\tilde{\chi}_\ell^\pm H^\pm}^R \mathcal{G}_{\tilde{\chi}_1^0\tilde{\chi}_\ell^\pm W^\pm}^{R*}~, \qquad\qquad \Xi_{RR} = \mathcal{G}_{h_i H^\pm W^\pm}\mathcal{G}_{\tilde{\chi}_1^0\tilde{\chi}_\ell^\pm H^\pm}^R \mathcal{G}_{\tilde{\chi}_1^0\tilde{\chi}_\ell^\pm W^\pm}^{L*}~.
\end{align}
\vspace{0.5cm}\newline
(4) $h_i=h_1/h_2$, $F=\tilde{\chi}_\ell^\pm$, $S=G^\pm$, $V=W^\pm$.
\begin{align}
\Xi_{LL} = \mathcal{G}_{h_i G^\pm W^\pm}\mathcal{G}_{\tilde{\chi}_1^0\tilde{\chi}_\ell^\pm G^\pm}^L \mathcal{G}_{\tilde{\chi}_1^0\tilde{\chi}_\ell^\pm W^\pm}^{R*}~, \qquad\qquad \Xi_{LR} = \mathcal{G}_{h_i G^\pm W^\pm}\mathcal{G}_{\tilde{\chi}_1^0\tilde{\chi}_\ell^\pm G^\pm}^L \mathcal{G}_{\tilde{\chi}_1^0\tilde{\chi}_\ell^\pm W^\pm}^{L*}~,\\
\Xi_{RL} = \mathcal{G}_{h_i G^\pm W^\pm}\mathcal{G}_{\tilde{\chi}_1^0\tilde{\chi}_\ell^\pm G^\pm}^R \mathcal{G}_{\tilde{\chi}_1^0\tilde{\chi}_\ell^\pm W^\pm}^{R*}~, \qquad\qquad \Xi_{RR} = \mathcal{G}_{h_i G^\pm W^\pm}\mathcal{G}_{\tilde{\chi}_1^0\tilde{\chi}_\ell^\pm G^\pm}^R \mathcal{G}_{\tilde{\chi}_1^0\tilde{\chi}_\ell^\pm W^\pm}^{L*}~.
\end{align}

In the above, we have used the following \cite{Drees:2004jm}:

\begin{align*}
    C_{\ell k}^{L} &= {N}_{\ell 2} {V}_{k1}^{*} -\frac{1}{\sqrt{2}} {N}_{\ell 4} {V}_{k2}^{*}~,
    \\
    C_{\ell k}^{R} &= {N}_{\ell 2}^{*} {U}_{k1} +\frac{1}{\sqrt{2}} {N}_{\ell 3}^{*} {U}_{k2}~,
    \\
    \mathcal{N}_{\ell n}^{L} &= \frac{1}{2}\big(-{N}_{\ell 3} {N}_{n 3}^{*} + {N}_{\ell 4} {N}_{n 4}^{*} \big)~,
    \\
    \mathcal{N}_{\ell n}^{R} &= - \big(\mathcal{N}_{\ell n}^{L}\big)^{*}~,
    \\
    Q_{k\ell} &= \frac{1}{2} {V}_{k1} {U}_{\ell 2}~,
    \\
    S_{k\ell} &= \frac{1}{2} {V}_{k2} {U}_{\ell 1}~,
    \\
    Q_{\ell k}^{\prime L} &= c_\beta\Big[{N}_{\ell 4}^{*} {V}^{*}_{k1}+\frac{1}{\sqrt{2}} {V}^{*}_{k2}\big({N}^{*}_{\ell 2} + t_{W} {N}^{*}_{\ell 1}\big)\Big]~,
    \\
    Q_{\ell k}^{\prime R} &= s_\beta\Big[{N}_{\ell 3}{U}_{k1}-\frac{1}{\sqrt{2}} {U}_{k2}\big({N}_{\ell 2} + t_{W} {N}_{\ell 1}\big)\Big]~,
    \\
    Q_{n\ell}^{\prime\prime} &= \frac{1}{2}\big[{N}_{n3}\big({N}_{\ell 2}-t_{W}{N}_{\ell 1}\big) + {N}_{\ell 3}\big({N}_{n 2}-t_{W} {N}_{n 1}\big)\big]~,
    \\
    S_{n\ell}^{\prime\prime} &= \frac{1}{2}\big[{N}_{n4}\big({N}_{\ell 2}-t_{W}{N}_{\ell 1}\big) + {N}_{\ell 4}\big({N}_{n 2}-t_{W}{N}_{n 1}\big)\big]~.
    \\
\end{align*}

\renewcommand{\theequation}{C.\arabic{equation}}
\setcounter{equation}{0}
\section*{Appendix C}
The counterterm Lagrangian for the $\tilde{\chi}_1^0-\tilde{\chi}_1^0-h_i$ 
interaction ($\mathscr{L}_{CT}$) is given as follows:  
\beq
\mathscr{L}_{CT}   \supset      -\dfrac{1}{2} h_1 \bar{\tilde{\chi}}_1^0 (\delta \mathscr{C}_1^R P_R 
+ \delta \mathscr{C}_1^L P_L) \tilde{\chi}_1^0  -  \dfrac{1}{2} h_2 \bar{\tilde{\chi}}_1^0 (\delta 
\mathscr{C}_2^R P_R + \delta \mathscr{C}_2^L P_L) \tilde{\chi}_1^0,\ ,  \nonumber \\
\eeq
where $ \delta \mathscr{C}_i^L =  \delta \mathscr{C}_i^{R*}$ for $i \in \{1,2\}$. 

In the above expression,  $\delta \mathscr{C}_1^R$ is given by, 
\beq
\delta \mathscr{C}_1^R =-\frac{e}{4 s_{\mathrm{W}}^2} \left(\frac{2}{c_{\mathrm{W}}^3}A_1+ B_1 ~\frac{s_{\mathrm{W}}}{c_{\mathrm{W}}}\right) \, . 
\eeq
where $A_1$ and $B_1$ are given by, 
\begin{eqnarray} 
A_1   & =   & - 2 \left(s_\alpha N_{13}+c_\alpha \mathrm{N}_{14}\right)\left((\delta \mathrm{Z}_{\mathrm{e}} s_{\mathrm{W}}-\delta s_{\mathrm{W}}) \mathrm{N}_{12} c_{\mathrm{W}}^3-
\mathrm{N}_{11}(\delta s_{\mathrm{W}} s_{\mathrm{W}}+\delta \mathrm{Z}_{\mathrm{e}} 
c_{\mathrm{W}}^2) s_{\mathrm{W}}^2\right) ,\ \nonumber\\
 B_1  & =   &\sum_{i=1}^4 \left(\left(s_\alpha N_{i3}+c_\alpha N_{i4}\right)(s_{\mathrm{W}} N_{11}-
 c_{\mathrm{W}} N_{12}) + (s_{\mathrm{W}} N_{i1}-c_{\mathrm{W}} N_{i2}) 
 (s_\alpha N_{13}+c_\alpha N_{14})\right). \nonumber \\
 &   & \left(\delta Z_{1i}^{*} + \delta Z_{i1}^{*}\right) + 2 \left((\delta Z^{\mathrm{H}}_{11} s_\alpha- \delta Z^{\mathrm{H}}_{12} c_\alpha) N_{13}+
 (\delta Z^{\mathrm{H}}_{11} c_\alpha+ \delta Z^{\mathrm{H}}_{12} s_\alpha) N_{14} \right).\nonumber \\
 &  & (s_{\mathrm{W}} N_{11}-c_{\mathrm{W}} N_{12}).
\end{eqnarray}
In these expressions $s_{\mathrm{W}} = \sin\theta_{\mathrm{W}} $, 
$c_{\mathrm{W}} = \cos\theta_{\mathrm{W}}$ where $\theta_{\mathrm{W}}$ is the 
Weinberg angle, $\delta s_{\mathrm{W}}$ and $\delta c_{\mathrm{W}}$ denote 
the respective counterterms. Further, $c_\alpha = \cos\alpha$, $s_\alpha = 
\sin\alpha$; where $\alpha$ is the mixing angle in the CP-even Higgs boson sector,  
$\delta \mathrm{Z}_{\mathrm{e}}$ denotes the counterterm corresponding to 
the charge $e$. 

We use the following abbreviations in this section: 
$\widetilde{\mathrm{Re}}$ takes the real part of loop integrals but does not affect the complex couplings. For the notations, we have closely followed \cite{Fritzsche:2013fta} and \cite{Fritzsche:2002bi}. The relevant counter-terms have been listed below.

For the Higgs sector, the following counterterms are relevant \cite{Fritzsche:2013fta}:
$$
\begin{aligned}
\delta Z_{{H}_1} & =-\left.\operatorname{Re} \Sigma_{h_2}^{\prime}(0)\right|_{\alpha=0, \mathrm{div}}, \\
\delta Z_{{H}_2} & =-\left.\operatorname{Re} \Sigma_{h_1}^{\prime}(0)\right|_{\alpha=0, \mathrm{div}}, \\
\delta t_\beta & =\frac{1}{2} t_\beta\left(\delta Z_{{H}_2}-\delta Z_{{H}_1}\right).
\end{aligned}
$$

In the gauge boson sector, the relevant counter-terms are as follows \cite{Fritzsche:2013fta}:

$$
\begin{aligned}
\delta M_Z^2 & =\widetilde{\operatorname{Re}} \Sigma_Z^T\left(M_Z^2\right), \\
\delta M_W^2 & =\widetilde{\operatorname{Re}} \Sigma_W^T\left(M_W^2\right),\\
\delta Z_{\gamma \gamma} & =-\widetilde{\operatorname{Re}} \Sigma_\gamma^{\prime T}(0),\\
\delta Z_{Z \gamma} & =\frac{2}{M_Z^2} \widetilde{\operatorname{Re}} \Sigma_{\gamma Z}^T(0),\\
\delta s_{\mathrm{w}}& =\frac{1}{2} \frac{c_{\mathrm{w}}^2}{s_{\mathrm{w}}}\left(\frac{\delta M_Z^2}{M_Z^2}-\frac{\delta M_W^2}{M_W^2}\right), \\
\delta Z_e & =\frac{1}{2}\left(\frac{s_{\mathrm{w}}}{c_{\mathrm{w}}} \delta Z_{Z \gamma}-\delta Z_{\gamma \gamma}\right) .
\end{aligned}
$$

The counterterms to the gaugino and higgsino mass parameters are determined from the 
chargino-neutralino sector. In the In the CCN(n) scheme, these are given by 
\cite{Fritzsche:2002bi,Baro2009,Fritzsche:2013fta,Chatterjee:2011wc}:  

\begin{center}
\begin{align}
\delta M_1= & \frac{1}{N_{n 1}^{* 2}}\left\{\delta m_{\tilde{\chi}_n^0}^{\mathrm{OS}}+\delta N_n-N_{n 2}^{* 2} \delta M_2+2 N_{n 3}^* N_{n 4}^* \delta \mu\right\},\\
\delta M_2= & \frac{1}{k_o r_o-k_d r_d}\biggl\{U_{12}^* V_{12}^* \delta m_{\tilde{\chi}_2^{ \pm}}^{\mathrm{OS}}-U_{22}^* V_{22}^* \delta m_{\tilde{\chi}_1^{ \pm}}^{\mathrm{OS}}+ \nonumber\\
& \sqrt{2}\left(c_\beta\left(k_d-k_o\right) V_{12}^* V_{22}^*+s_\beta\left(r_o-r_d\right) U_{12}^* U_{22}^*\right) M_W c_\beta^2 \delta t_\beta+ \nonumber\\
& \left(s_{\beta}\left(k_d - k_o\right) V_{12}^* V_{22}^*-c_\beta\left(r_o-r_d\right) U_{12}^* U_{22}^*\right) \frac{\delta M_W^2}{\sqrt{2} M_W}\biggr\}, 
\end{align}
\end{center}
\begin{center}
\begin{align}
\delta \mu= & \frac{1}{k_o r_o-k_d r_d}\biggl\{U_{21}^* V_{21}^* \delta m_{\tilde{\chi}_1^{ \pm}}^{\mathrm{OS}}-U_{11}^* V_{11}^* \delta m_{\tilde{\chi}_2^{ \pm}}^{\mathrm{OS}}- \nonumber\\
& \sqrt{2}\left(s_\beta\left(k_d-k_o\right) V_{11}^* V_{21}^*+c_\beta\left(r_o-r_d\right) U_{11}^* U_{21}^*\right) M_W c_\beta^2 \delta t_\beta+ \nonumber\\
& \left(c_\beta\left(k_d-k_o\right) V_{11}^* V_{21}^*-s_\beta\left(r_o-r_d\right) U_{11}^* U_{21}^*\right) \frac{\delta M_W^2}{\sqrt{2} M_W} \biggr\},
\end{align}
\end{center}
where we have used the following notations 
$$
k_d=U_{11}^* U_{22}^*, \quad k_o=U_{12}^* U_{21}^*, \quad r_d=V_{11}^* V_{22}^*, \quad r_o=V_{12}^* V_{21}^*,
$$
and,
\begin{center}
\begin{align}
\delta N_n= & 2 c_\beta^2 \delta t_\beta\left(s_\beta N_{n 3}^*+c_\beta N_{n 4}^*\right)\left(M_W N_{n 2}^*-M_Z s_{\mathrm{w}} N_{n 1}^*\right)\nonumber\\
 & + \left(c_\beta N_{n 3}^*-s_\beta N_{n 4}^*\right)\left(N_{n 1}^*\left[\frac{\delta M_Z^2}{M_Z} s_{\mathrm{w}}+2 M_Z \delta s_{\mathrm{w}}\right]-N_{n 2}^* \frac{\delta M_W^2}{M_W}\right), \nonumber\\
  \delta m_{\tilde{\chi}_n^0}^{\mathrm{OS}}= & \widetilde{\operatorname{Re}}\left[m_{\tilde{\chi}_n^0} \Sigma_{\tilde{\chi}^0}^L\left(m_{\tilde{\chi}_n^0}^2\right)+\Sigma_{\tilde{\chi}^0}^{S L}\left(m_{\tilde{\chi}_n^0}^2\right)\right]_{n n},\nonumber\\  
\delta m_{\tilde{\chi}_c^{ \pm}}^{\mathrm{OS}}=& \widetilde{\operatorname{Re}}\left[\frac{m_{\tilde{\chi}_c^{ \pm}}}{2}\left(\Sigma_{\tilde{\chi}^{-}}^L\left(m_{\tilde{\chi}_c^{ \pm}}^2\right)+\Sigma_{\tilde{\chi}^{-}}^R\left(m_{\tilde{\chi}_c^{ \pm}}^2\right)\right)+\Sigma_{\tilde{\chi}^{-}}^{S L}\left(m_{\tilde{\chi}_c^{ \pm}}^2\right)\right]_{c c} .
\end{align}
\end{center}

We have used $n=4$ for BP1-4 and $n=3$ for BP5-6.
A more thorough analysis can be found in \cite{Fritzsche:2002bi} and \cite{Fritzsche:2013fta}.\\
We have used an on-shell renormalization scheme, which has been implemented in the \texttt{Formcalc} 
\cite{Fritzsche:2013fta}, to evaluate these counterterms for the benchmark scenarios. 

Next,  $\delta \mathscr{C}_2^R$ is given by,  
\beq
\delta \mathscr{C}_2^R=\frac{e}{4 s_{\mathrm{W}}^2}\left(\frac{2}{c_{\mathrm{W}}^3}A_2+B_2 
\frac{s_{\mathrm{W}}}{c_{\mathrm{W}}} \right). 
\eeq
In the above expression $A_2$ and $B_2$ are given by, 
\begin{eqnarray} 
 A_2  &= & - 2 \left(c_\alpha Z_{13}-s_\alpha Z_{14}\right)\left(\left(\delta Z_{\mathrm{e}} 
s_{\mathrm{W}}-\delta s_{\mathrm{W}}\right)Z_{12} c_{\mathrm{W}}^3-Z_{11}\left(\delta 
s_{\mathrm{W}} s_{\mathrm{W}}+\delta Z_{\mathrm{e}} c_{\mathrm{W}}^2\right) s_{\mathrm{W}}^2\right)\nonumber\\
B_2 & = & \sum_{i=1}^4\left[\left(c_\alpha N_{i 3}-s_\alpha N_{i 4}\right)\left(s_W N_{11}-c_W N_{12}
\right)+\left(s_W N_{i 1}-c_W N_{i 2}\right)\left(c_\alpha N_{13}-s_\alpha N_{14}\right)\right]. \nonumber \\
& & \left(\delta \bar{Z}_{i 1}+\delta Z_{i1}^{*}\right) +2 (\delta Z_{22}^{H}-\delta Z_{12}^{H})\left[s_\alpha (N_{13}-N_{14})+c_\alpha (N_{13}+ N_{14})
\right]\left(s_W N_{11}-c_W N_{12}\right). \nonumber\\
\end{eqnarray}
The wavefunction renormalization counterterms for the neutralino sector and the CP-even Higgs 
sector have been determined using the on-shell renormalization scheme \cite{Fritzsche:2002bi, 
Frank:2006yh}, following the implementation in \texttt{FormCalc}\cite{Fritzsche:2013fta}. For 
different benchmark scenarios, variants of the on-shell renormalization scheme have been 
adopted, a detailed discussion may be found in \cite{Chatterjee:2011wc}.

\bigskip
\bibliographystyle{JHEPCust.bst}
\bibliography{Higgsino}

\end{document}